\pdfoutput=1  

\documentclass[aps,prd,superscriptaddress,floatfix,nofootinbib,preprintnumbers,eqsecnum,twocolumn]{revtex4-1}

\usepackage{amsmath,float,graphicx,placeins,amssymb,multirow,pgfplots,pgfplotstable}
\usepackage{empheq}
\usepackage{tikz}
\usepackage{comment}
\usepackage{enumerate,slashed,cancel,comment,color,bbm,graphicx,rotating,placeins,mathtools,multirow,hhline}
\usepackage[normalem]{ulem}
\usepackage{array}

\usepackage{hyperref}
\usepackage{color}
 \hypersetup
 {
   colorlinks,
   linkcolor={blue!80!black},
   citecolor={blue!70!black},
   urlcolor={blue!70!black}
 }

\usepackage{lipsum}
\usepackage{diagbox}
\usetikzlibrary{shapes.geometric,arrows}
\graphicspath{}

\def\beq{\begin{equation}}
\def\eeq{\end{equation}}
\def\beqn{\begin{eqnarray}}
\def\eeqn{\end{eqnarray}}
\def\half{\mbox{\small ${\frac{1}{2}}$}}

\def\quarter{\mbox{\small ${\frac{1}{4}}$}}

\newcommand{\newc}{\newcommand}
\def\calZ{{\cal Z}}
\def\calM{{\cal M}}
\def\calV{{\cal V}}
\def\calF{{\cal F}}

\def\bQ{{\bf Q}}
\def\bT{{\bf T}}

\def\Qs{{\bf q}}

\def\barOmega{{\overline{\Omega}}}

\def\half{{\textstyle{1\over 2}}}
\def\quarter{{\textstyle{1\over 4}}}
\def\ie{{\it i.e.}\/}
\def\eg{{\it e.g.}\/}
\def\etc{{\it etc}.\/}
\def\inbar{\,\vrule height1.5ex width.4pt depth0pt}
\def\IR{\relax{\rm I\kern-.18em R}}
 \font\cmss=cmss10 \font\cmsss=cmss10 at 7pt
\def\IQ{\relax{\rm I\kern-.18em Q}}
\def\IZ{\relax\ifmmode\mathchoice
 {\hbox{\cmss Z\kern-.4em Z}}{\hbox{\cmss Z\kern-.4em Z}}
 {\lower.9pt\hbox{\cmsss Z\kern-.4em Z}}
 {\lower1.2pt\hbox{\cmsss Z\kern-.4em Z}}\else{\cmss Z\kern-.4em Z}\fi}
\begin{document}

\title{Stasis in an Expanding Universe:\\   A Recipe for Stable Mixed-Component Cosmological Eras }

\def\andname{\hspace*{-0.5em}} 

\author{Keith R. Dienes}
\email[Email address: ]{dienes@arizona.edu}
\affiliation{Department of Physics, University of Arizona, Tucson, AZ 85721 USA}
\affiliation{Department of Physics, University of Maryland, College Park, MD 20742 USA}
\author{Lucien Heurtier}
\email[Email address: ]{lucien.heurtier@durham.ac.uk}
\affiliation{IPPP, Durham University, Durham, DH1 3LE, United Kingdom}
\author{Fei Huang}
\email[Email address: ]{huangf4@uci.edu}
\affiliation{CAS Key Laboratory of Theoretical Physics, Institute of Theoretical Physics,\\ 
Chinese Academy of Sciences, Beijing 100190, China}
\affiliation{Department of Physics and Astronomy, University of California, Irvine, CA  92697  USA}
\author{Doojin Kim}
\email[Email address: ]{doojin.kim@tamu.edu}
\affiliation{Mitchell Institute for Fundamental Physics and Astronomy,\\
Department of Physics and Astronomy, Texas A\&M University, College Station, TX  77843 USA}
\author{\\ Tim M.P. Tait}
\email[Email address: ]{ttait@uci.edu}
\affiliation{Department of Physics and Astronomy, University of California, Irvine, CA  92697  USA}
\author{Brooks Thomas}
\email[Email address: ]{thomasbd@lafayette.edu}
\affiliation{Department of Physics, Lafayette College, Easton, PA  18042 USA}

\begin{abstract}
One signature of an expanding universe is the time-variation of the cosmological abundances of 
its different components.  For example, a radiation-dominated universe inevitably gives way 
to a matter-dominated universe, and critical moments such as matter-radiation equality are fleeting.
In this paper, we point out that this lore is not always correct, and that it is possible to obtain a 
form of ``stasis'' in which the relative cosmological abundances $\Omega_i$ of the different components 
remain unchanged over extended cosmological epochs, even as the universe expands.   
Moreover, we demonstrate that such situations are not fine-tuned,
but are actually global attractors within certain cosmological frameworks, with the universe
naturally evolving towards such long-lasting periods of stasis for a wide variety of initial conditions.
The existence of this kind of stasis therefore gives rise to a host of new theoretical possibilities 
across the entire cosmological timeline, ranging from potential implications for primordial density 
perturbations, dark-matter production, and structure formation all the way to early reheating, 
early matter-dominated eras, and even the age of the universe.  
\end{abstract}
\maketitle

\tableofcontents

\def\ie{{\it i.e.}\/}
\def\eg{{\it e.g.}\/}
\def\etc{{\it etc}.\/}
\def\taubar{{\overline{\tau}}}
\def\qbar{{\overline{q}}}
\def\kbar{{\overline{k}}}
\def\bQ{{\bf Q}}
\def\calT{{\cal T}}
\def\calN{{\cal N}}
\def\calF{{\cal F}}
\def\calM{{\cal M}}
\def\calZ{{\cal Z}}

\def\beq{\begin{equation}}
\def\eeq{\end{equation}}
\def\beqn{\begin{eqnarray}}
\def\eeqn{\end{eqnarray}}
\def\apo{\mbox{\small ${\frac{\alpha'}{2}}$}}
\def\half{\mbox{\small ${\frac{1}{2}}$}}
\def\sqapo{\mbox{\tiny $\sqrt{\frac{\alpha'}{2}}$}}
\def\sqap{\mbox{\tiny $\sqrt{{\alpha'}}$}}
\def\sqapxtwo{\mbox{\tiny $\sqrt{2{\alpha'}}$}}
\def\aptwo{\mbox{\tiny ${\frac{\alpha'}{2}}$}}
\def\apofour{\mbox{\tiny ${\frac{\alpha'}{4}}$}}
\def\bosqtwo{\mbox{\tiny ${\frac{\beta}{\sqrt{2}}}$}}
\def\btosqtwo{\mbox{\tiny ${\frac{\tilde{\beta}}{\sqrt{2}}}$}}
\def\apofour{\mbox{\tiny ${\frac{\alpha'}{4}}$}}
\def\sqaptwo{\mbox{\tiny $\sqrt{\frac{\alpha'}{2}}$}  }
\def\apoeight{\mbox{\tiny ${\frac{\alpha'}{8}}$}}
\def\sapoeight{\mbox{\tiny ${\frac{\sqrt{\alpha'}}{8}}$}}

\newc{\gsim}{\lower.7ex\hbox{{\mbox{$\;\stackrel{\textstyle>}{\sim}\;$}}}}
\newc{\lsim}{\lower.7ex\hbox{{\mbox{$\;\stackrel{\textstyle<}{\sim}\;$}}}}
\def\calM{{\cal M}}
\def\calV{{\cal V}}
\def\calF{{\cal F}}
\def\bQ{{\bf Q}}
\def\bT{{\bf T}}
\def\Qs{{\bf q}}

\def\half{{\textstyle{1\over 2}}}
\def\quarter{{\textstyle{1\over 4}}}
\def\ie{{\it i.e.}\/}
\def\eg{{\it e.g.}\/}
\def\etc{{\it etc}.\/}
\def\inbar{\,\vrule height1.5ex width.4pt depth0pt}
\def\IR{\relax{\rm I\kern-.18em R}}
 \font\cmss=cmss10 \font\cmsss=cmss10 at 7pt
\def\IQ{\relax{\rm I\kern-.18em Q}}
\def\IZ{\relax\ifmmode\mathchoice
 {\hbox{\cmss Z\kern-.4em Z}}{\hbox{\cmss Z\kern-.4em Z}}
 {\lower.9pt\hbox{\cmsss Z\kern-.4em Z}}
 {\lower1.2pt\hbox{\cmsss Z\kern-.4em Z}}\else{\cmss Z\kern-.4em Z}\fi}


\section{Introduction, motivation, and basic idea}

One of the earliest and most profound discoveries of modern cosmology is that we live in an expanding universe. 
With this one discovery, the age-old paradigm of an everlasting static universe was overthrown, replaced by a universe whose fundamental characteristics are time-dependent.  
Chief among these characteristics are the abundances of the different components which
contribute to its energy density.
Indeed,  it is traditional to refer to the different epochs 
through which the universe evolves in terms of the abundances which dominate during those epochs,
with the current paradigm positing that the universe passed from an 
initial inflationary epoch dominated by vacuum energy to a 
reheating epoch dominated by the energy of an oscillating inflaton
to a radiation-dominated post-reheating epoch to a matter-dominated post-reheating epoch --- one 
 which is only now giving way to a second epoch dominated by vacuum energy.

This passage from epoch to epoch is almost inevitable.   Indeed, according to the Friedmann equations, cosmological expansion induces a redshifting effect that causes the abundances of the different energy components of the universe to scale with time in different ways.  It can therefore happen that the smallest one now will later be vast (just as the present now will later be past);  these abundances are constantly changing.   

As an example, let us focus on the energy densities $\rho_M$ and $\rho_\gamma$
associated with matter and radiation respectively, along with their corresponding abundances $\Omega_M\sim \rho_M/H^2$ and
$\Omega_\gamma \sim \rho_\gamma/H^2$, where $H(t)=\dot a/a$ is the Hubble parameter and $a(t)$ is the scale factor.
In general, $\rho_M$ and $\rho_\gamma$ evolve as $a^{-3}$ and
$a^{-4}$ respectively, but the time-dependence of the scale factor $a(t)$ in turn depends (through the Friedmann equations) on the instantaneous mix of components in the associated cosmology.
We thus obtain an evolving non-linear system in which 
the  values of the abundances $\Omega_M$ and $\Omega_\gamma$
at any moment influence their own instantaneous rates of change.
For example, 
we might start in a radiation-dominated universe with $\Omega_\gamma\gg \Omega_M$,
but in such a universe
$\Omega_\gamma$ remains approximately constant
while $\Omega_M\sim t^{+1/2}$.
As a result we find that $\Omega_M$ grows and eventually becomes 
significant compared to $\Omega_\gamma$.   Indeed, 
with $\Omega_M\approx \Omega_\gamma$ we  find
$\Omega_M \sim t^{+2/7}$ while
$\Omega_\gamma\sim t^{-2/7}$.
Thus $\Omega_M$ continues to grow even beyond $\Omega_\gamma$.
Eventually we enter a matter-dominated epoch
with $\Omega_M\gg \Omega_\gamma$, whereupon
$\Omega_M$ remains approximately constant while $\Omega_\gamma\sim t^{-2/3}$.
Our radiation-dominated epoch has thus become a matter-dominated epoch, simply as a result
of cosmological expansion.    
In a similar way, a matter-dominated epoch generally gives way to a vacuum-energy-dominated epoch.  
 Indeed, special moments such as those exhibiting matter-radiation equality are fleeting,
since even at the instant when 
$\Omega_M=\Omega_\gamma$ we see that
$\Omega_M$ is growing while $\Omega_\gamma$ is shrinking.
The lesson, then, seems clear:   In an expanding universe, 
the relative sizes of the different contributions are continually in flux.  As a result,
epochs containing non-trivial mixtures of energy components 
are generally unstable, with component ratios such as $\Omega_M/\Omega_\gamma$ perpetually 
evolving in time 
     regardless of where such epochs might be situated along the cosmological timeline.

In this paper, we shall demonstrate that this general expectation 
need not always hold true.
 In particular, we shall
demonstrate that it is possible to construct scenarios in which such mixed-component cosmological eras can be stable over extended epochs lasting as many $e$-folds as desired, with values of 
$\Omega_M$ and $\Omega_\gamma$ remaining strictly constant despite cosmological expansion.
In other words, we shall demonstrate that it is 
    generally
possible to have long-lasting epochs which are not matter-dominated
or radiation-dominated, and not necessarily dominated by any particular component at all!
 We shall refer to 
such epochs as periods of ``stasis''.   For example, we shall provide an explicit model that gives rise to an extended stasis epoch exhibiting strict matter-radiation equality, with $\Omega_M=\Omega_\gamma$ holding throughout.
Extended epochs lasting arbitrary numbers of $e$-folds can also be constructed exhibiting other ratios
between $\Omega_M$ and $\Omega_\gamma$.

As we shall demonstrate, these scenarios emerge naturally in realistic scenarios of physics
beyond the Standard Model.    Moreover, we shall demonstrate that these stasis states are not fine-tuned, and are 
indeed {\it global dynamical attractors}\/ within these scenarios.  Thus, within these scenarios, 
the universe need not begin within a period of stasis in order for stasis to arise ---
the universe will necessarily evolve into a stasis state
for a wide variety of initial conditions.
Finally, we shall find in all cases  that
the stasis state also has a natural ending after which normal cosmological evolution resumes.   Thus
our stasis epoch has both a beginning and an end --- a feature which potentially allows it to be
``spliced'' into various points along the standard cosmological timeline.

It may initially seem impossible to arrange such periods of stasis between matter and
radiation.   After all, for the reasons discussed above, matter inevitably dominates over radiation;
   this is so intrinsic a prediction of the Friedmann equations that this
conclusion seems unavoidable.
On the other hand, matter can {\it decay}\/ back into radiation.
This then might provide a natural counterbalance to the effects of cosmological expansion, 
causing the matter abundance $\Omega_M$ to shrink while the radiation abundance $\Omega_\gamma$
grows.
Our idea, then, is a simple one:  {\it can these two effects be balanced against each other?
 More specifically, can particle decay
be balanced against cosmological expansion in order to induce an extended time interval of stasis during which the matter and radiation abundances
each remain constant?}\/    

Of course, particle decay is a relatively short process, localized in time.
In order to have an extended period of stasis we would therefore require an extended period during which
particle decays are continually occurring.
This would be the case if we had a large tower of matter states  $\phi_\ell$ ($\ell=0,1,...,N-1$), with each state sequentially  decaying directly (or preferentially) into radiation.   
As is well known from the Dynamical Dark Matter framework~\cite{Dienes:2011ja,Dienes:2011sa,Dienes:2012jb}, 
many scenarios for physics beyond the Standard Model
give rise to precisely such towers of dark-matter states.   The question is then whether 
these sequential decays down the tower could be exploited in order to sustain an extended period of cosmological stasis.

\begin{figure*}[t!]
\centering
\includegraphics[keepaspectratio, width=0.8\textwidth]{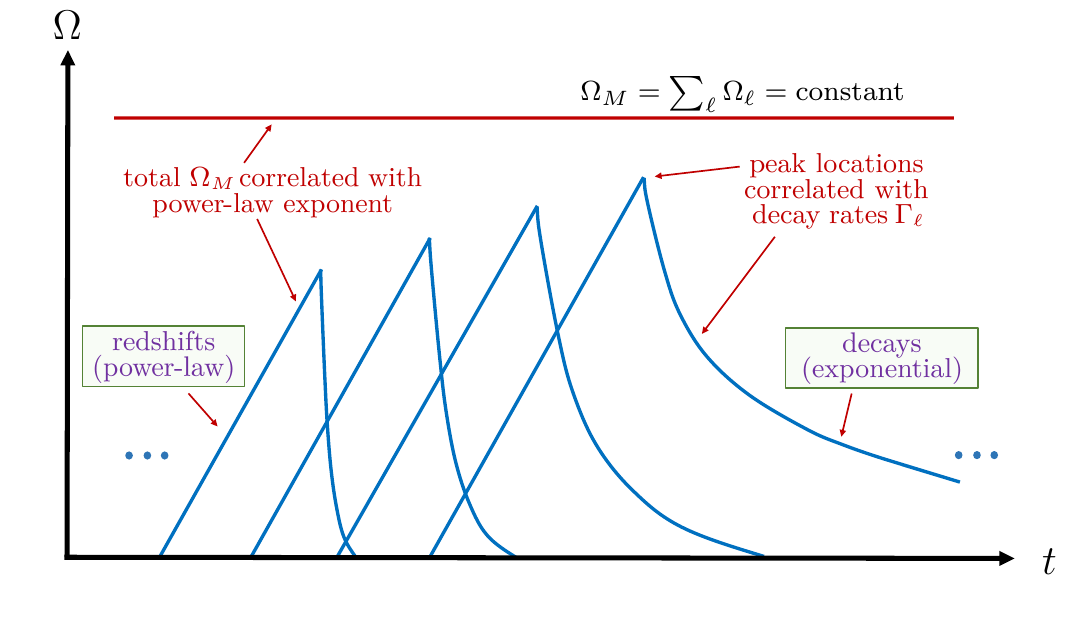}
\caption{A sketch of the basic stasis mechanism:   the abundances $\Omega_\ell(t)$ of individual matter components (blue) each experience power-law growth
due to cosmological expansion before eventually experiencing exponential decay.  
For simplicity in this idealized sketch, this power-law growth is shown as linear while the transition to exponential decay is sketched as sharp.   
Nevertheless this process conspires to keep the total matter abundance $\Omega_M\equiv \sum_\ell \Omega_\ell(t)$ constant (red), thereby producing an extended epoch of cosmological stasis.   Note that this process is highly non-trivial:  the power-law growth for some $\Omega_\ell$ must be balanced against the exponential decays of the other $\Omega_\ell$ while the exponent of the power-law growth is correlated with the value of the total $\Omega_M$ through the Friedmann equations and
the decay rate $\Gamma_\ell$ of each component is correlated with the time $\tau_\ell$ at which $\Omega_\ell$ hits a maximum and the decay begins to dominate.   As a result this system does not exhibit time-translation invariance even though the total $\Omega_M$ remains constant.    Despite its complexity, we shall demonstrate that the stasis state  not only occurs naturally in well-motivated physics scenarios but is actually a {\it global dynamical attractor}\/ in such situations, with the system generically evolving towards stasis even if it does not begin in stasis.}
\label{fig1}
\end{figure*}

This is clearly a tall order, and at first glance such a balancing might seem to be impossible.
In order to appreciate the difficulties involved, let us consider how such an idealized scenario might work.
A sketch of such a scenario appears in Fig.~\ref{fig1}, where we have illustrated the behavior of the
individual abundances $\Omega_\ell(t)$ of each of the matter fields within the tower (blue).
  In general, as discussed above,
 each abundance $\Omega_\ell(t)$ initially grows according to a common power-law as
the result of cosmological expansion;  for convenience and simplicity this power-law growth is sketched within
Fig.~\ref{fig1} as linear.
However, once the appropriate decay time $\tau_\ell$ is reached for each component
(idealized in Fig.~\ref{fig1} as a sharp transition), 
the behavior of $\Omega_\ell(t)$ changes and now reflects an exponential decay.
Of course, our goal is for all of this to occur in such a way that
the sum $\sum_\ell \Omega_\ell(t)$ --- \ie, 
the total matter abundance $\Omega_M$  also shown in Fig.~\ref{fig1} (red) --- remains constant.

Given this sketch, we can immediately see the complications involved.
First, in order to keep the sum $\Omega_M$ constant, at any moment we need to somehow be cancelling the {\it power-law}\/ growth of the abundances $\Omega_\ell$ of the lighter components which have not yet decayed against the 
{\it exponential}\/ decays of the abundances $\Omega_\ell$ of the heavier states which have.
Second, we see that each successive $\Omega_\ell$ must reach a greater maximum value before decaying than did the previous abundance $\Omega_{\ell+1}$, since with each decay we have fewer remaining matter states contributing to $\Omega_M$.
Third, it follows from the Friedmann equations that the exponent of the common power-law growth experienced by
each $\Omega_\ell$ prior to decay must be correlated with the total matter abundance $\Omega_M$ --- an observation
which 
provides a non-linear ``feedback'' constraint on our system.
Finally, as our decays proceed down the tower, the decay lifetimes $\tau_\ell$ are continually increasing.   This implies that 
the corresponding decay widths $\Gamma_\ell$ are continually decreasing, which means that the successive exponential decays must occur with slower and slower rates.   All of these features are illustrated in Fig.~\ref{fig1}.

With all of these tight constraints, it is remarkable that such a stasis state with constant $\Omega_M$ 
can ever emerge.   
However, we shall demonstrate that this is exactly what occurs.
As a result, the existence of this kind of stasis state gives rise to a host of new theoretical possibilities across the entire cosmological timeline,
ranging from potential implications for primordial density perturbations, dark-matter production, and structure formation all the way to early 
reheating, early matter-dominated eras, 
and the age of the universe.

This paper is organized as follows.
In Sect.~\ref{deriving_constraints}, we start by studying the stasis 
  phenomenon
itself and derive a set of mathematical conditions that must be satisfied within any period of stasis.   At this stage of our analysis, the very notion of a stasis state
implicitly requires that such a state be truly eternal, without beginning or end.
In Sect.~\ref{model}, we then present a model of stasis --- \ie, a general model 
which arises naturally in many extensions of the Standard Model and which generally satisfies these conditions.   Thus, 
our model gives rise to stasis.  However, we shall find that our model contains certain ``edge'' (or ``boundary'') effects
that cause the system to deviate from true stasis at times which are extremely early or late compared with the time
at which our tower of states is originally produced.   Thus, within our model, we shall find that our stasis epoch 
is actually a finite one in which there exist both a natural {\it entrance into}\/ as well as {\it exit from}\/ stasis.   This is ultimately a beneficial feature, implying that our stasis state is ultimately of finite duration, after which normal cosmological evolution resumes.
In Sect.~\ref{attractor}, we then study what happens when such systems are not originally in stasis, and demonstrate that the stasis state is a global dynamical attractor.   Thus, regardless of the initial conditions, our system will always eventually enter into a stasis state and remain in stasis until all decays have concluded.   

Taken together, these results demonstrate that the stasis state is both stable and robust.
In Sect.~\ref{vacuum}, we then consider various extensions of our results.  In particular, we study the behavior that 
emerges when additional energy components 
beyond radiation and matter are introduced into the cosmology. 
Finally, in Sect.~\ref{implications}, we summarize our results and consider various possible theoretical and phenomenological implications of stasis across the cosmological timeline.   We also outline ideas for future research.

    We emphasize that our main interest in this paper is the stasis phenomenon itself --- 
    \ie, the theoretical possibility that such stable mixed-component eras can be realized within an expanding universe.
    Needless to say, phenomenological constraints may make it difficult 
    to introduce stasis epochs into certain portions of the standard
    cosmological timeline.  For example, those portions of the timeline after nucleosynthesis are deeply constrained by observational data and therefore cannot be significantly modified. 
    Such stasis epochs nevertheless 
     represent a viable phenomenological possibility during earlier periods along the cosmological 
    timeline, such as during earlier portions of radiation-domination or even during reheating.   
    We shall therefore study stasis as a general theoretical phenomenon
    throughout most of this paper, 
      and defer our discussion of its phenomenological implications to Sect.~\ref{implications}.

\FloatBarrier
\section{Stasis:  General considerations\label{deriving_constraints}}

Throughout this paper, ``stasis'' will refer to any extended period during which the total matter and radiation abundances 
$\Omega_M$ and $\Omega_\gamma$ remain constant despite cosmological expansion.
In this section, we provide an analytical discussion of stasis, with
the goal of obtaining mathematical conditions  
that characterize this state and must therefore be satisfied therein.
In particular, we shall do this in two separate steps:
\begin{itemize}
\item  we shall first determine a condition that characterizes stasis at any moment in time --- \ie, a minimal
         condition necessary for stasis to {\it exist}\/;  and 
\item  we shall then determine two additional conditions that must hold in order for stasis
           to {\it persist}\/ over an extended period.  
\end{itemize}
We shall now address each of these issues in turn.

\subsection{Minimal condition for the existence of stasis}

Let us begin by assuming a flat Friedmann-Robertson-Walker (FRW) universe containing only 
\begin{itemize}
 \item    a tower of matter states $\phi_\ell$ 
               where the indices $\ell=0,1,2,....$ are assigned 
              in order of increasing mass;  and 
\item     radiation (collectively denoted $\gamma$) into which the $\phi_\ell$ can decay.
\end{itemize}
We shall let $\rho_\ell$ and $\rho_\gamma$ 
denote the corresponding energy densities
and $\Omega_\ell$ and $\Omega_\gamma$  the corresponding abundances.
We shall also let $\Gamma_\ell$ denote the decay rates for the $\phi_\ell$.

Recall that for any energy density $\rho_i$ (where $i=\ell,\gamma$), 
the corresponding abundance $\Omega_i$ is given by 
\beq
      \Omega_i   ~\equiv~ \frac{8\pi G}{3H^2} \rho_i
\eeq
where $H$ is the Hubble parameter and $G$ is Newton's constant.
From this it follows that
\beq
    \frac{d\Omega_i}{dt} ~=~ \frac{8\pi G}{3} \left( \frac{1}{H^2} \frac{d\rho_i}{dt} - 2 \frac{\rho_i}{H^3} \frac{dH}{dt} \right)~.
\label{stepone}
\eeq
We can simplify this expression 
through the use of the Friedmann ``acceleration'' equation for $dH/dt$, which in this universe takes the form
\beqn
  \frac{dH}{dt}  
  ~&=&~ -H^2 - \frac{4\pi G}{3} \left(  \sum_i \rho_i + 3 \sum_i p_i\right)\nonumber\\
  ~&=&~ -H^2 - \frac{4\pi G}{3} \left(  \sum_\ell \rho_\ell + 2 \rho_\gamma \right)\nonumber\\
  ~&=&~ - \half H^2 \left( 2+ \Omega_M + 2 \Omega_\gamma\right)\nonumber\\
  ~&=&~ - \half H^2 \left( 4-\Omega_M \right)~.
\label{acceleq}
\eeqn
Note that in passing to the second line of Eq.~(\ref{acceleq}) we have recognized that matter and radiation have $w=0$ and $w=1/3$ respectively, where $w_i\equiv p_i/\rho_i$ is the
equation-of-state parameter for component $i$.
Thus $p_\ell=0$ and $p_\gamma = \rho_\gamma/3$.
Likewise, in passing to the third line we have defined the total matter abundance
          $\Omega_M \equiv \sum_\ell \Omega_\ell$,
and in passing to the fourth line we have imposed the constraint $\Omega_M + \Omega_\gamma=1$.
Substituting Eq.~(\ref{acceleq}) into Eq.~(\ref{stepone}) we then obtain
\beq
    \frac{d\Omega_i}{dt} 
     ~=~ \frac{8\pi G}{3H^2} \frac{d\rho_i}{dt} + H \Omega_i \left( 4-\Omega_M\right)~,
\label{convert}
\eeq
yielding
\beqn
    \frac{d\Omega_M}{dt} 
     ~&=&~ \frac{8\pi G}{3H^2} \sum_\ell \frac{d\rho_\ell}{dt} 
     + H \Omega_M \left( 4-\Omega_M \right) ~ \nonumber\\
    \frac{d\Omega_\gamma}{dt} 
     ~&=&~ \frac{8\pi G}{3H^2} \frac{d\rho_\gamma}{dt}  
    + H \Omega_\gamma \left( 4-\Omega_M\right)~.
\label{convert2}
\eeqn
These are thus general relations for the time-evolution of $\Omega_M$ and $\Omega_\gamma$ in terms of 
$d\rho_\ell/dt$ and $d\rho_\gamma/dt$.
Of course, since $\Omega_M+ \Omega_\gamma =1$, it follows that $d\Omega_M/dt = -d\Omega_\gamma/dt$.
From Eq.~(\ref{convert2}) we therefore obtain the self-consistency constraint 
\beq
     \frac{8\pi G}{3H^2} \left( \sum_\ell \frac{d\rho_\ell}{dt} +  \frac{d \rho_\gamma}{dt}\right) 
     ~=~ H \left(\Omega_M-4\right)~, 
\label{selfcon}
\eeq
which is tantamount to asserting that $\rho_\ell$ and $\rho_\gamma$ are the only contributions to the total
energy density of the universe.

Given the relations in Eq.~(\ref{convert2}), our final step is to insert appropriate equations of motion for 
$d\rho_\ell/dt$ and $d\rho_\gamma/dt$.   It is here that we introduce the idea that the production of radiation $\gamma$
comes from the decays of the $\phi_\ell$.   Since each $\phi_\ell$ is assumed to decay into radiation $\gamma$ with
rate $\Gamma_\ell$, and given that each decay process conserves energy, these equations of motion are given by
\beqn
         \frac{d\rho_\ell}{dt} ~&=&~ -3 H \rho_\ell - \Gamma_\ell \rho_\ell~ \nonumber\\
         \frac{d\rho_\gamma}{dt} ~&=&~ -4 H \rho_\gamma + \sum_\ell \Gamma_\ell \rho_\ell~ .
\label{eoms}
\eeqn
Note that these equations of motion indeed satisfy the constraint in Eq.~(\ref{selfcon}).
The results in Eq.~(\ref{convert2}) then take the form
\beq
    \frac{d\Omega_M}{dt} 
     ~=~ -\sum_\ell \Gamma_\ell \Omega_\ell  
     + H \left( \Omega_M - \Omega_M^2\right)  
\label{convert3}
\eeq
with $d\Omega_\gamma/dt = - d\Omega_M/dt$.
Note that $\Omega_M-\Omega_M^2 = \Omega_M (1-\Omega_M) = \Omega_M\Omega_\gamma$.

The differential equation for $\Omega_M$ in Eq.~(\ref{convert3}) is completely general, describing the
complete time-evolution of $\Omega_M$ and $\Omega_\gamma$.
It is important to realize that these differential equations do 
not imply that $\Omega_M$ or $\Omega_\gamma$ are monotonic functions of time.
In general, each $\Omega_\ell$ comprising $\Omega_M$ has a complicated time dependence:   
as evident from Eq.~(\ref{convert3}),
the decay process tends 
to push $\Omega_\ell$ downward,
while the existence of an induced $\Omega_\gamma>0$ in the background cosmology tends to affect the Hubble expansion
in such a way as to push $\Omega_\ell$ upwards.
Thus, each $\Omega_\ell$ can either rise or fall as a function of time, implying that $\Omega_M$ --- and therefore $\Omega_\gamma$ ---
can likewise either rise or fall 
as a function of time.
This is ultimately the result of the competition between
the two terms on the right side of Eq.~(\ref{convert3}).
Of course, for situations in which the decay widths $\Gamma_\ell$ are
all significantly greater than $H$, 
the effects of the decays 
will dominate and in that case $d\Omega_M/dt$ will be
strictly negative, resulting in a 
monotonically falling $\Omega_M$.

Given the result in Eq.~(\ref{convert3}), we now seek a 
steady-state ``stasis'' solution in which $\Omega_M$ and $\Omega_\gamma$ are constant.
Clearly such a solution will arise if the effects of the $\phi_\ell$ decays are 
precisely counterbalanced by the Hubble expansion.
While there are many ways of seeking such a solution, we shall do this in two steps.   First, we shall impose
the condition that $d\Omega_M/dt=0$.   
This then yields the constraint
\beq
\boxed{
    ~~\sum_\ell^{\phantom{x}} \Gamma_\ell \Omega_\ell  ~=~ H (\Omega_M - \Omega_M^2)~.
}
\label{condition}
\eeq
This is clearly a necessary (but not sufficient) condition for stasis.
In the next subsection, we shall determine the additional conditions under which 
Eq.~(\ref{condition}), once satisfied at some time $t_\ast$, actually {\it remains}\/ satisfied 
over an extended time interval.

Note that since $0\leq \Omega_M\leq 1$, both sides of Eq.~(\ref{condition}) are necessarily non-negative.
Indeed, the right side of this equation can equivalently be written as $H \Omega_M\Omega_\gamma$ or
$H(\Omega_\gamma-\Omega_\gamma^2)$.
It also follows from Eq.~(\ref{condition}) that no solution for $d\Omega_M/dt=0$ is even possible
at a given time unless 
\beq
    \sum_\ell \Gamma_\ell \Omega_\ell  ~\leq~  \frac{H}{4}~.
\label{upperlim}
\eeq
This provides an upper limit on the possible decay widths $\Gamma_\ell$.
When this inequality is saturated with $d\Omega_M/dt=0$ we necessarily have $\Omega_M=\Omega_\gamma=1/2$. 
Otherwise, if the inequality in Eq.~(\ref{upperlim}) is satisfied but not saturated,
other values of $\Omega_M$ (both bigger and smaller than $1/2$) are in principle possible.

\subsection{Conditions for the persistence of stasis}

As we have seen, 
Eq.~(\ref{condition}) furnishes us with a minimal condition for stasis.
In general, however, such a condition will be satisfied only at a particular instant of time $t_\ast$.
By itself, this would clearly not lead to a true stasis state in which $\Omega_M$ is constant.
For the purposes of understanding stasis, we are 
therefore interested in determining the additional condition(s), if any,
that will allow Eq.~(\ref{condition}) 
to {\it remain}\/ satisfied over an extended period of time. 
Indeed, it is only in this way that we can obtain a true period of stasis.

Starting from Eq.~(\ref{condition}), 
there are two ways in which we might demand  
that $\Omega_M$ actually remain constant at some fixed stasis value $\barOmega_M$.
First, at $t=t_\ast$, we could impose not only $d\Omega_M/dt=0$, but also $d^n\Omega_M/dt^n=0$ for all integers $n>1$.
This would then guarantee the absence of any time evolution for $\Omega_M$.
However, rather than impose this infinite set of constraints, we can do this in a much quicker way:
assuming that the equality in Eq.~(\ref{condition}) has been achieved at some time $t_\ast$, we now simply need to demand that 
both sides of this equation evolve with time in the same manner.
This would ensure that Eq.~(\ref{condition}) remains satisfied
under time evolution.

Before proceeding, we note two important implications of imposing such an additional requirement.
First, by demanding that both sides of Eq.~(\ref{condition}) behave identically under time evolution,
we are actually demanding an {\it eternal}\/ stasis in which $\Omega_M$ is fixed, without beginning or end.
Of course, this sort of eternal stasis is only an idealized abstraction
which cannot be representative of a realistic cosmology
and which requires, in particular, a correspondingly perpetual decay process.
Nevertheless, 
understanding the mathematics of such an idealized stasis will ultimately prove useful in allowing
us to understand how to achieve a more realistic period of stasis in which $\Omega_M$ remains constant at some value $\barOmega_M$
over an extended but finite time interval.  

The second implication of demanding that both sides of Eq.~(\ref{condition}) evolve identically with respect
to time is that $t_\ast$, which we have identified as the time at which Eq.~(\ref{condition}) is instantaneously
satisfied, now becomes nothing more than a {\it fiducial}\/ reference time, \ie, an arbitrary choice which cannot carry physical
significance.   Indeed, although we shall find it useful to assume that Eq.~(\ref{condition}) is satisfied at $t=t_\ast$
before working our way towards a solution for all $t$,
no physical condition for stasis can ultimately depend on the choice of $t_\ast$.

In order to determine the extra conditions we require for stasis,
let us now study how each term in Eq.~(\ref{condition}) evolves with time
during a supposed period of stasis.
First, we observe that we can solve for the Hubble parameter directly via Eq.~(\ref{acceleq}). 
Since $\Omega_M$ is presumed  constant with some fixed value $\barOmega_M$ during
stasis, we obtain the exact solution
\beq
    H(t) = \left(\frac{2}{4-\barOmega_M}\right) \,\frac{1}{t}~ ~~~\Longrightarrow~~~~
   \kappa =  \frac{6}{4-\barOmega_M}~ 
\label{Hubble}
\eeq
where $\kappa$ corresponds to the parametrization $H(t)= \kappa/(3t)$.
Indeed, from Eq.~(\ref{Hubble}) we verify the standard results that $\kappa=2$ 
for $\barOmega_M=1$ (\ie, a matter-dominated universe),
while $\kappa=3/2$ for $\barOmega_M=0$ (\ie, a radiation-dominated universe).
We emphasize that Eq.~(\ref{Hubble}) is an exact result only under the stasis assumption,
guaranteeing that $\Omega_M$ --- and therefore $\kappa$ --- remain strictly constant.
This solution for $H(t)$ in turn implies that during stasis, the scale factor grows as 
\beq
         a(t) ~=~ a_\ast \left( \frac{t}{t_\ast} \right)^{\kappa/3} ~=~ 
            a_\ast \left( \frac{t}{t_\ast}\right)^{2/(4-\barOmega_M)}~
\label{scalefactor}
\eeq
with $t_\ast$ representing an arbitrary fiducial time and 
 the `$\ast$' subscript indicating that the relevant quantity is evaluated at $t=t_\ast$.
Moreover, from the first equation
within Eq.~(\ref{eoms}) we
find the solution
\beqn
    \rho_\ell(t) ~&=&~ \rho_{\ell}^\ast \left[ \frac{a(t)}{a_\ast}\right]^{-3} e^{-\Gamma_\ell (t-t_\ast)}~\nonumber\\
                 ~&=&~ \rho_{\ell}^\ast \left( \frac{t}{t_\ast}\right)^{-6/(4-\barOmega_M)}  e^{-\Gamma_\ell (t-t_\ast)}~.
\label{rhol}
\eeqn
This in turn implies that
\beq
    \Omega_\ell(t) ~=~ \Omega_\ell^\ast  
                 \left( \frac{t}{t_\ast}\right)^{2-6/(4-\barOmega_M)}  e^{-\Gamma_\ell (t-t_\ast)}~.
\label{Omegal}
\eeq

Given these results, we thus have two conditions which must be satisfied simultaneously 
in order to have an extended period of stasis:
\begin{empheq}[box=\fbox]{align}
\sum_\ell \Omega_\ell(t) ~&=~ \barOmega_M~~\nonumber\\
     ~~ \sum_\ell  \Gamma_\ell \Omega_\ell(t) ~&=~ 
     \frac{ 2\barOmega_M(1-\barOmega_M)}{4-\barOmega_M} \, \frac{1}{t} ~~
\label{masterconstraints}
\end{empheq}
where $\Omega_\ell(t)$ is given in Eq.~(\ref{Omegal}).  
Indeed, these two conditions ensure that Eq.~(\ref{condition}) is satisfied not only instantaneously, at one particular time $t_\ast$, but eternally, for all times.
Note that these results together imply the constraint
\beq
      \frac{\sum_\ell \Gamma_\ell \Omega_\ell}{\sum_\ell \Omega_\ell} ~=~  
       \frac{  2 (1-\barOmega_M)}{4-\barOmega_M} \,\frac{1}{t}~.
\label{quotient}
\eeq
Going forward, our goal will be to find systems for which both constraints in Eq.~(\ref{masterconstraints})
are satisfied as exactly as possible.

At first glance, it may seem surprising that the constraints in Eq.~(\ref{masterconstraints}) can ever be satisfied.
Indeed, the forms of these constraints present two immediate challenges. 
First, while the right side of the second constraint in Eq.~(\ref{masterconstraints}) drops like $1/t$, 
the individual component abundances $\Omega_\ell(t)$ never drop as $1/t$:
as indicated in Eq.~(\ref{Omegal}),
they either grow as a power-law during early times $t\ll 1/\Gamma_\ell$
(when the exponential decay is not yet dominant),
or they fall exponentially as $t\sim 1/\Gamma_\ell$ (ultimately overriding the power-law growth).
However, what Eq.~(\ref{masterconstraints}) is telling us
is that these two effects must somehow cancel within the sum $\sum_\ell \Gamma_\ell \Omega_\ell(t)$, leaving behind an overall $1/t$ dependence.
Second, all of this must happen for each individual $\Omega_\ell(t)$ while somehow simultaneously keeping their sum $\sum_\ell \Omega_\ell(t)$ fixed, so that the rising values of $\Omega_\ell$ from what are presumably lighter modes
with smaller widths (experiencing later decays) perfectly compensate for the exponential decays of the heavier modes (which experience earlier decays).
Note that this second challenge is indeed different from the first:
the second challenge concerns the {\it unweighted}\/ sum of the individual $\Omega_\ell$,
while the first challenge is sensitive to the sum in which each contribution $\Omega_\ell$ is 
weighted by the corresponding width $\Gamma_\ell$.
However, as we shall now demonstrate, very accurate simultaneous solutions to both constraints in 
 Eq.~(\ref{masterconstraints}) can nevertheless indeed be found.

\FloatBarrier
\section{A model of stasis \label{model}}

In Sect.~\ref{deriving_constraints}, we obtained two conditions, as listed within Eq.~(\ref{masterconstraints}),
which together yield an eternal stasis existing for all times.
However, in reality, these conditions cannot be strictly satisfied for all times.
For example, regardless of whether the tower of $\phi_\ell$ states is finite or infinite, there is an early time 
immediately after these states are produced  during which the decay process is just beginning. 
However, because the universe is expanding even at this early time, the required balancing between decay and cosmological expansion 
cannot yet have been achieved, and consequently we do not yet expect to have realized stasis.
Likewise, there will eventually come a time at which all of the decays will have essentially concluded.
At this point we expect our period of stasis to end.

Despite these observations, the critical issue is whether there exist solutions for 
the spectrum of decay widths $\lbrace \Gamma_\ell\rbrace$ and abundances $\lbrace \Omega_\ell\rbrace$ across our tower of states
which will at least lead to a period of stasis {\it during}\/ the sequential decay process.
Given the discussion at the end of Sect.~\ref{deriving_constraints},
it might seem that such solutions for $\lbrace \Gamma_\ell\rbrace$ and $\lbrace \Omega_\ell\rbrace$
must be very carefully arranged.
Remarkably, however, we shall now demonstrate that there exist 
solutions for $\lbrace \Gamma_\ell\rbrace$ and $\lbrace \Omega_\ell\rbrace$  
which not only satisfy these constraints but which are also relatively simple
and which emerge naturally in realistic scenarios for physics beyond the Standard Model.

Towards this end, let us consider a spectrum of 
decay widths $\lbrace \Gamma_\ell\rbrace$ and abundances $\lbrace \Omega_\ell\rbrace$ which satisfy
the general scaling relations
\beq
       \Gamma_\ell = \Gamma_0\left(\frac{m_\ell}{m_0}\right)^\gamma~,~~~~~
       \Omega_\ell^{(0)} = \Omega_0^{(0)} \left(\frac{m_\ell}{m_0}\right)^\alpha~
\label{scalings}
\eeq
where $\alpha$ and $\gamma$ are general scaling exponents, 
where the mass spectrum takes the form 
\beq
          m_\ell ~=~ m_0 + (\Delta m) \ell^\delta~
\label{massform}
\eeq
with $m_0\geq 0$, $\Delta m>0$, and $\delta>0$
treated as general free parameters,
and where
the superscript `0' within Eq.~(\ref{scalings}) denotes the time $t=t^{(0)}$ at which the $\phi_\ell$ are initially produced
(thereby setting a common clock for the subsequent $\phi_\ell$ decays).
Our goal is then to determine those values --- if any --- of the eight parameters
\beq
 \lbrace \alpha,\gamma,\delta, m_0, \Delta m, \Gamma_0, \Omega_0^{(0)}, t^{(0)}\rbrace
\label{parameterlist}
\eeq
for which the constraints in Eq.~(\ref{masterconstraints}) can be satisfied.
Of course, within the context of this model, we have $\Omega_M=1$ at $t=t^{(0)}$ before the decay process has begun.   
This in turn requires that we choose
$\Omega_0^{(0)} = \left[\sum_{\ell=0}^{N-1} (m_\ell/m_0)^\alpha\right]^{-1}$.
 
Before proceeding further, our choice of the scaling relations in Eqs.~(\ref{scalings}) and (\ref{massform}) deserves comment.
It may initially seem that we have adopted these relations for the sole purpose of achieving stasis.
However, {\it these exact relations actually have an independent history}\/~\cite{Dienes:2011ja,Dienes:2011sa}
{\it as characterizing the towers of states that naturally emerge within a variety of actual models of physics beyond the
Standard Model.}\/
  For example, taking the $\phi_\ell$ as
  the Kaluza-Klein (KK) excitations of a five-dimensional scalar
  field compactified on a circle of radius $R$ (or a $\mathbb{Z}_2$ orbifold thereof)
  results in 
  either $\lbrace m_0,\Delta m,\delta\rbrace = \lbrace m, 1/R, 1\rbrace$
  or $\lbrace m_0,\Delta m,\delta\rbrace =\lbrace m, 1/(2 m R^2), 2\rbrace$,
  depending on whether $m R \ll1$ or $mR\gg 1$,
  respectively, where $m$ denotes the four-dimensional scalar mass~\cite{Dienes:2011ja, Dienes:2011sa}.
  Alternatively, taking the $\phi_\ell$ as the bound states of a strongly-coupled
  gauge theory yields $\delta = 1/2$, where $\Delta m$ and $m_0$ are determined by
  the Regge slope and intercept of the strongly-coupled theory, respectively~\cite{Dienes:2016vei}.
  Thus $\delta=\lbrace 1/2, 1,2\rbrace$ serve as compelling ``benchmark'' values.
  Likewise, $\gamma$ is generally 
  governed by the particular $\phi_\ell$ 
  decay mode.  For example, if $\phi_\ell$ decays to photons
  through a dimension-$d$ contact operator of the form ${\cal O}_\ell \sim c_\ell \phi_\ell {\cal F}/\Lambda^{d-4}$ 
  where $\Lambda$ is an appropriate mass scale and where ${\cal F}$ is an operator built from photon fields, we have $\gamma= 2d - 7$.
  Thus values such as $\gamma=\lbrace 3,5,7\rbrace$ can serve as relevant benchmarks.
  Finally, $\alpha$ is governed 
  by the original production mechanism for the $\phi_\ell$ fields.
  For example, one typically finds that $\alpha<0$ for misalignment production~\cite{Dienes:2011ja, Dienes:2011sa}, while
  $\alpha$ can generally be of either sign for thermal freeze-out~\cite{Dienes:2017zjq}.

Given this general model, we can now evaluate the sums which appear on the left sides of 
our constraint equations in Eq.~(\ref{masterconstraints}).   
Our goal, of course, is to avoid assuming stasis and to find the conditions
under which our model nevertheless satisfies these stasis constraints.

We begin by focusing on the behavior of the abundance of any individual component $\Omega_\ell(t)$.
In Eq.~(\ref{Omegal}), we derived the time-dependence of $\Omega_\ell(t)$, but this derivation assumed 
that we were already within stasis.  Indeed, within the calculation leading to Eq.~(\ref{Omegal}), 
the assumption of stasis entered into the form of the gravitational 
redshift factor $(t/t^{(0)})^{2-6/(4-\barOmega_M)}$;  
without assuming stasis, this factor would be much more complicated.
For simplicity and generality, we shall therefore let $h(t_i,t_f)$ denote the net gravitational redshift 
factor that accrues between any two times $t_i$ and $t_f$. 
We thus have
\beq
    \Omega_\ell(t) ~=~ \Omega_\ell^{(0)}\,   
                 h(t^{(0)},t)  \, e^{-\Gamma_\ell (t-t^{(0)})}~.
\label{Omegalnew}
\eeq
Note that this $h$-factor is necessarily $\ell$-independent since the gravitational redshift affects all 
components equally.
We then find that
\beq
 \sum_\ell \Omega_\ell(t) ~=~ 
 \Omega_0^{(0)} \, h(t^{(0)},t)  \,\sum_\ell \left( \frac{m_\ell}{m_0} \right)^\alpha \!
         e^{ - \Gamma_0 \left( \frac{m_\ell}{m_0}\right)^\gamma (t-t^{(0)})} 
\label{bigsum1}
\eeq
where we have used the scaling relations in Eq.~(\ref{scalings}).

In order to evaluate this sum, we shall make three approximations.
First, we shall take the continuum limit
\beq
     \Delta m\to 0~,~~~
         N\to \infty
\label{continuum}
\eeq
such that
$m_{\rm max}\equiv m_0+ (\Delta m) (N-1)^\delta$
is  held constant.
In this limit, the  masses $m_\ell$ become a continuous variable $m$ ranging from $m_0$ to $m_{\rm max}$, so that for any function $f(m_\ell/m_0)$ we can replace
\beqn
     &&   \sum_{\ell=0}^{N-1} f\left( \frac{m_\ell}{m_0} \right)\nonumber\\ 
      &&~~\to ~  \frac{1}{\delta} \int_{m_0}^{m_{\rm max}}  \frac{dm}{m-m_0}\,  \left( \frac{m-m_0}{\Delta m}\right)^{1/\delta}
          \!  f\left( \frac{m}{m_0}\right) \nonumber\\
      && ~~=~ \frac{1}{\delta} \int_{0}^{m_{\rm max}-m_0} \frac{dm}{m}\,  \left( \frac{m}{\Delta m}\right)^{1/\delta} 
         \! f\left(\frac{m}{m_0} + 1\right)~
\label{contsum}
\eeqn
where the additional integrand factor is the Jacobian $d\ell/dm$. 
 Simultaneously, we shall also take
 \beq
          m_0\to 0~,~~~m_{\rm max}\to \infty
 \label{edge}
\eeq
in such a way that $\Delta m/m_0$ --- and therefore all values of $m_\ell/m_0\sim m/m_0$  in Eq.~(\ref{scalings})  --- are kept constant.
This limit extends the range of $m$-integration in Eq.~(\ref{contsum}) from $0$ to $\infty$.
Finally, we shall also approximate
\beq
  f\left(\frac{m}{m_0} + 1\right) ~\approx~  f\left(\frac{m}{m_0}\right)~
\label{warp}
\eeq
within Eq.~(\ref{contsum}).
This represents a ``warping'' of our integrand which is relevant only for small values of $m/m_0$.

We shall later verify that all three of these approximations are relatively harmless, with effects that can easily be understood and interpreted. 
However, the net effect of these three approximations is that we can
convert our $\ell$-sum in Eq.~(\ref{bigsum1}) into an $m$-integral via
\beq
     \sum_{\ell=0}^{N-1} f\left( \frac{m_\ell}{m_0}\right)  ~\to~ 
        \frac{1}{\delta} \int_{0}^\infty   \frac{dm}{m}\,  \left( \frac{m}{\Delta m}\right)^{1/\delta}\!  f\left( \frac{m}{m_0}\right)~,
\eeq
whereupon
Eq.~(\ref{bigsum1}) becomes
\beqn
 && \sum_\ell \Omega_\ell(t) ~=~ \frac{ \Omega_0^{(0)}}{\delta (\Delta m)^{1/\delta}}
  \, h(t^{(0)},t)  \nonumber\\
  && ~~~ \times \int_{0}^\infty  dm\, 
     	 m^{1/\delta-1} \left(\frac{m}{m_0}\right)^{\alpha}\! 
  e^{ -\Gamma_0 \left(\frac{m}{m_0}\right)^\gamma \!(t-t^{(0)})}\,.~\nonumber\\
\label{integralform}
\eeqn
For $\gamma>0$ and $\alpha+1/\delta> 0$,
we then find that this integral can be evaluated in closed form, yielding the result
\beqn
 && \sum_\ell \Omega_\ell(t) ~=~ \frac{ \Omega_0^{(0)}}{\gamma\delta}
     \left( \frac{m_0}{\Delta m}\right)^{1/\delta} \, \Gamma\left( \frac{\alpha+1/\delta}{\gamma}\right)
          \nonumber\\
  && ~~~ \times~
    h(t^{(0)},t)  \,
     \left[ \Gamma_0 (t-t^{(0)}) \right]^{ -(\alpha+1/\delta)/\gamma}~~~\nonumber\\
\label{result1}
\eeqn
where $\Gamma(x)$ denotes the Euler gamma-function.
Likewise, repeating the same steps for $\sum_\ell \Gamma_\ell \Omega_\ell(t)$, we obtain
\beqn
 && \sum_\ell \Gamma_\ell \Omega_\ell(t) ~=~ \frac{ \Gamma_0 \Omega_0^{(0)}}{\gamma\delta}
     \left( \frac{m_0}{\Delta m}\right)^{1/\delta} \, \Gamma\left( \frac{\alpha+\gamma+1/\delta}{\gamma}\right)
          \nonumber\\
  && ~~~ \times~
     h(t^{(0)},t) \,
     \left[ \Gamma_0 (t-t^{(0)}) \right]^{ -(\alpha+\gamma+1/\delta)/\gamma}~.~~\nonumber\\
\label{result2}
\eeqn
Thus, dividing Eq.~(\ref{result2}) by Eq.~(\ref{result1}) and recalling that $\Gamma(x+1)/\Gamma(x) = x$,
we obtain
\beq
   \frac{   \sum_\ell \Gamma_\ell \Omega_\ell(t) } {\sum_\ell \Omega_\ell(t)} ~=~ 
    \frac{\alpha+1/\delta}{\gamma}\,
     \frac{1}{t-t^{(0)}}~.
\label{quotient_prediction}
\eeq
    
Comparing this result with the constraint equation in Eq.~(\ref{quotient}), 
the first thing we notice is that our model
has produced a power-law time dependence in the time {\it difference}\/ $t-t^{(0)}$ rather than in $t$ itself.
In principle, this therefore does not satisfy the criterion in Eq.~(\ref{quotient}).
However, this criterion can be {\it approximately}\/ satisfied 
so long as
\beq
        t^{(0)}~ \ll~ t~. 
\label{modelcond1}
\eeq
In other words, as we originally anticipated,
for stasis to emerge within our model we must restrict our attention to periods of time which
are sufficiently far beyond the $\phi_\ell$ production time that the initial ``edge'' effects have died away.
Indeed, the precision with which this power-law scaling requirement is satisfied
only increases the further we are from these initial edge effects.
Eq.~(\ref{modelcond1}) is thus our first condition for this model, indicating that
we do not expect stasis to develop in this model until considerably after $t^{(0)}$.
This thereby provides a natural {\it beginning}\/ to the stasis period.

Let us now assume that Eq.~(\ref{modelcond1}) is satisfied. 
Comparing the overall coefficients in Eqs.~(\ref{quotient}) and (\ref{quotient_prediction}) we then obtain a constraint on suitable values of $(\alpha,\gamma,\delta)$:
\beq
   \frac{1}{\gamma} \left( \alpha + \frac{1}{\delta}\right) ~=~ \frac{2 (1-\barOmega_M)}{4-\barOmega_M} ~.
\label{alphagammaconstraint}
\eeq
Equivalently, for any tower of states parametrized by $(\alpha,\gamma,\delta)$, 
this can be inverted in order to obtain the corresponding predicted value of $\barOmega_M$ during stasis, yielding
\beq
\boxed{
       ~~\barOmega_M ~=~  \frac{  2 \gamma\delta - 4(1+\alpha\delta) }{2 \gamma\delta - (1+ \alpha \delta) }~.
}
\label{stasisOmegaM}
\eeq 
For example,
with $(\alpha,\gamma,\delta)=(1,5,1)$, Eq.~(\ref{stasisOmegaM}) yields $\barOmega_M=1/4$.
However, for $(\alpha,\gamma,\delta)=(1,7,1)$, Eq.~(\ref{stasisOmegaM}) yields $\barOmega_M=1/2$.
{\it This is then a case of stasis with matter-radiation equality}\/!
In general, from Eq.~(\ref{stasisOmegaM}) we see that stasis with matter-radiation equality will occur provided
\beq
              \frac{1+\alpha \delta}{\gamma \delta} ~=~ \frac{2}{7}~.
\eeq

The fact that we must have $0\leq \barOmega_M\leq 1$ places bounds on 
the possible values of $\alpha$, $\gamma$, and $\delta$ that can give rise to stasis.
For example, in accordance with our expectation that the heavier $\phi_\ell$ 
will decay more rapidly than the lighter $\phi_\ell$, we can restrict our attention to cases with $\gamma>0$.
In such cases, 
Eq.~(\ref{stasisOmegaM}) in conjunction with $0\leq \barOmega_M\leq 1$ and the requirement that $\alpha+1/\delta >0$ [as indicated above Eq.~(\ref{result1})] immediately leads to a restricted range for $\alpha$:
\beq
               -\frac{1}{\delta} ~< ~ \alpha ~\leq~ \frac{\gamma}{2} - \frac{1}{\delta}~.
\label{range}
\eeq

The conditions  in Eqs.~(\ref{alphagammaconstraint}) and (\ref{stasisOmegaM})
emerged from demanding that our model satisfy Eq.~(\ref{quotient}).
However, Eq.~(\ref{quotient}) only emerged as the quotient of the two more fundamental constraints
in Eq.~(\ref{masterconstraints}).   We must therefore also demand that these constraints are each individually
satisfied.  Of course, having already satisfied Eq.~(\ref{quotient}), we need only concentrate on
one of these constraints.  
Let us assume that our model indeed gives rise to stasis for $t\gg t^{(0)}$, and then verify
that this assumption leads to a self-consistent result.
Assuming stasis for $t\gg t^{(0)}$, we
can write 
\beqn
              h(t^{(0)},t) ~&=&~  h(t^{(0)},t_\ast)\, h(t_\ast,t) \nonumber\\
               ~&=&~ h(t^{(0)},t_\ast)\, \left( \frac{t}{t_\ast} \right)^{2-6/(4-\barOmega_M)}
\label{factorizedh}
\eeqn
where $t_\ast\gg t^{(0)}$ is some fiducial time beyond which stasis has developed.
Inserting this into 
our result in Eq.~(\ref{result1}) then yields
\beqn
  &&  \sum_\ell \Omega_\ell (t) 
    ~=~  \frac{ \Omega_0^{(0)}}{\gamma\delta}
          \left( \frac{m_0}{\Delta m}\right)^{1/\delta} \, \Gamma\left( \frac{\alpha+1/\delta}{\gamma}\right)
          \nonumber\\
  && ~~~ \times\,
  h(t^{(0)}, t_\ast)  \left( \frac{t}{t_\ast} \right)^{2-6/(4-\barOmega_M)} \!
     \left[ \Gamma_0 (t-t^{(0)}) \right]^{ -(\alpha+1/\delta)/\gamma}\,.\nonumber\\
\label{result1new}
\eeqn
However, given the conditions in Eqs.~(\ref{modelcond1}) and (\ref{alphagammaconstraint}),
we see that 
the rather complicated time-dependence in Eq.~(\ref{result1new}) cancels!
This by definition verifies that we are indeed within 
a period of stasis, with a constant $\Omega_M\equiv \sum_\ell \Omega_\ell(t)$.
We are thus left with only one additional self-consistency constraint on our model:
\beqn
    \barOmega_M ~&=&~   \frac{\Omega_0^{(0)}}{\gamma\delta} \left( \frac{m_0}{\Delta m}\right)^{1/\delta}
          \Gamma\left( \frac{\alpha+1/\delta}{\gamma}\right) \nonumber\\  
        && ~~~~~~~ \times~ h(t^{(0)},t_\ast) \, \left( \frac{1}{\Gamma_0 t_\ast}\right)^{(\alpha+1/\delta)/\gamma}~.~~~~~
\label{residualconstraint}
\eeqn
At first glance, this constraint equation is not particularly illuminating.  
However, via Eqs.~(\ref{Omegalnew}), (\ref{alphagammaconstraint}), and (\ref{factorizedh}), we see that
\beqn
 \Omega_0^{(0)} \,h(t^{(0)},t_\ast)\, \left( \frac{1}{\Gamma_0 t_\ast}\right)^{(\alpha+1/\delta)/\gamma}\! &=&~ 
      \Omega_0(\tau_0) e^{\Gamma_0 (\tau_0-t^{(0)})}~~~\nonumber\\
   &\approx&~ e\, \Omega_0(\tau_0)
\label{relationn}
\eeqn
where $\tau_0\equiv 1/\Gamma_0$ and where in passing to the second line we have assumed that
$\tau_0\gg t^{(0)}$.
Substituting Eq.~(\ref{relationn}) into Eq.~(\ref{residualconstraint}) we thus obtain the constraint
\beq
  \barOmega_M ~\approx~  X \, \Omega_0(\tau_0) ~
\label{Cconstraint2}
\eeq
where
\beq
     X ~\equiv~ 
 \frac{e}{\gamma\delta} \left( \frac{m_0}{\Delta m}\right)^{1/\delta}
      \Gamma\left( \frac{\alpha+1/\delta}{\gamma}\right) ~.
\eeq
Note that this proportionality constant $X$ does not include any of the potentially large 
ratios of time intervals
that originally appeared in Eq.~(\ref{residualconstraint}).
Thus, as long as $\Delta m\sim m_0$, we find that $X\sim {\cal O}(1)$.

It is not difficult to interpret this result.
Ordinarily, $\barOmega_M$ receives significant contributions $\Omega_\ell(t)$
from each of the individual components.    However, by the time we reach $t\approx \tau_0$, 
 the lightest component is just about to begin decaying while all of the heavier components have already decayed to various extents.   Thus the dominant contribution to $\barOmega_M$ at $t=\tau_0$ comes from $\Omega_0(\tau_0)$, while the contributions $\Omega_\ell(\tau_0)$ with $\ell\geq 1$ are exponentially suppressed.     We then expect that $\barOmega_M$ will be approximately equal to $\Omega_0(\tau_0)$, with
 the proportionality coefficient $X$ in Eq.~(\ref{Cconstraint2})
including (and the difference $X-1$ quantifying) the residual contributions from all of the heavier states as well as the approximations made in passing from the exact (discrete) sum in Eq.~(\ref{bigsum1})
 to the integral in Eq.~(\ref{integralform}).   Of course, for $t\gsim \tau_0$, our decay process ends and our system exits from the stasis state.

We therefore conclude that our system satisfies the requirements for an extended period of stasis during the decay
process so long as the conditions in Eqs.~(\ref{modelcond1}), (\ref{stasisOmegaM}), and (\ref{Cconstraint2})
are satisfied.  
Alternatively, comparing with the parameter list in Eq.~(\ref{parameterlist}), we see that
Eq.~(\ref{modelcond1}) constrains $t^{(0)}$ (or simply requires that this production time be significantly earlier 
than any ensuing period of stasis), while
Eq.~(\ref{stasisOmegaM}) gives the resulting value of $\barOmega_M$
in terms of $\lbrace \alpha,\gamma,\delta\rbrace$
and Eq.~(\ref{Cconstraint2}) involves all of the remaining parameters and can be viewed as constraining the overall scale $\Omega_0(\tau_0)$ [or equivalently $\Omega_0^{(0)}$].
Although it might seem that the constraints in Eqs.~(\ref{stasisOmegaM}) and (\ref{Cconstraint2})
represent fine-tunings that we must impose on the parameters of our model,
we shall find in Sect.~\ref{attractor} that
no such fine-tuning is required, and that stasis ultimately emerges within this model
even if these constraints are not originally satisfied.

As evident from the above derivation, we have made a number of approximations in obtaining these results.
In particular, in order to evaluate the sum in Eq.~(\ref{bigsum1}), we made three approximations:
we treated our tower of states as a continuum, as in Eq.~(\ref{continuum});
we then took the limits in Eq.~(\ref{edge}), thereby essentially disregarding ``edge'' effects at the top and bottom of our tower;
and finally we made the approximation in Eq.~(\ref{warp}).
These approximations indicate that the precise power-law time-dependence that is required for stasis according
to the constraints in  Eq.~(\ref{masterconstraints}) is 
at best only approximate for a realistic discrete tower of states $\phi_\ell$.
However, it is easy to understand the situations in which these approximations might fail, thereby disturbing the 
power-law behavior and consequently disrupting the resulting stasis.
In general, for $\gamma>0$, the decay widths $\Gamma_\ell$ increase as a function of the masses $m_\ell$, 
implying that the heavier $\phi_\ell$ states tend to decay first while the lighter $\phi_\ell$ decay later.
For this reason, we expect that the approximations in Eqs.~(\ref{edge}) and (\ref{warp})  --- approximations
which primarily come into play only at the tops or bottoms of our towers --- will primarily
affect the behavior of our system only at extremely early and/or late times, respectively. 
This is consistent with our further condition in Eq.~(\ref{modelcond1}).
Indeed, as a particularly dramatic example of these edge effects,
we observe that  with $\gamma>0$ and $\alpha+1/\delta >0$, our integral results in Eqs.~(\ref{result1}) and
(\ref{result2})
actually diverge as $t$ approaches the initial production time $t^{(0)}$.  Of course, this divergence is completely spurious, since our actual model has a finite number of states within the decaying tower and
thus contains no such divergences.    This is clear illustration of the fact that our integral approximation
is highly inaccurate at such early times.

It is precisely the failure of our approximations at extremely early and/or late times which explains why our stasis
(which would otherwise have been strictly eternal, \ie, time-independent, as in Sect.~\ref{deriving_constraints})
actually has a beginning and an end, emerging in full force only after the first few decays have already occurred and ending as the system approaches the final decays.
We therefore regard these edge effects as beneficial features, 
indicating that there will necessarily exist both an  entrance into, as well as an  exit from, our stasis epoch, such as would
be required in any realistic cosmological scenario.
At other times far from these ``edge'' effects, we shall nevertheless find numerically that this stasis is quite robust.

\begin{figure*}[t!]
\centering
\includegraphics[keepaspectratio, width=0.49\textwidth]{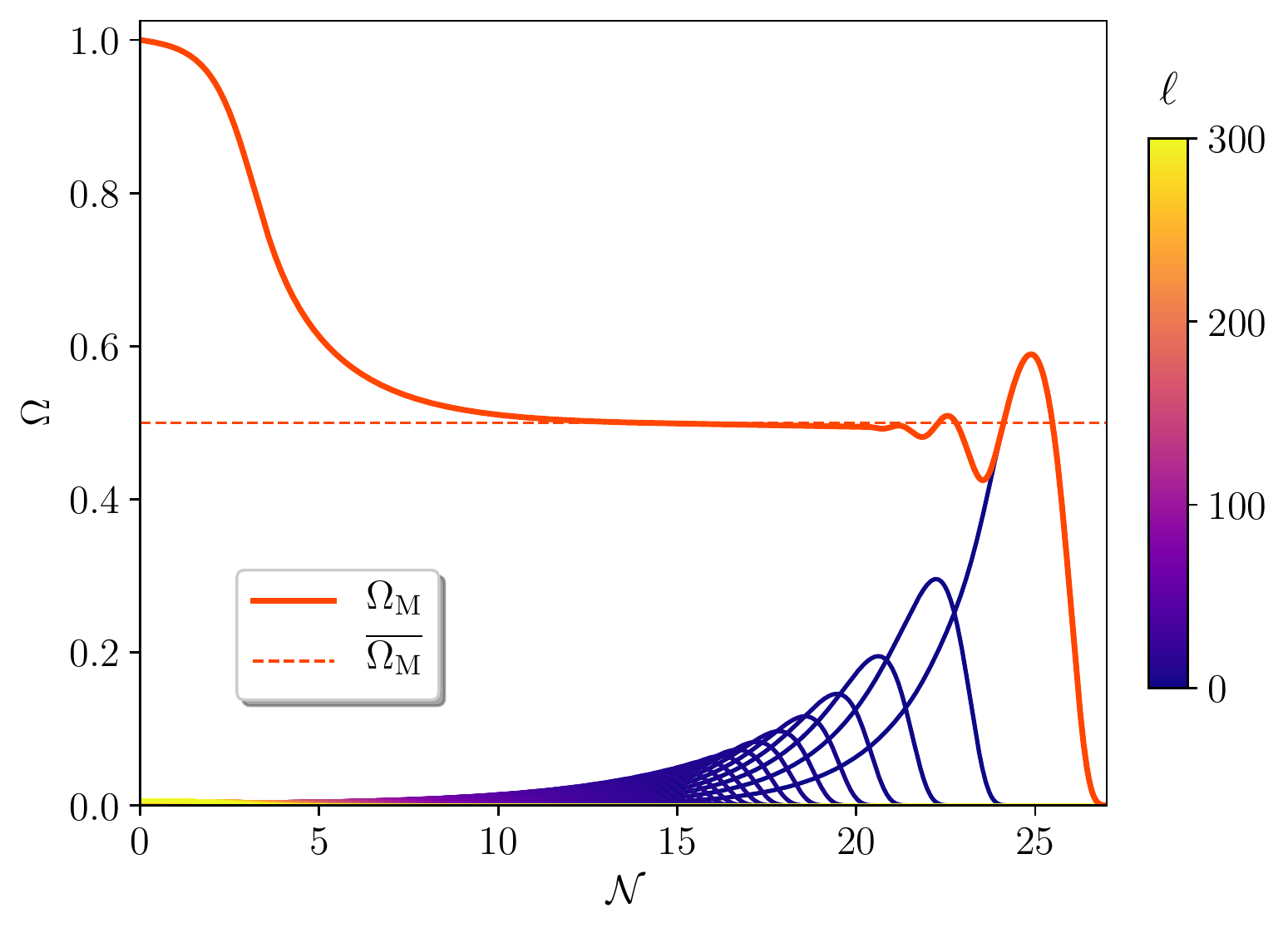}
\hfill
\includegraphics[keepaspectratio, width=0.49\textwidth]{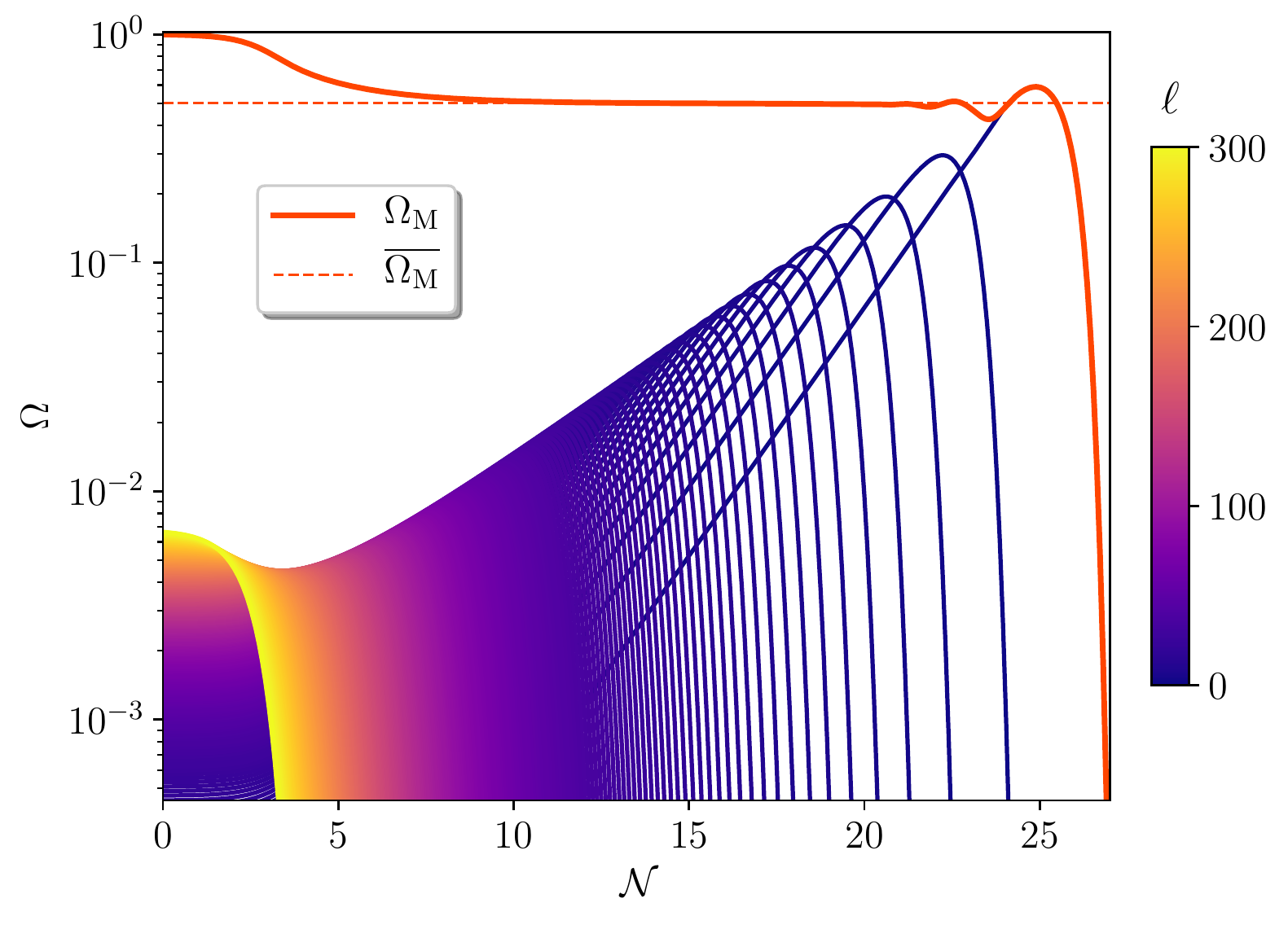}
\caption{The individual matter abundances $\Omega_\ell$ (orange/blue) and the corresponding total matter abundance $\Omega_M$ (red), plotted as functions of the number $\calN$ of $e$-folds since the initial $\phi_\ell$ production.   These curves were generated through a direct numerical solution of the relevant Boltzmann equations for our discrete tower of decaying states without invoking any approximations, and correspond to the parameter choices
 $(\alpha,\gamma,\delta)=(1,7,1)$ [for which $\barOmega_M=1/2$],  with $\Delta m = m_0$, $N=300$, and $\Gamma_{N-1}/H^{(0)}=0.01$.
In the left panel the abundances are plotted on a linear scale, while in the right panel these same abundances are plotted on a logarithmic scale.
We see that our system begins with $\Omega_M=1$ at $t=t^{(0)}$, with each individual $\Omega_\ell$ component exhibiting a non-trivial behavior, first
growing as a power-law due to cosmological redshifting before ultimately decaying exponentially.   
Despite this complexity, their sum $\Omega_M$ nevertheless evolves towards a stasis epoch in which $\Omega_M$ remains essentially constant
for approximately $15$ $e$-folds before exiting stasis.   Even longer periods of stasis can be produced if $N$ is increased.  This thereby provides a concrete realization of the basic stasis mechanism sketched in Fig.~\ref{fig1}.   Eventually the stasis ends as we approach the final decays of the lightest modes. }
\label{fig2}
\end{figure*}

In order to illustrate the stasis epoch --- together with its beginning and end --- we can perform a direct numerical study of this system, 
with the time evolution determined through exact numerical solutions of the relevant Boltzmann equations for our discrete tower of decaying states without 
any approximations.   In Fig.~\ref{fig2} we illustrate the behavior of the individual abundances $\Omega_\ell(t)$ as well as
 the resulting behavior for their sum $\Omega_M(t)$,
where for concreteness we have chosen the benchmark values
 $(\alpha,\gamma,\delta)=(1,7,1)$ [for which $\barOmega_M=1/2$],  with $\Delta m = m_0$, $N=300$, and $\Gamma_{N-1}/H^{(0)}=0.01$.
As anticipated, we see that we not only have a robust period of stasis lasting $\approx 15$ $e$-folds, but we also have a clear entrance into this epoch as well as an exit from it.
During the stasis epoch, we nevertheless find that  $\Omega_M$ holds steady at the value $\barOmega_M$ predicted in Eq.~(\ref{stasisOmegaM}).
As we shall shortly see, similar results hold for other values of our benchmarks as well,
    leading to other values for $\barOmega_M$.

Two final comments are in order.
First, we observe from the right panel of Fig.~\ref{fig2} that
although the individual abundance contributions $\Omega_\ell(t)$ have fairly complicated behaviors, they
each attain a maximum value at approximately $t\approx \tau_\ell\equiv 1/\Gamma_\ell$ before beginning their exponential decays.   [More precisely, the maximum value in each case occurs at 
$t_\ell = \zeta \tau_\ell$ where $\zeta\equiv 2-6/(4-\barOmega_M)$, but these extra factors of $\zeta$ will
cancel below and can thus be ignored.]
Moreover, when plotting $\log\Omega$ versus ${\cal N}\sim\log t$ (such as in this panel), we see that these maximum values all lie along a straight line which we may consider to be the ``envelope'' function for the individual $\Omega_\ell(t)$.
This linear envelope function is a critical ingredient in producing the stasis state.

It is easy to see how this envelope function emerges
during stasis.  
From Eqs.~(\ref{Omegalnew}) and (\ref{factorizedh}) 
we see that
\beq
 \Omega_\ell(\tau_\ell)~=~ \Omega_\ell^{(0)} \,h(t^{(0)},t_\ast)\, 
             \left( \frac{\tau_\ell}{t_\ast}\right)^{(\alpha+1/\delta)/\gamma} e^{-1}
\eeq
where $t_\ast$ is any fiducial time within stasis and where 
we have approximated $e^{-\Gamma_\ell(\tau_\ell -t^{(0)})} \sim e^{-1}$ for $\tau_\ell\gg t^{(0)}$.    We then find
\beqn
        \frac{\Omega_\ell(\tau_\ell)}{\Omega_0(\tau_0)} 
~&=&~ \frac{\Omega_\ell^{(0)}}{\Omega_0^{(0)}} 
                    \left( \frac{\tau_\ell}{\tau_0}\right)^{(\alpha+1/\delta)/\gamma}\nonumber\\
~&=&~ \left( \frac{m_\ell}{m_0} \right)^\alpha
                    \left( \frac{\tau_\ell}{\tau_0}\right)^{(\alpha+1/\delta)/\gamma}\nonumber\\
~&=&~ \left( \frac{\Gamma_\ell}{\Gamma_0} \right)^{\alpha/\gamma}
                    \left( \frac{\tau_\ell}{\tau_0}\right)^{(\alpha+1/\delta)/\gamma}\nonumber\\
~&=&~   \left( \frac{\tau_\ell}{\tau_0}\right)^{1/(\gamma \delta)} ~,
\eeqn
indicating that this envelope line has constant positive slope $1/(\gamma\delta)$.

The existence of this envelope line provides an added perspective regarding the constraint in Eq.~(\ref{Cconstraint2}).   It is clear that this rising envelope line must eventually intersect the horizontal $\barOmega_M={\rm constant}$ line.   What Eq.~(\ref{Cconstraint2}) tells us is that this intersection point occurs near $t\approx \tau_0$, as illustrated in the right panel of Fig.~\ref{fig2}.

Our second comment is that we are now also in a position to 
estimate the {\it duration}\/ of the stasis state.
In general, if we disregard the ``edge effects'' at the beginning and end of the decay process,
we can roughly identify the stasis state as stretching from the decay of the 
heaviest state in the tower at $t\approx \tau_{N-1}$ until the decay of the lightest state at $t\approx \tau_0$, where we have treated each of these quantities as significantly greater than $t^{(0)}$.  
We then find that the number of $e$-folds during stasis is given by
\beqn
     \calN_s ~&\equiv& ~ \log\left[ \frac{a(t=\tau_0)}{a(t=t_{N-1})}\right] ~=~ \frac{2}{4-\barOmega_M} \log \left( \frac{\Gamma_{N-1}}{\Gamma_0}\right)\nonumber\\
       ~&=&~    \frac{2\gamma }{4-\barOmega_M} \log\left( \frac{m_{N-1}}{m_0}\right)\nonumber\\
           ~&=&~    \frac{2\gamma }{4-\barOmega_M} \log\left[ 1+ \frac{\Delta m}{m_0} (N-1)^\delta\right]\nonumber\\ 
~&\approx&~    \frac{2\gamma \delta}{4-\barOmega_M} \log N
\label{NN}
\eeqn
where we have used Eq.~(\ref{scalefactor}) in the first equality and where in passing
 to the final line we have taken $N\gg 1$ and $\Delta m/m_0\sim {\cal O}(1)$.
We thus see that we can adjust the number of $e$-folds associated with the stasis epoch simply by
adjusting the number of states in the tower.

\FloatBarrier
\section{Stasis as a global attractor \label{attractor}}

In previous sections we have studied the properties of the stasis state and developed a model in which this state naturally arises.
Indeed, in Fig.~\ref{fig2} we demonstrated this numerically for a particular choice of parameters in our model.
However, it may seem that this choice was somehow  fine-tuned.
To address this issue, we shall now return to the basic dynamical equations that underlie this system and demonstrate that
the stasis state is actually a {\it global attractor}\/ for this system.
Thus, regardless of the particular parameter choices we might make within our model, we are inevitably drawn into a stasis epoch.

We begin our analysis with Eq.~(\ref{convert3}).  Indeed, this equation serves as the fundamental equation of motion for our system and thus
governs its dynamics.
Our goal, then, is to demonstrate that all solutions to this equation within our model framework inevitably head towards the stasis solution $\Omega_M\to \barOmega_M$,
where $\barOmega_M$ is the value of $\Omega_M$ during stasis.
Unfortunately, Eq.~(\ref{convert3}) contains two quantities whose general connections to $\Omega_M$ and $\barOmega_M$ are not obvious:
these are the $\ell$-sum $\sum_\ell \Gamma_\ell \Omega_\ell$ and the Hubble parameter $H(t)$.
Our first task will therefore be to derive general expressions for each of these quantities within our model,
{\it but without assuming stasis}\/.

Let us first consider $\sum_\ell \Gamma_\ell \Omega_\ell$.   
We have already evaluated this quantity in Sect.~\ref{model},
obtaining the result in Eq.~(\ref{result2}).
Indeed, like the corresponding result in Eq.~(\ref{result1}), this result is completely general
and does not rely on any assumption of stasis.
As a result, the quotient of these two results in Eq.~(\ref{quotient_prediction}) is
also completely general.    We therefore have the result
\beqn
\sum_\ell \Gamma_\ell \Omega_\ell ~&=&~ \left( \frac{\alpha+ 1/\delta}{\gamma} \right) \frac{\Omega_M}{t-t^{(0)}}\nonumber\\
       &=&~ \left\lbrack \frac{2(1-\barOmega_M)}{4-\barOmega_M} \right\rbrack \frac{\Omega_M}{t-t^{(0)}}~,
\label{ellsumeval}
\eeqn
where in passing to the second line we have used the result in Eq.~(\ref{alphagammaconstraint}).  Note, in particular, that
the result in Eq.~(\ref{alphagammaconstraint}) also holds independently of stasis since it is nothing more than a rewriting of the $(\alpha,\gamma,\delta)$ parameters in terms of the eventual stasis value $\barOmega_M$.
We thus see from Eq.~(\ref{ellsumeval}) that in general $\sum_\ell \Gamma_\ell \Omega_\ell$ depends on 
both $\Omega_M$ and $\barOmega_M$.

Let us now turn to the Hubble parameter $H(t)$.   We previously evaluated $H(t)$ in Eq.~(\ref{Hubble}), but that derivation assumed stasis.
We must therefore now proceed more generally by integrating Eq.~(\ref{acceleq}).
This immediately yields the relation
\beq
 \frac{1}{H} - \frac{1}{H^{(0)}}  ~=~ (t-t^{(0)}) \left[ \frac{4-\langle \Omega_M\rangle}{2}\right]
\label{intermedHubble}
\eeq
where $H^{(0)}$ is the Hubble value at $t=t^{(0)}$ and where
 $\langle \Omega_M\rangle$ at any time $t$ is the time-averaged value of $\Omega_M$ since $t=t^{(0)}$:
\beq
               \langle \Omega_M\rangle ~\equiv~ \frac{1}{t-t^{(0)}} \int_{ t^{(0)} }^t dt' \, \Omega_M(t')~.
\label{avgHubble}
\eeq
Eq.~(\ref{intermedHubble}) then immediately yields
\beq
               H(t) ~=~  \frac{2}{4-\langle\Omega_M\rangle}\,  \frac{1}{t-t^{(0)}} ~,
\label{Hubble2}
\eeq
where we have assumed $H^{(0)} (t-t^{(0)}) \gg 1$.   As expected, this result reduces to the result in Eq.~(\ref{Hubble}) if we are within
an eternal stasis, but otherwise depends on the complete time-history of the Hubble parameter since $t=t^{(0)}$ and thus makes absolutely no assumptions about the actual time-evolution of $\Omega_M$.

Inserting our results for $\sum_\ell \Gamma_\ell \Omega_\ell$ and $H(t)$ from Eqs.~(\ref{ellsumeval}) and (\ref{Hubble2}) into  Eq.~(\ref{convert3}), we  find that our equation of motion for this system now takes the form
\beq
           \frac{d\Omega_M}{dt} ~=~  \frac{\Omega_M}{t-t^{(0)}}
       \left [  \frac{ 2(1-\Omega_M)}{4-\langle \Omega_M\rangle } -  \frac{2(1-\barOmega_M)}{4-\barOmega_M}\right]~.
\label{neweom}
\eeq
Of course, this result immediately allows us to verify that $d\Omega_M/dt=0$ when $\Omega_M=\langle \Omega_M\rangle = \barOmega_M$,
 consistent with our original (eternal) stasis solution.   
However, in general, we see that this equation --- although first-order in time-derivatives --- actually depends on {\it two}\/ time-dependent variables, $\Omega_M$ and $\langle \Omega_M\rangle$, which need not have any direct relation to each other.

One way to analyze the dynamics of this system is to recognize that the definition of $\langle \Omega_M\rangle$ in Eq.~(\ref{avgHubble}) actually provides us with another first-order differential equation, this one for $\langle \Omega_M \rangle$:
\beq
            \frac{d \langle \Omega_M\rangle }{dt} ~=~  \frac{1}{t-t^{(0)}} \left[  \Omega_M - \langle \Omega_M\rangle \right]~.
\label{eomavg}
\eeq
Indeed, this equation is nothing but the time-derivative of the definition in Eq.~(\ref{avgHubble}).
The two coupled first-order equations (\ref{neweom}) and (\ref{eomavg}) could then be combined into a single {\it second-order}\/ differential equation for $\Omega_M$.
However, it will prove simpler (and more conceptually transparent) to retain these two first-order equations, treating $\Omega_M$ and $\langle \Omega_M\rangle$ as independent variables, and then study the behavior of the corresponding
 two-variable  dynamical system
\beq
          \begin{cases}
           \displaystyle  \frac{d\Omega_M}{dt} &\!=~ \displaystyle  \frac{1}{t-t^{(0)}} \, f(\Omega_M, \langle \Omega_M\rangle) \phantom{\Biggl |}\\
       \displaystyle \frac{d\langle \Omega_M\rangle}{dt} &\!=~  \displaystyle \frac{1}{t-t^{(0)}} \, g(\Omega_M, \langle \Omega_M\rangle)~,
         \end{cases}
\label{system}
\eeq
where
\beqn
    ~~ f(\Omega_M, \langle \Omega_M\rangle) ~&\equiv&~ \Omega_M \left[
                 \frac{ 2(1-\Omega_M)}{4-\langle \Omega_M\rangle } -  \frac{2(1-\barOmega_M)}{4-\barOmega_M}\right]\phantom{\Biggl |} \nonumber\\
     g(\Omega_M, \langle \Omega_M\rangle) ~&\equiv&~ \Omega_M - \langle\Omega_M\rangle~.\nonumber\\
\label{fgdef}
\eeqn
We note in passing that we are always free to shift our independent variable for this system from $t$ to $\log t$, thereby rendering the right sides of Eq.~(\ref{system}) independent of time.   This system is therefore effectively autonomous.

It is clear from these equations that the stasis solution corresponds to $\Omega_M=\langle \Omega_M\rangle = \barOmega_M$.
Moreover, this solution will be a local attractor if both of the eigenvalues of the corresponding Jacobian matrix $J$ are negative when $J$ is evaluated at the stasis point.   
In our case, the Jacobian matrix is given by  $J= (t-t^{(0)})^{-1} \widehat J$ where $\widehat J$ is the time-independent Jacobian matrix
\beq
       \widehat  J ~=~ \begin{pmatrix}
               \partial_{\Omega_M}  f & 
               \partial_{\langle \Omega_M \rangle} f \\
               \partial_{\Omega_M}  g & 
               \partial_{\langle \Omega_M  \rangle} g 
       \end{pmatrix}~.
\label{Jacobian}
\eeq
Evaluated at the stasis point, this matrix takes the form
\beq
          \widehat J \,\bigl |_s ~=~ 
          \begin{pmatrix}
         A & B \\
         1 & -1 \\
         \end{pmatrix}
\eeq
where the symbol $|_{s}$ indicates that the expression is evaluated 
within stasis
and where
\beq
A ~\equiv ~ -\frac{2\barOmega_M}{4-\barOmega_M}~,~~~~
B ~\equiv~ \frac{2 \barOmega_M (1-\barOmega_M)}{(4-\barOmega_M)^2}~.~
\label{ABdef}
\eeq
The corresponding eigenvalues are therefore given by
\beq
\lambda_\pm  ~=~ \frac{   -(4+\barOmega_M) \pm \sqrt{\barOmega_M^2 - 16 \barOmega_M + 16}}
         {2(4-\barOmega_M)} ~,~~
\eeq
whereupon we see that
\beq
      \lambda_\pm <0  ~~~~{\rm for~all}~~  0\leq \barOmega_M\leq 1~.
\eeq
We therefore conclude that the stasis state is (at least) a local attractor, stable against small deviations $\delta \Omega_M$ and $\delta \langle \Omega_M \rangle$.   Thus, if perturbed, our system will necessarily return to the stasis state.

\begin{figure}[t!]
\centering
\includegraphics[keepaspectratio, width=0.45\textwidth]{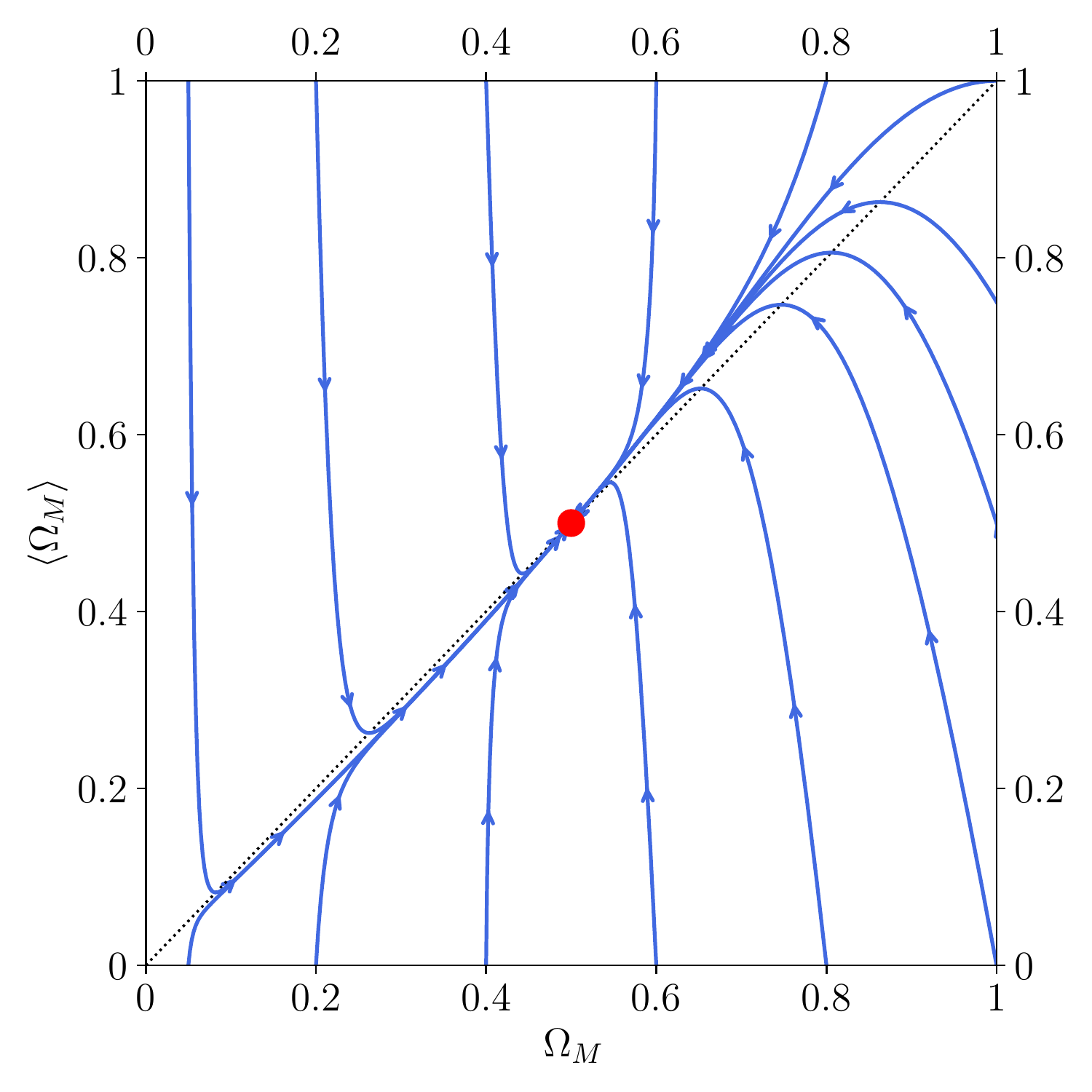}
\caption{Trajectories (blue curves) within the $(\Omega_M,\langle\Omega_M\rangle)$-plane for the system defined in Eq.~(\ref{system}).  For concreteness we have shown the case with stasis values 
$\Omega_M = \langle \Omega_M\rangle = \barOmega_M=1/2$ (central red dot), but similar results emerge for all stasis values $\barOmega_M$.   Because this system is effectively autonomous, any point along a given trajectory can be taken as a starting point without affecting the subsequent trajectory.   Given these trajectories, we see that the  stasis state serves as a global attractor for this system.}
\label{attractorfig}
\end{figure}

Of course, the question remains as to whether our system will flow to the stasis state if we are originally {\it far}\/ from it.    
In such cases, $\Omega_M$ and 
$\langle \Omega_M\rangle$ need not be close to $\barOmega_M$  and need not even be close to each other.   The best way to answer this question is therefore to assume arbitrary initial values for $\Omega_M$ and $\langle \Omega_M\rangle$ within the range $0\leq \lbrace \Omega_M,\langle \Omega_M\rangle\rbrace \leq1 $, and then examine how these two variables evolve under the time-evolution specified in Eq.~(\ref{system}).    In other words, we seek to determine the {\it trajectories}\/ that our system maps out in the $(\Omega_M,\langle\Omega_M\rangle)$-plane according to Eq.~(\ref{system}).    
In Fig.~\ref{attractorfig}, we illustrate these trajectories for the case in which the
stasis solution is given by $\Omega_M = \langle\Omega_M\rangle = \barOmega_M= 1/2$.
As we see, all trajectories for this system ultimately flow towards this stasis point.
Moreover, similar trajectory maps emerge regardless of the chosen stasis point.
We therefore conclude that the stasis state is not only a local attractor, but actually a {\it global}\/ one.

At first glance,
given that our initial conditions at the production time $t=t^{(0)}$ are always
 $\Omega_M= \langle \Omega_M\rangle =1$,
it might seem that most of the trajectories plotted in Fig.~\ref{attractorfig} are irrelevant for our situation.
However, we must recall that in deriving the differential equations in Eq.~(\ref{system}) 
that govern our system
--- particularly in establishing Eq.~(\ref{ellsumeval}) ---
we made a number of approximations when passing
from the discrete sum in Eq.~(\ref{bigsum1}) to the integral form in Eq.~(\ref{integralform}).   
These approximations were discussed in detail in Sect.~\ref{model}, and are listed in Eqs.~(\ref{continuum}),
(\ref{edge}), and (\ref{warp}).
Some of these approximations turn out to be of little consequence for the current situation, such as taking the continuum limit as in Eq.~(\ref{continuum}).
However, the $m_{\rm max}\to \infty$ approximation within Eq.~(\ref{edge})
can be significant, since this approximation essentially eliminates the transient ``edge'' effects that arise  at early times immediately after the production time $t^{(0)}$, when the heaviest $\phi_\ell$ states are just beginning to decay.   Since the integral approximation ignores these edge effects, the corresponding dynamical equations in Eq.~(\ref{system}) are valid only after these edge effects have died away.

These edge effects can nevertheless have significant impacts on the dynamics of our system.   For example, if the decay widths are relatively large, the most massive $\phi_\ell$ states will decay extremely promptly, thereby inducing a significant initial depletion of $\Omega_M$ and $\langle \Omega_M\rangle$ that occurs before Eq.~(\ref{system}) becomes valid.    We shall see explicit examples of this phenomenon below.
Thus, while the differential equations in Eq.~(\ref{system}) accurately describe the dynamics of our system {\it after}\/ the initial edge effects have died away, these initial edge effects are capable of shifting $\Omega_M$ and $\langle \Omega_M\rangle$ to new locations $\Omega'_M$ and $\langle \Omega_M\rangle'$ in the $(\Omega_M,\langle\Omega_M\rangle)$-plane 
which are quite far 
from their original $t^{(0)}$ location
$\Omega_M=\langle\Omega_M\rangle=1$.
It is then these new values $\Omega_M'$ and $\langle \Omega_M\rangle'$ which constitute the ``initial'' point for the subsequent trajectory in Fig.~\ref{attractorfig}.
Thus, to the extent that we regard our dynamical system as governed by the differential equations in Eq.~(\ref{system}), we should properly regard the initial conditions for this dynamics to be those associated with $\Omega'_M$ and $\langle \Omega_M\rangle'$.    In other words, the initial conditions for the trajectories in Fig.~\ref{attractorfig} should be taken to be those that exist not at the production time $t^{(0)}$, but rather at a subsequent time after which the initial edge effects have died away.    These edge effects can thus can place our system on an entirely different trajectory than that beginning at $\Omega_M=\langle \Omega_M\rangle =1$.

Fortunately, this observation does not affect our conclusion that the stasis state is a global attractor.    No matter what behaviors are induced within our system by the initial edge effects, we know that they must ultimately yield values $\Omega_M'$ and $\langle \Omega_M\rangle'$ which remain within the plane shown in Fig.~\ref{attractorfig}.   Indeed, we have seen in Fig.~\ref{attractorfig} that {\it any}\/ such trajectory within this plane eventually leads to the stasis state.    Thus our stasis state remains a global attractor even when the initial edge effects are included.

\begin{figure*}
\centering
\includegraphics[keepaspectratio, width=0.6\textwidth]{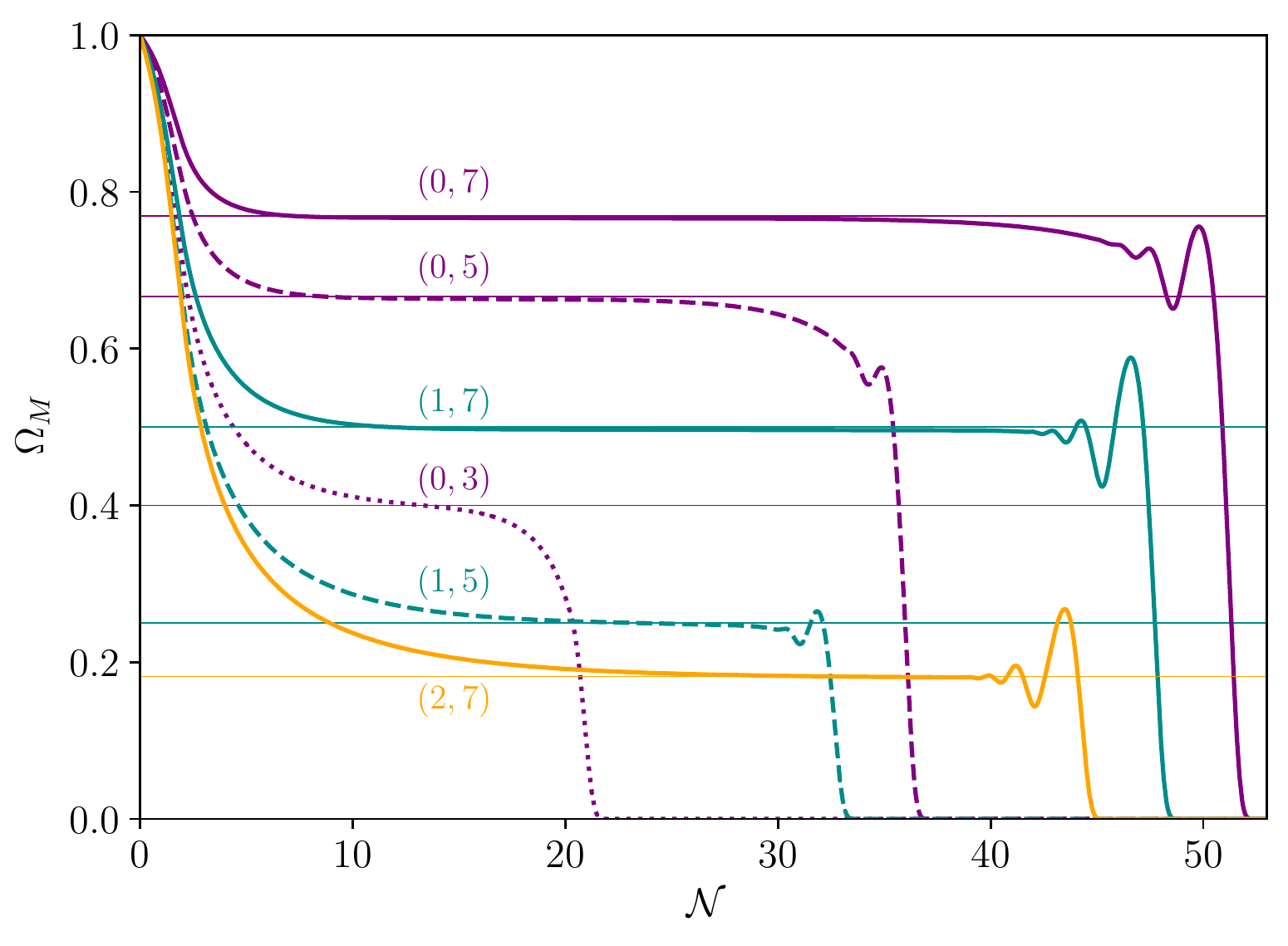}
\vskip -0.07 truein
\caption{The total matter abundance $\Omega_M$, plotted as a function of the number $\calN$ of $e$-folds since the initial $\phi_\ell$ production 
time $t^{(0)}$ for a variety of different benchmark values of $(\alpha,\gamma)$ satisfying
the constraints in Eq.~(\ref{range}). 
For each plot we have taken $\delta=1$,
  $\Delta m=m_0$, $N=10^5$, and $H^{(0)}/\Gamma_{N-1} = 0.1$
as relevant benchmarks.
Each curve begins at $\Omega_M=1$ at the initial time $t^{(0)}$ and eventually falls to $\Omega_M\to 0$ as $t\to \infty$.   However, we see that in each case there is a prolonged epoch lasting many $e$-folds during which $\Omega_M(t)$ settles into a stasis state with $\Omega_M(t)=\barOmega_M$;  indeed the duration of the stasis state can be increased at will simply by increasing $N$.    For each value of $(\alpha,\gamma)$, the corresponding stasis values $\barOmega_M$ are indicated with horizontal dashed lines.   
Ultimately in each case, the period of stasis ends as we reach the final $\phi_\ell$ decays;  the resulting oscillations in $\Omega_M$ reflect the discretization effects associated with the successive final decays.}
\label{fig4}
\bigskip
\smallskip
\centering
 ~\hskip 0.4 truein \includegraphics[keepaspectratio, width=0.65\textwidth]{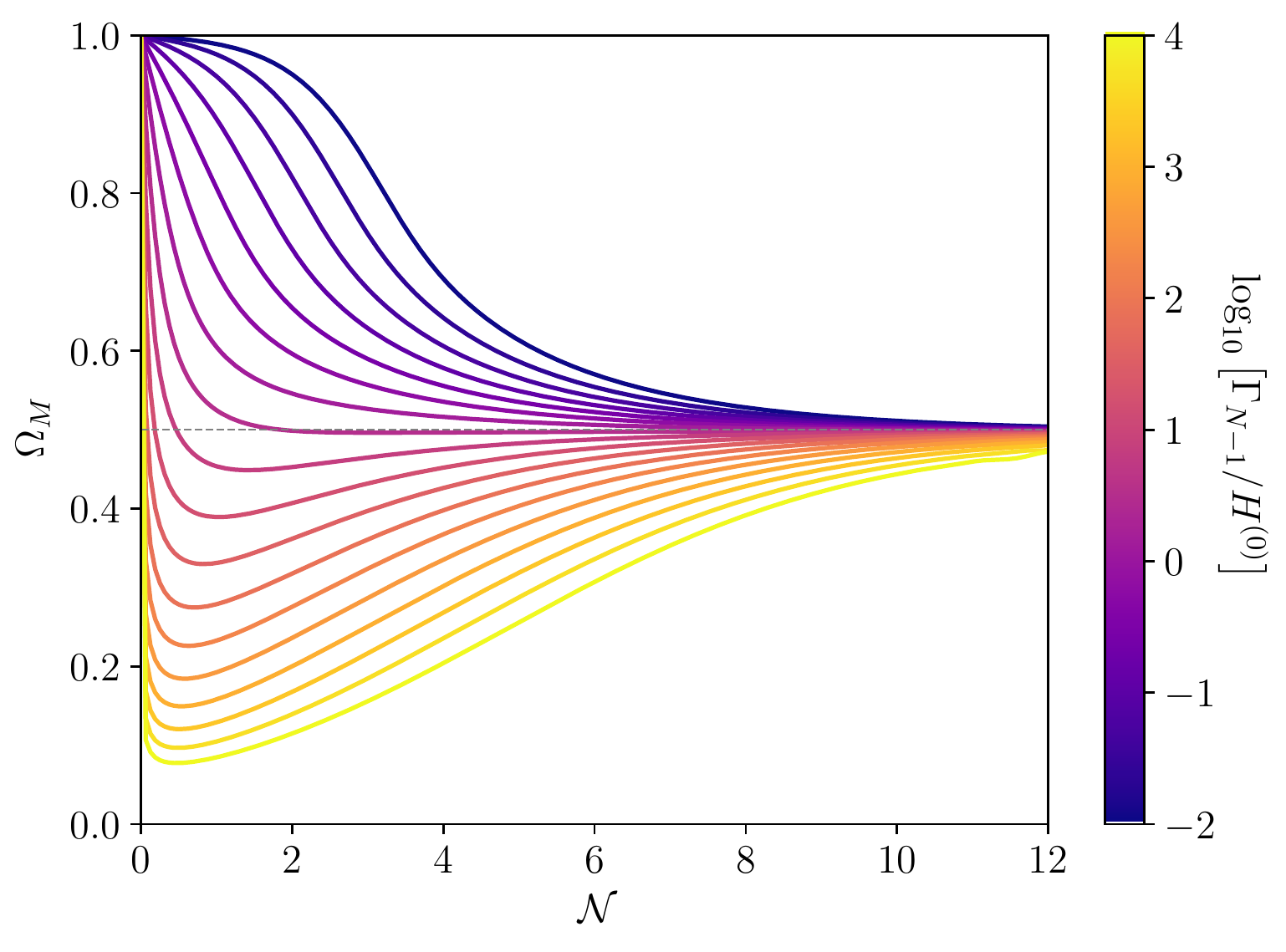}
\vskip -0.07 truein
\caption{The total matter abundance $\Omega_M$, plotted as a function of the number $\calN$ of $e$-folds since the initial $\phi_\ell$ production 
time $t^{(0)}$ for a variety of different values of $\Gamma_{N-1}/H^{(0)}$, where $H^{(0)}$ is the value of the Hubble parameter at the $\phi_\ell$ production time $t^{(0)}$.    For this plot we have chosen the benchmark values $(\alpha,\gamma,\delta)= (1,7,1)$ and taken $\Delta m=m_0$ and $N=300$.
We see that as $\Gamma_{N-1}/H^{(0)}$ increases, our decays occur more rapidly and $\Omega_M$ therefore drops more rapidly from its initial value $1$ after $t^{(0)}$.    Indeed, for $\Gamma_{N-1}/H^{(0)}\gsim 3$, the initial drop in $\Omega_M$ is so rapid that $\Omega_M$ initially drops {\it below}\/ the stasis value $\barOmega_M$ before rebounding due to cosmological expansion;  $\Omega_M$ therefore ultimately approaches the stasis value $\barOmega_M$ from below.
Such curves then correspond to trajectories in Fig.~\ref{attractorfig} for which stasis is approached from the $\Omega_M< 1/2$ region.
However,  in all cases our system is inevitably drawn towards the same stasis configuration, illustrating that the stasis state is indeed a global attractor for this system.}
\label{fig5}
\end{figure*}

Changing the parameters of our model can alter not only the initial edge effects but also the resulting stasis abundance $\barOmega_M$.   However, as discussed above, the emergence of a stasis state at $\barOmega_M$ is a robust phenomenon which persists even as the values of the parameters of our model are changed.  For example, in Fig.~\ref{fig4}, we show the time-evolution of $\Omega_M$ for a variety of values of $\alpha$ and $\gamma$, taking $\delta=1$ as a fixed benchmark.   In each case, we see that $\Omega_M$ immediately begins to fall from 1 but eventually enters into a stasis epoch in which $\Omega_M$ is essentially fixed at the corresponding stasis value $\barOmega_M$ predicted by Eq.~(\ref{stasisOmegaM}).   Indeed, as evident within Fig.~\ref{fig4}, this stasis epoch can persist for many $e$-folds (and can be extended indefinitely by increasing $N$)  before dissipating when the lightest $\phi_\ell$ states decay.  We emphasize that these plots were generated as direct numerical solutions of the relevant Boltzmann equations without invoking any approximations.   They thus accurately represent the actual behavior of our system.

Even more dramatically, the emergence of the stasis state is also robust against changes in the overall decay rate for the $\phi_\ell$ particles.  This feature is illustrated in Fig.~\ref{fig5}.   In general, for any tower of $\phi_\ell$ states whose decay rates $\Gamma_\ell$ are connected through the scaling relations in Eq.~(\ref{scalings}), 
we can parametrize this overall decay rate through the dimensionless quantity $\Gamma_{N-1}/H^{(0)}$ where
 $\Gamma_{N-1}$ is the decay rate $\Gamma_\ell$ for the most massive particle (\ie, that with  $\ell=N-1$) and where
$H^{(0)}$
denotes the value of the Hubble parameter at the production time $t=t^{(0)}$.   
When $\Gamma_{N-1}/H^{(0)}\ll 1$, the decays begin relatively slowly after the production time $t^{(0)}$ and the initial edge effects are correspondingly mild.  As a result, the approximations leading to Eq.~(\ref{system}) become valid rather quickly, with $\Omega_M'$ and $\langle \Omega_M\rangle'$ still fairly close to $1$.
By contrast, when $\Gamma_{N-1}/H^{(0)}\gg 1$, a significant number of the heaviest $\phi_\ell$ states decay promptly after $t^{(0)}$, leading to a rapid depletion in $\Omega_M$ and $\langle \Omega_M\rangle$.
Indeed, in such cases $\Omega_M$ may even initially fall {\it below}\/ the eventual stasis value $\barOmega_M$,
as illustrated in Fig.~\ref{fig5}.  Nevertheless, the stasis state still serves as an attractor in such cases;  the only difference is that $\Omega_M$ now approaches its stasis value $\barOmega_M$ from {\it below}\/ rather than above.
Indeed, in such cases these edge effects have produced an ``initial'' value $\Omega_M'$ which lies {\it below}\/
$\barOmega_M$, so that our system follows a trajectory within Fig.~\ref{attractorfig} along which  the value of $\Omega_M$ {\it increases}\/ rather than decreases.

We conclude, then, that our stasis state is a global attractor, emerging regardless of the initial conditions and regardless of the values of the parameters in our model.   Our stasis state is thus the Rome of our dynamical system, and all roads lead to it.


\FloatBarrier
\section{Stasis in the presence of additional energy components\label{vacuum}}

Thus far, we have considered the emergence of stasis within universes consisting of only matter and radiation.
Given this, a
 natural question is to understand how this picture is modified if our universe also contains
an additional energy component $X$ beyond matter and radiation,
 with general equation-of-state parameter $w_X$.
To study this, we can repeat our derivations, only now allowing for an  initial abundance
$\Omega_X$ in 
addition to $\Omega_M$ and $\Omega_\gamma$.

Our derivation proceeds exactly as before.
Including an energy contribution for which $p_X =w_X \rho_X$, 
we find that Eq.~(\ref{acceleq}) now becomes
\beqn
  \frac{dH}{dt}  
  ~&=&~ - \half H^2 \left[ 2+ \Omega_M + 2 \Omega_\gamma + (1+ 3 w_X) \Omega_X\right]\nonumber\\
  ~&=&~ - \half H^2 \left[ 4 -\Omega_M  + (3 w_X-1)  \Omega_X\right]~
\label{acceleqLam}
\eeqn
where in passing to the second line we have now identified
$\Omega_\gamma = 1-\Omega_M -\Omega_X$.
This in turn implies that Eq.~(\ref{convert}) 
becomes
\beq
    \frac{d\Omega_i}{dt} 
     ~=~ \frac{8\pi G}{3H^2} \frac{d\rho_i}{dt} + H \Omega_i \left[ 4-\Omega_M + (3w_X-1) \Omega_X \right]~.
\label{convertLam}
\eeq
While Eq.~(\ref{eoms}) continues to apply without modification,
we now additionally have $d \rho_X/dt  = -3 (1+w_X) H \rho_X$.
This of course assumes that $X$ is uncoupled from matter or radiation.
Substituting these results into Eq.~(\ref{convertLam}) we then find that the time-evolutions of our
three abundances $\Omega_M$, $\Omega_\gamma$, and $\Omega_X$ 
are described by  a system of three coupled differential equations:  
\beqn
         \frac{d\Omega_M}{dt} ~&=&~ -\sum_\ell \Gamma_\ell \Omega_\ell  + H \Omega_M (\Omega_\gamma +3w_X \Omega_X) \nonumber\\ 
         \frac{d\Omega_\gamma}{dt} ~&=&~ \sum_\ell \Gamma_\ell \Omega_\ell  - H \Omega_\gamma [\Omega_M +(1-3w_X) \Omega_X] \nonumber\\ 
         \frac{d\Omega_X}{dt} ~&=&~ H \Omega_X \left[ \Omega_\gamma - 3w_X (\Omega_M+\Omega_\gamma)\right]~.
\label{threediffeqs}
\eeqn
Of course,  $d\Omega_M/dt + d\Omega_\gamma/dt+ d\Omega_X/dt=0$, so only two of these equations are independent.   

As expected, these equations reduce to Eq.~(\ref{condition}) when $\Omega_X=0$ 
and when we can therefore identify $\Omega_\gamma= 1-\Omega_M$.
Moreover, we see from the third equation within Eq.~(\ref{threediffeqs}) that
if $\Omega_X$ vanishes at any initial time, then $\Omega_X$ remains vanishing for all times, thereby 
reproducing our previous results.
However, we are now interested in stasis configurations in which $\Omega_M$, $\Omega_\gamma$, and $\Omega_X$ 
are all constant but non-zero,  with values 
$\barOmega_M$,  $\barOmega_\gamma$, and $\barOmega_X\equiv 1-\barOmega_M-\barOmega_\gamma$ respectively.

Since stasis requires a non-zero constant $\Omega_X$, a minimal condition for stasis is $d\Omega_X/dt=0$.
From the third line of Eq.~(\ref{threediffeqs}) we then see that this will only happen if
\beq
\boxed{
      ~~ w_X ~=~  \frac{ \barOmega_\gamma}{3(\barOmega_M+\barOmega_\gamma)}~.~
}
\label{wX}
\eeq
Equivalently, inverting this relation, we see that for
any value of $\barOmega_X$ with equation-of-state parameter $w_X$, 
the  corresponding stasis solutions for this system must all take the general form
\beq
\begin{cases}
~\barOmega_M &\!=~  (1- 3w_X) (1-\barOmega_X)\\
~\barOmega_\gamma &\!=~ 3 w_X (1-\barOmega_X)~. \\
\end{cases}
\label{stasissolnX}
\eeq

It is easy to interpret the condition in Eq.~(\ref{wX}).
In general, the quantity in Eq.~(\ref{wX}) is nothing but the equation-of-state parameter for 
the combined matter$\,+\,$radiation subsystem during stasis.    
Thus, Eq.~(\ref{wX}) tells us that we can append any additional component $\Omega_X$ onto a combined 
matter$\,+\,$radiation subsystem without destroying its stasis property 
so long as the equation-of-state parameter $w_X$ of this additional component
matches the stasis equation of state of the original matter$\,+\,$radiation subsystem.
This ensures that the total system --- with the $X$-component included --- 
continues to have the same stasis equation of state $w_X$ as the original matter$\,+\,$radiation subsystem
and thus remains $w_X$-dominated.
The $X$-abundance $\Omega_X$ then remains constant under cosmological redshifting --- even though it is unaffected by the
decays of the matter components --- simply  as a result of the general property
that the abundance of any quantity with a given equation-of-state parameter $w$ always remains constant in 
a fully $w$-dominated universe.

Eq.~(\ref{wX}) came from the final equation in Eq.~(\ref{threediffeqs}) and thus
represents only one condition for stasis. 
The other remaining condition comes from the first two equations in Eq.~(\ref{threediffeqs}).
Indeed, demanding $d\Omega_M/dt = d\Omega_\gamma/dt=0$ 
we obtain the additional condition
\beq
    \sum_\ell \Gamma_\ell \Omega_\ell ~=~  H \,\frac{\barOmega_M \barOmega_\gamma}{\barOmega_M+\barOmega_\gamma}~=~
           3 w_X H \barOmega_M~.
\label{mainX}
\eeq
This result is the analogue of Eq.~(\ref{condition}).

Of course, we wish to ensure that Eq.~(\ref{mainX}) holds 
not only at one instant but over an extended stasis time interval.
Within such an interval we see from Eqs.~(\ref{acceleqLam}) and (\ref{wX}) 
that the Hubble parameter $H$ now takes the simple form
\beq
      H(t) ~=~ \frac{2}{3(1+w_X)}\, \frac{1}{t}~, 
\label{HubbleX}
\eeq
whereupon we find that 
\beq
    \Omega_\ell(t) ~=~ \Omega_\ell^\ast  
                 \left( \frac{t}{t_\ast}\right)^{2 w_X/(1+w_X)}
                 e^{-\Gamma_\ell (t-t_\ast)}
\label{OmegalX}
\eeq
for any fiducial time $t_\ast$ during stasis.
From Eq.~(\ref{mainX})
we thus have two additional conditions beyond that in Eq.~(\ref{wX})
which must also be satisfied simultaneously 
in order to have an extended period of stasis: 
\begin{empheq}[box=\fbox]{align}
\sum_\ell \Omega_\ell(t) ~&=~ \barOmega_M~~\nonumber\\
     ~~ \sum_\ell  \Gamma_\ell \Omega_\ell(t) ~&=~ 
     \frac{ 2 w_X \barOmega_M}{ 1+ w_X} \, \frac{1}{t} ~~
\label{masterconstraintsX}
\end{empheq}
where $\Omega_\ell(t)$ is given in Eq.~(\ref{OmegalX}).
Eq.~(\ref{masterconstraintsX}) is of course the analogue of Eq.~(\ref{masterconstraints}),
and leads to the condition
\beq 
      \frac{\sum_\ell \Gamma_\ell \Omega_\ell}{\sum_\ell \Omega_\ell}  ~=~ \frac{2 w_X}{1+w_X} \,\frac{1}{t}~,
\label{newquotientX}
\eeq
which is the analogue of 
Eq.~(\ref{quotient}).

It turns out that our model from Sect.~\ref{model} --- in conjunction with an additional energy component $X$ ---
 furnishes us with a realization of this three-component stasis as well.
Indeed, the only required modification to our model 
is that we no longer assert $\Omega_M=1$ as an initial condition at the production time $t^{(0)}$.
Because of the assumed presence of the additional $X$-component within our system,
we shall instead leave the initial value of $\Omega_M$ arbitrary.
However, proceeding exactly as in Sect.~\ref{model}, 
we once again obtain the result given in Eq.~(\ref{quotient_prediction}) ---
a result which did not depend on the initial value of $\Omega_M$.
Comparing with Eq.~(\ref{newquotientX}) we thus identify
\beq
       \frac{2 w_X}{1+ w_X} ~=~ \frac{\alpha + 1/\delta}{\gamma}~,
\eeq
or equivalently
\beq
          w_X ~=~ \frac{   1+\alpha \delta}{2 \gamma \delta - (1+ \alpha \delta)}~.
\eeq
We thus see that the parameters $(\alpha,\gamma,\delta)$ of our model must be chosen
appropriately for the equation-of-state parameter $w_X$ of the desired $X$-component that we wish to add.
Our model then yields constant stasis values $\lbrace\barOmega_M,\barOmega_\gamma,\barOmega_X\rbrace$
satisfying Eq.~(\ref{stasissolnX}), with the same provisos as discussed in Sect.~\ref{model} for the
two-component stasis.

Note that our model does not yield specific values for $\barOmega_M$, $\barOmega_\gamma$, or 
$\barOmega_X$ until specific initial values of $\Omega_M$ and $\Omega_\gamma$ are 
chosen at the production time $t^{(0)}$.   In principle this is
no different from the simpler two-component case we have already considered, given that even in the two-component case
we also chose a specific initial value $\Omega_M=1$ (with an implied corresponding initial choice $\Omega_\gamma=0$).   
Indeed, this choice precluded any room for an initial additional energy 
component $\Omega_X$.   In this sense, allowing more general initial values 
 $\Omega_M<1$ with $\Omega_\gamma=0$
is tantamount to allowing
an initial value $\Omega_X>0$.   After the initial transient edge effects have died away,
this then ultimately leads to a particular non-zero stasis value $\barOmega_X$, whereupon we then obtain
specific predicted values for $\barOmega_M$ and $\barOmega_\gamma$ satisfying Eq.~(\ref{stasissolnX}).

Following the steps in Sect.~\ref{attractor}, we can also demonstrate that our three-component stasis continues to be an attractor like its two-component cousin.  Of course our phase space is now described by {\it four}\/ independent dynamical variables,
namely $\Omega_M$, $\langle \Omega_M\rangle$, $\Omega_\gamma$, and $\langle \Omega_\gamma\rangle$.  One potentially
surprising feature is that the corresponding $4\times 4$ Jacobian matrix [analogous to the Jacobian matrix in Eq.~(\ref{Jacobian})]
actually has only {\it three}\/ negative eigenvalues {\it and one zero eigenvalue}\/. 
However, this is completely in keeping with our expectation that our stasis solution is no longer an attractor {\it point}\/ in the corresponding phase space, but
rather an attractor {\it line}\/.   This attractor line is analogous to a flat direction in the sense
that all points along the line correspond to equally valid solutions.  Indeed, moving along this ``flat direction'' 
corresponds to shifting the value of $\barOmega_X$, with 
the corresponding values of $\barOmega_M$ and $\barOmega_\gamma$ tracking the line of stasis solutions 
given in Eq.~(\ref{stasissolnX}).   
It is therefore not a surprise that it ultimately 
comes down to a particular choice of initial conditions
for $\Omega_X$ (or equivalently for $\Omega_M$ and $\Omega_\gamma$) which determines where along this line our resulting stasis 
is eventually realized.

We see, then, that the two-component stasis which has been the focus of this paper is not 
an isolated phenomenon,
existing for universes containing only matter and radiation.
{\it Rather, we now see that our two-component stasis is actually 
the endpoint of an entire {\it line}\/ of possible stasis solutions in which a variety of additional energy components $X$
with varying abundances $\Omega_X$ and equations of state $w_X$ are also possible.}\/
This observation once again reinforces our conclusion
that stasis is a generic feature in these sorts of theories.

Of course, not all types of energy components $X$ may be introduced.
From Eq.~(\ref{wX}) it follows that
\beq
            0 ~<~ w_X ~<~ 1/3~.
\eeq
As discussed in Ref.~\cite{Turner:1983he},
this includes, for example,
the energies associated with scalar theories in which the
scalars $\phi$  oscillate coherently within monomial potentials $V(\phi)\sim |\phi|^n$ where 
$2< n <4$.   
However, this precludes energy components with $w_X<0$, such as vacuum energy $\Lambda$ with $w_\Lambda= -1$.
This result implies that stasis is not possible when $\Omega_\Lambda\not=0$.
However, even with $\Omega_\Lambda\not=0$, it is possible that the {\it ratio}\/ of $\Omega_M$ and $\Omega_\gamma$ might 
nevertheless remain constant, thereby yielding what might be considered a weaker form of stasis. 
Of course, even when $\Omega_\Lambda\not =0$, 
an extended period of approximate stasis 
can exist during cosmological epochs 
for which $\Omega_\Lambda$ --- though growing --- is still small.
Indeed, as we shall discuss in Sect.~\ref{implications},
this is the most likely context in which any phenomenologically realistic model of stasis might appear.
Moreover, as long as $\Omega_\Lambda\not=0$ at some initial time,
the inevitable growth of $\Omega_\Lambda$ 
provides an additional natural mechanism for exiting the period of stasis
and resuming a more traditional period of cosmological evolution.

At first glance, given the result in Eq.~(\ref{wX}), it might seem that our extra $X$-component could itself be  
{\it composite}\/,
consisting of two subcomponents $A$ and $B$, so long as this composite $X$-sector
has the required total abundance $\barOmega_X= 1-\barOmega_M-\barOmega_\gamma$ 
as well as the required total equation-of-state parameter $w_X$ given in Eq.~(\ref{wX}).   
If so, one could imagine that as $\Omega_A$ grows, $\Omega_B$
would shrink to compensate and thereby keep $\Omega_A+\Omega_B$ and $w_X$ fixed.
One could even further speculate that the growing component could be vacuum energy $\Lambda$. 
However, it is easy to verify that 
even though we might carefully set $\Omega_X$ and $w_X$ to the required values at a given time,
the ensuing $A/B$ dynamics will generally keep neither the total abundance of the $X$-sector fixed at $\barOmega_X$
nor the total equation-of-state parameter fixed at $w_X$, as required for stasis
within the matter$\,+\,$radiation sector.
Indeed, the only way to have our $X$-sector retain both its abundance $\barOmega_X$ {\it and}\/ its equation-of-state parameter $w_X$ 
is to have the $X$-sector consist of only a single component $X$
whose abundance is then naturally fixed as a result of its equation-of-state parameter $w_X$ 
matching that of the matter$\,+\,$radiation sector.

\FloatBarrier
\section{Discussion and cosmological implications \label{implications}}

In this paper we have
demonstrated the existence of a new theoretical possibility for early-universe cosmology:
epochs of cosmological {\it stasis}\/
during which the relative abundances $\Omega_i$ of the different energy components
remain constant despite cosmological expansion.
Such stasis epochs therefore need not be radiation-dominated {\it or}\/ matter-dominated,
and need not be dominated by any particular component at all. 
We demonstrated that such epochs emerge naturally in many extensions to the Standard Model
and that the stasis state even serves as
a global attractor within the associated cosmological frameworks.
As a result, 
within these frameworks,
the universe will naturally evolve
towards such periods of stasis for a wide variety of initial conditions,
even if the system does not begin in stasis.
Moreover, as we have seen, each period of stasis comes equipped with 
not only a natural beginning but also a natural ending.
Depending on the parameters of the underlying theory,
such stasis epochs can nevertheless persist for arbitrary lengths of time.

Needless to say, our results give rise to a host of new theoretical possibilities for 
physics across the entire cosmological timeline. 
Indeed, an epoch of cosmological stasis can be expected to provide 
non-trivial modifications to the evolution of primordial density perturbations 
as well as the dynamics of cosmic reheating. 
The existence of a stasis epoch can also affect dark-matter production, structure formation, and even 
estimates of the age of the universe.
However, in order to study such possibilities, we must first understand 
where and how
our stasis epoch might arise within what might otherwise be considered the standard cosmological timeline.

Within the standard $\Lambda$CDM cosmology, 
the universe first undergoes a phase of accelerated expansion known as cosmic inflation.  
Immediately after this inflationary epoch, the energy density of the universe is 
typically dominated by the coherent oscillations 
of the inflaton field.  The subsequent decays of this field then reheat the universe, 
thereby giving rise to a radiation-dominated (RD) era.  
Since the energy density of radiation is diluted by cosmic expansion more rapidly than that of matter, 
the relative abundance of matter rises over time, reaches parity with the abundance of radiation at 
the point of matter-radiation equality (MRE), and  then exceeds the abundance of radiation, 
eventually coming to dominate the universe.  
The ensuing matter-dominated (MD) 
era then persists until very late times, at which point vacuum energy becomes dominant.

Although this timeline is relatively simple and compelling,
there is considerable room for modification without running afoul of experimental or observational data.  In particular, it is possible to imagine ``splicing'' an epoch of stasis
into this timeline, either as an additional segment inserted into the timeline or as the replacement for 
a segment which is removed.

To see how this might occur, let us first recall that our stasis scenario is one in which
the universe passes from a matter-dominated epoch into a period of stasis and then finally into
a radiation-dominated epoch. 
In general, this initial MD epoch begins at the time $t^{(0)}$ 
at which our $\phi_\ell$ states are produced 
--- provided the $\phi_\ell$ are produced non-relativistically and with a sufficiently large abundance. 
Of course, if these fields are relativistic 
at the production time $t^{(0)}$, 
or if there exists a significant abundance of radiation at that time,
then the universe might be radiation-dominated for a few $e$-folds after $t^{(0)}$.
However, even in such cases, our  $\phi_\ell$ states will eventually come to dominate the universe 
and the universe will enter a matter-dominated epoch.
Thus, in either case, our stasis scenario generically begins with a pre-stasis MD epoch.
However, once our $\phi_\ell$ particles begin to decay, we then
enter the stasis period.
Indeed, as we have demonstrated in Sect.~\ref{attractor}, the stasis state is a global attractor 
regardless of the particular initial conditions.  
Thus, as long as the $\phi_\ell$ states are produced with appropriately scaled abundances
and lifetimes as in Sect.~\ref{model},  we will necessarily enter into a period of stasis.
Finally, as we reach the time at which the last $\phi_\ell$ states decay, we then exit the stasis era and enter a 
radiation-dominated epoch in which no $\phi_\ell$ particles remain.

Given these observations, there are many locations along the standard $\Lambda$CDM 
timeline during which a stasis epoch might occur.  However, for concreteness, we shall highlight two scenarios that naturally stand out when constructing a cosmological model of stasis. 

\begin{itemize}

\item {\bf\underbar{Stasis spliced into RH}\/:}~
  A minimal approach to splicing a period of stasis into the standard cosmological timeline involves choosing a location
  where there is already a transition from a MD universe to a RD universe, accompanied by an episode of particle production during which our $\phi_\ell$ states might be initially populated.    Within the standard cosmology, there is only one such location:   during reheating.  Indeed, assuming that the inflaton oscillates coherently within a nearly quadratic potential, the universe is effectively matter-dominated during this period, evolving with an equation-of-state parameter $w\approx 0$. The decay of the inflaton, which is usually assumed to reheat the universe, would instead produce the $\phi_\ell$ states.   Such states would then quickly become non-relativistic (if they were not already produced non-relativistically), thereby extending the reheating period into a longer MD epoch. The subsequent stasis epoch and the $\phi_\ell$ decays therein would then ultimately provide a new environment for reheating~\cite{visogenesis}.  Finally, once the $\phi_\ell$ decays have concluded, the universe will be radiation-dominated, with a traditional $\Lambda$CDM evolution beyond that point.   
  Thus, schematically, this scenario amounts to an insertion into the RH epoch of the form
\beq
 {\rm RH} ~\longrightarrow~ {\rm RH} + {\rm stasis} ~.
\label{insertion1}
\eeq

\item {\bf\underbar{Stasis spliced into RD}\/:}~
  Given that our stasis  scenario leads to a radiation-dominated epoch, another possibility is to splice the stasis scenario directly into the usual RD era.
This would require that $t^{(0)}$, the production time for our $\phi_\ell$ states, occur at a time when the universe is already radiation-dominated.   If the $\phi_\ell$ states behave as massive matter,
then their energy density will eventually dominate the total
energy density of the universe even though the universe was radiation-dominated at $t^{(0)}$.
This scenario thus induces a new early matter-dominated epoch (EMDE) immediately  prior to the onset of stasis, with the subsequent transition to stasis only occurring once the $\phi_\ell$ states begin to decay.  During the ensuing stasis epoch, of course, the universe consists of 
an admixture of matter and radiation with unchanging relative abundances.  Finally, after the last $\phi_\ell$ particles have decayed, the universe is once again radiation-dominated, with subsequent time evolution proceeding in the traditional $\Lambda$CDM manner.   
Thus, this scenario schematically amounts to an insertion into the RD epoch of the form
\beq
~~~~~{\rm RD} ~\longrightarrow~ {\rm RD} + {\rm EMDE} + {\rm stasis} +
                           {\rm RD}~.
\label{insertion2}
\eeq

\end{itemize}

In either of these scenarios,
the equation-of-state parameter $w$ for the stasis epoch can take essentially any value within the
range $0<w<1/3$.     Indeed, within the stasis scenario this is achieved through the emergence of a stable mixed state
rather than through the introduction of a pure state involving a new type of cosmological fluid.
As a result 
of this unorthodox value of $w$, the evolution of the universe during the stasis era is unlike its evolution during  any other cosmological epoch, expanding more rapidly 
than it does during a RD epoch but more slowly than it does during a MD epoch.


\begin{figure*}   
\centering
 ~\hskip -0.35 truein
\includegraphics[keepaspectratio, width=0.70\textwidth]{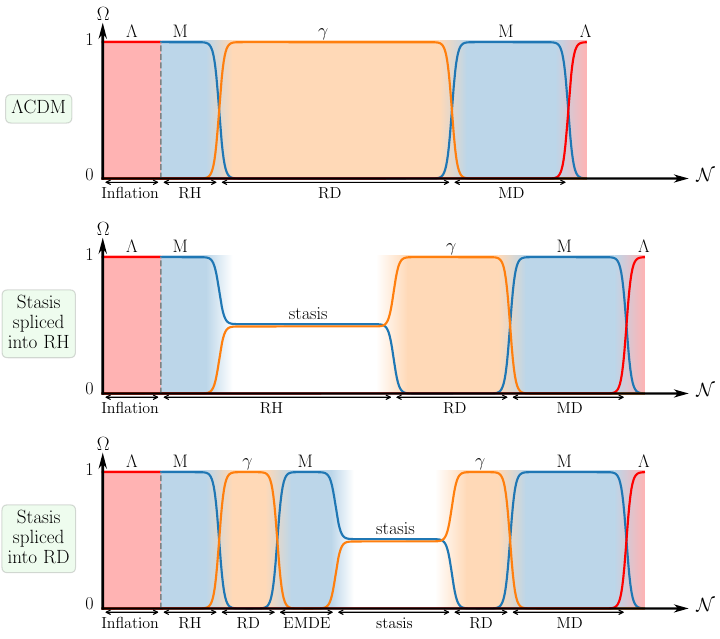}
\vskip -0.02 truein
\caption{Sketch of the traditional $\Lambda$CDM cosmology (top panel) as well
as the two scenarios itemized in the text 
and described schematically in Eqs.~(\ref{insertion1}) and (\ref{insertion2}) 
in which a stasis epoch with $\barOmega_M=\barOmega_\gamma=1/2$ 
is inserted into 
the cosmological timeline (middle and bottom panels).
Abundances associated with vacuum energy (red), matter (blue), and radiation (orange)
are plotted as functions of the number $\calN$ of $e$-folds,
with corresponding background shadings indicating the dominant component in each epoch.
In the ``stasis spliced into RH'' scenario (middle panel),
reheating occurs during the stasis epoch and results
from the decays of the $\phi_\ell$ states~\cite{visogenesis}.  By contrast,
in the ``stasis spliced into RD'' scenario (bottom panel), reheating has already concluded but
the insertion of the stasis scenario 
induces the existence of an early matter-dominated era (labeled `EMDE').}
\label{fig6}
\bigskip
\bigskip
%
%
%
\centering
\hfill
\includegraphics[keepaspectratio, width=0.48\textwidth]{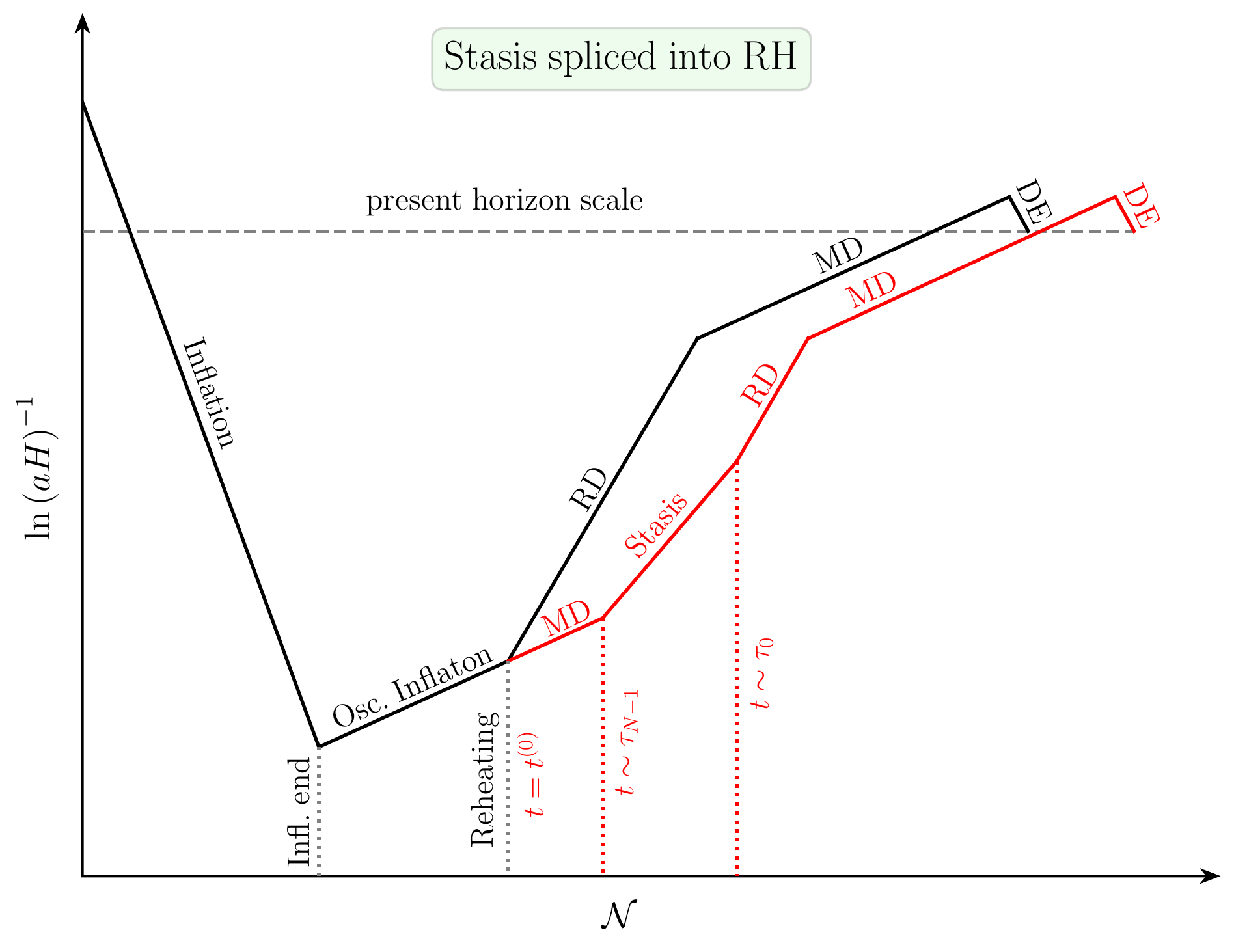}
\hfill
\includegraphics[keepaspectratio, width=0.48\textwidth]{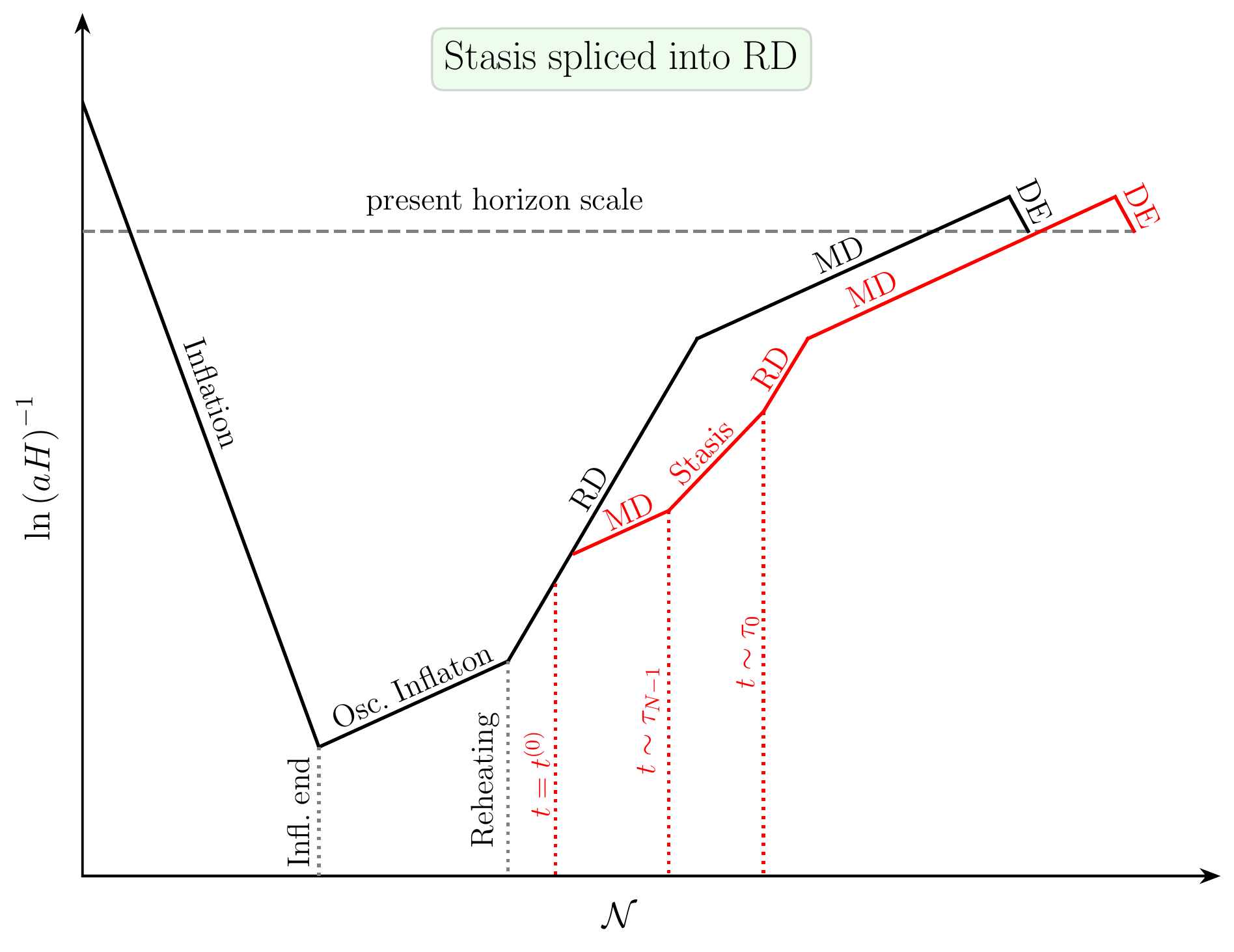}
\hfill
\vskip -0.02 truein
\caption{
The same scenarios as illustrated in Fig.~\protect\ref{fig6},
only now sketched in terms of the evolution of the comoving Hubble radius $(aH)^{-1}$.
The present-day horizon scale is also indicated.
In each panel the black lines indicate the standard $\Lambda$CDM cosmology,
while the red lines indicate the new cosmologies that result in the ``stasis spliced into RH'' and ``stasis spliced into RD'' scenarios (left and right panels, respectively).
In each case we have explicitly indicated the time $t^{(0)}$ at which the $\phi_\ell$ states
are initially produced as well as the times $\tau_{N-1}$ and $\tau_0$ which respectively approximate
the onset and cessation of the stasis state.
In both cases we see that the insertion of a stasis period delays the subsequent timeline relative
to traditional $\Lambda$CDM expectations.}
\label{fig7}
\end{figure*}


In Fig.~\ref{fig6} we illustrate the traditional $\Lambda$CDM cosmology 
as well as the two alternative scenarios itemized above.  
In the top panel, we show the traditional $\Lambda$CDM cosmology, sketching 
the relative cosmological abundances $\Omega_i$ associated with vacuum energy ($i= \Lambda$, in red), 
matter ($i=M$, in blue), and radiation ($i=\gamma$, in orange) as they evolve through an initial 
inflationary epoch followed by a reheating epoch, 
a radiation-dominated epoch, and ultimately
a matter-dominated epoch that is only now giving way to an epoch dominated again by vacuum energy.
We have also shaded each region according to the dominant energy component during that epoch.
By contrast, in the middle and bottom panels, we sketch the new scenarios in which a stasis interval with $\barOmega_M=\barOmega_\gamma=1/2$ occurs within the reheating epoch (middle panel) or the radiation-dominated epoch (lower panel).    
As shown, the latter possibility requires the introduction of a new early matter-dominated epoch (EMDE) immediately prior to the emergence of the stasis state.    
    Of course, it is only for simplicity that we have chosen to illustrate these
    stasis states as having $\barOmega_M=\barOmega_\gamma=1/2$ in Fig.~\ref{fig6};
    any stasis configuration with $\barOmega_M = 1-\barOmega_\gamma$ would have been equally valid 
    for either of the two cases shown.

In either case, we see from Fig.~\ref{fig6} that the intervening stasis period has the effect of delaying the entire cosmological timeline.   
Thus, while the universe eventually returns to the traditional $\Lambda$CDM script in each case, it does so  at an age which is advanced relative to that normally ascribed to it within the $\Lambda$CDM framework.   
We also observe that during these stasis epochs, the universe is not dominated by any particular energy component.
It is for this reason that the stasis epochs are not shown in Fig.~\ref{fig6} with any shaded background.
Normally it is not possible to have such an extended unshaded epoch;   the different abundances $\Omega_i$ 
are constantly in flux and it is only during the relatively brief transition periods between epochs that the universe
might fail to have a dominant component.
However, during stasis, this situation can persist across an arbitrary number of $e$-folds.

In Fig.~\ref{fig7}, we illustrate these same two alternate scenarios, 
this time plotting not the corresponding cosmological abundances $\Omega_i$ 
but rather the corresponding comoving Hubble radius $(aH)^{-1}$.
Within each panel of Fig.~\ref{fig7}, we show not only
the standard $\Lambda$CDM  
cosmology (black lines) but also the corresponding modified cosmology (red lines) 
in which a stasis epoch arises.
Note that the sketches in Figs.~\ref{fig6} and \ref{fig7} correspond 
to the regime in which $H^{(0)} \gg \Gamma_{N-1}$, 
thereby ensuring that all of the $\phi_\ell$ states behave like matter before they start decaying. 

One important feature illustrated in these figures
is that the introduction of a stasis period either prior to or during the traditional RD era
requires a shortening of the duration of the RD era as a whole.
For example, we observe from the sketches in Fig.~\ref{fig6} that 
the period during which $\Omega_\gamma$ dominates (shaded orange) 
is far longer in the $\Lambda$CDM case than it is in either of two cases that involve a prior period of stasis.
We stress that this shortening of the radiation-dominated era 
is {\it not}\/ imposed in order to preserve the age of the universe;   indeed, as already noted
above, both of the cosmologies that include the stasis epoch 
reach the present day only after a longer time interval has elapsed.
Rather, this shortening of the subsequent $\Omega_\gamma$-dominated epoch 
is required in order to guarantee not only 
a fixed horizon scale today,  but also a fixed number 
of $e$-folds since MRE.~
Indeed, both of these quantities are measured through observational data,
implying that the final $\Lambda$CDM-like portions of the cosmological timelines after MRE 
may be shifted horizontally in Fig.~\ref{fig7} but never vertically.

Introducing an epoch of stasis into the standard cosmological timeline has a number of effects.
First, within modified cosmologies involving a stasis epoch, 
the comoving Hubble radius grows {\it more slowly}\/ than it does in the standard cosmology. 
This in turn can have 
numerous consequences for observational cosmology.  For example, perturbations in the 
matter density reenter the horizon at a later time in cosmologies involving a stasis epoch 
than they do in the standard cosmology.  Consistency with observational data
therefore typically requires that the number of $e$-folds between horizon crossing and the 
end of inflation must be smaller in such cosmologies.  This in turn affects the predictions 
for inflationary 
observables~\cite{Easther:2013nga, Das:2015uwa, Allahverdi:2018iod, Heurtier:2019eou}. 

Another potential observational consequence of stasis stems from its effect on structure
formation.
During a RD epoch, primordial density perturbations which have already 
entered the horizon grow only logarithmically with the scale factor.  
By contrast, during a stasis epoch, these perturbations could potentially grow much more rapidly, as is the case 
within an EMDE~\cite{Erickcek:2011us,Fan:2014zua,Georg:2016yxa}.  Such rapid growth 
could therefore likewise result in the formation of compact objects such as  
primordial black holes~\cite{Green:1997pr, Georg:2016yxa} 
or ultra-compact 
minihalos~\cite{Erickcek:2011us,Barenboim:2013gya}.

\begin{figure}[t!]
\centering
\includegraphics[keepaspectratio, width=0.48\textwidth]{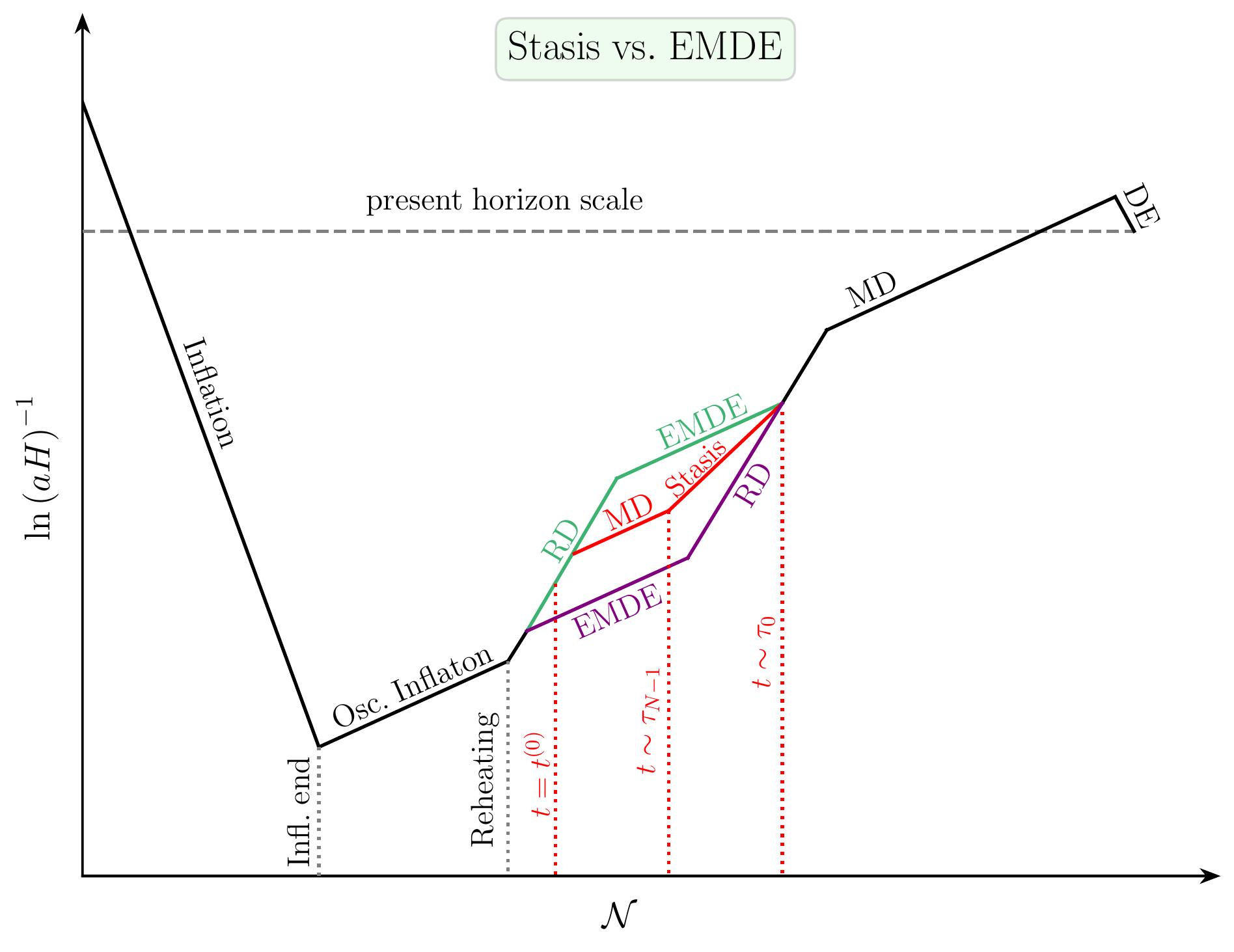}
\caption{Sketch of a scenario 
in which the insertion of a period of stasis (red) 
into the standard $\Lambda$CDM cosmology (black)
replaces either of two possible EMDE insertions (green or purple).
This replacement 
nevertheless leaves the universe on the same subsequent cosmological timeline.
Such a scenario therefore provides a testing ground for comparing the 
phenomenological effects of an EMDE insertion
versus those of a stasis insertion.}
\label{fig8}
\end{figure}

Unlike the scenarios illustrated in Figs.~\ref{fig6} and \ref{fig7},
it is also 
possible to imagine scenarios in which a period of stasis replaces an unorthodox modification
to the standard cosmology but otherwise places us back on the {\it same}\/ cosmological timeline.
For example, as illustrated in Fig.~\ref{fig8}, 
one particularly well-known modified cosmology consists of introducing an EMDE
entirely within the usual RD era.
Such an EMDE may be inserted either earlier within the RD era (as shown in purple in Fig.~\ref{fig8})
or later (as shown in dark cyan).
Nevertheless, 
as sketched in Fig.~\ref{fig8},    
it is possible to imagine splicing a stasis epoch into this region instead (red). 
Even though the stasis scenario and the EMDE scenario place the universe
on the same cosmological timeline for all subsequent times all the way to the present day,
it would be interesting to explore the extent to which one can use present-day observational 
information in order to determine which path the universe ultimately followed.
The scenario in Fig.~\ref{fig8} thus provides a framework within which one can 
directly compare the effects of an EMDE insertion against those of a stasis insertion.
Density perturbations with different wavenumbers enter the horizon at different times during the 
different cosmologies depicted in Fig.~\ref{fig8}.  Moreover, once they enter the horizon, they do not scale the same way 
with the scale factor $a$ during a period of stasis as they do in a MD epoch.  As a result, one would in general expect these different cosmologies to yield different perturbation spectra, with possible implications for small-scale structure.

A stasis epoch could also impact particle-physics processes in the early universe in a number of ways.  
Indeed, any model involving the out-of-equilibrium production of particles would be 
affected by the corresponding modification of the expansion history.  
The abundance of dark matter in our present universe typically depends on the equation 
of state of the universe both during and after the epoch in which that abundance is initially generated (see, {\it e.g.}\/, Refs.~\cite{DEramo:2017gpl,Hamdan:2017psw,Drees:2017iod,DEramo:2017ecx,Allahverdi:2018aux,Arias:2019uol,Allahverdi:2019jsc,StenDelos:2019xdk,Garcia:2020eof,Cheek:2021cfe}).
Likewise, the manner in which a lepton or baryon asymmetry evolves with time within 
the early universe is often sensitive to the expansion history of the universe 
and the extent to which energy can be transferred between different cosmological components.  
For example, within the context of electroweak baryogenesis, the washout of the baryon asymmetry 
by sphaleron processes is mitigated in scenarios in which the universe 
expands faster during the electroweak phase transition than it does during radiation
domination~\cite{Joyce:1996cp,Joyce:1997fc,Davidson:2000dw,Servant:2001jh,
Barenboim:2012nh}.  
Moreover, entropy production from the late decays of heavy fields,
such as those needed to sustain a stasis epoch, can also serve to dilute a pre-existing 
baryon or lepton asymmetry~\cite{Nardini:2011hu}.  This can be 
advantageous~\cite{Bramante:2017obj} in models
such as Affleck-Dine baryogenesis~\cite{Affleck:1984fy,Dine:1995kz,Dine:2003ax} 
in which the initial baryon asymmetry is typically too large.  
Modifying the expansion history  
of the universe can also modify the dynamics which governs the evolution of light scalar
fields such as the QCD axion and other, axion-like particles.  For example, it has 
been shown that an EMDE 
can render phenomenologically viable 
certain regions of parameter space for low-mass axion dark matter which would otherwise have 
been excluded~\cite{Nelson:2018via}.   It is reasonable to expect that 
a period of cosmological stasis would have similar consequences for the dynamics of 
such fields.   
Similarly, if a stochastic gravitational background can be produced prior to or during the stasis epoch,
its power spectrum is also expected to be modified 
due to the change in the expansion history and the injection of entropy from successive $\phi_\ell$ 
decays~\cite{Bernal:2019lpc, Allahverdi:2020bys, Guo:2020grp}.

The splicing of a stasis epoch into the cosmological timeline could also have implications for dark-matter physics.  In this paper, we have not assumed any particular relation between the dark matter and the dynamics involved in cosmic stasis.  However, most canonical mechanisms for establishing a cosmological abundance of dark-matter particles turn out to be compatible with modified cosmologies which include a stasis epoch.

The extent to which the presence of a stasis epoch impacts the properties of the dark matter ultimately depends on the time at which the dark-matter abundance is established.  One possibility is that the dark-matter abundance is generated only after stasis has concluded.  In this case, the stasis epoch has essentially no impact on the dark matter.  Another possibility is that the dark-matter abundance is established prior to the stasis epoch but remains negligible throughout this epoch and therefore does not disrupt the stasis itself.  In this case, the rate at which the dark-matter abundance redshifts as a function of time is modified because of the different background cosmology --- in particular the different equation-of-state parameter $w$ --- involved in stasis.

Yet another possibility is that the dark-matter abundance is established {\it during}\/ the stasis epoch itself.  This possibility is particularly intriguing, as the dynamics of dark-matter production can be modified in several ways if this production occurs during a stasis epoch.  
For example, when thermal freeze-out occurs during a stasis epoch, the annihilation cross-section required in order to achieve an appropriate late-time abundance for a given dark-matter particle is smaller than it is when freeze-out occurs in the standard $\Lambda$CDM cosmology.  This is because the production of entropy from $\phi_\ell$ decays dilutes the dark-matter relic abundance.  Indeed, this is similar to the case when freeze-out occurs during an EMDE~\cite{Giudice:2000ex,Gelmini:2006pw,Gelmini:2006pq,Erickcek:2015jza,Berlin:2016vnh,Berlin:2016gtr,Heurtier:2019eou}.
It is also possible that a non-thermal population of dark-matter particles could be produced directly from the decays of the $\phi_\ell$.  The late-time velocities of such dark-matter particles can be non-negligible, leading to a suppression of power on small scales.  Indeed, this is also known to occur in situations in which a significant number of dark-matter particles are produced by the decays of the oscillating scalar or massive particle species which dominates the universe during an EMDE~\cite{Miller:2019pss}.~   However, the dark-matter velocity distributions which arise from $\phi_\ell$ decays during a stasis epoch can be expected to be even more complicated than they would be within an EMDE, as these velocity distributions receive contributions from a large number of states decaying at different times.  In such scenarios, methods such as those developed in Ref.~\cite{Dienes:2020bmn} could potentially be employed in order to extract information about the $\phi_\ell$. 

While the dark matter need not be related to the $\phi_\ell$, it is also interesting to consider the possibility that the particle species which constitutes the dark matter is in fact simply the most long-lived of these $\phi_\ell$ tower states.  Consistency with the standard cosmology at late times would of course require that the lifetime of this lightest state be parametrically longer and its initial abundance parametrically smaller than those of the other $\phi_\ell$.  However, such separations of scales can occur naturally in a number of new-physics scenarios, including scenarios wherein the $\phi_\ell$ are the Kaluza-Klein modes of a field which propagates in the bulk in a theory involving extra spacetime dimensions.

Given the results of this paper, many avenues for future research suggest themselves.
For example, in addition to the issues discussed above, 
our work in this paper implicitly rested on certain assumptions which can be relaxed.
One assumption 
which has been implicit throughout 
our analysis is that the states $\phi_\ell$ are all non-relativistic when they are produced at 
$t^{(0)}$. 
While this is possible if these states are produced through a freeze-out mechanism, 
this would typically not be the case if the different states are produced through the decays of a heavier particle. 
As discussed in Ref.~\cite{Dienes:2020bmn},
the phase-space distribution 
of the $\phi_\ell$ particles could therefore be rather non-trivial at the time $t^{(0)}$.
Indeed, this distortion of the phase-space distribution could be viewed as introducing an additional
 initial ``edge'' effect that would also have to be taken into account in our analysis. 

Another assumption implicitly made throughout our work 
is that the different $\phi_\ell$ states remain out of equilibrium until 
they completely decay.  Within a specific particle-physics model, this assumption would need to be verified. 
In particular, if stasis takes place while the temperature of the thermal bath is sufficiently large, 
thermal processes could induce the thermalization of the lighter states within the tower and thereby 
modify their relative abundances,
while the heavier states remain decoupled and act as massive matter.

In our work we also assumed that the $\phi_\ell$ states only decay into radiation. 
However, it is possible that the heavy states might decay into lighter $\phi_\ell$ states within the tower
or other light particles beyond the tower. 
Of course, if the daughter states are sufficiently light, or more generally if the
mass differences between parents and daughters
are sufficiently large,
such decays will be highly exothermic.
The daughters will then be produced  
relativistically and continue to act as radiation.
As such, these decays will continue to effectively transfer
energy density from matter back to radiation, and can therefore continue to serve as
the counterbalance to cosmological expansion which tends to push the relative abundances the other way and induce stasis.
However, as the universe continues to cool, these daughter states will eventually become non-relativistic ---
an effect which then flows in the opposite direction, effectively transferring radiation back to matter.
Indeed, this reversal might serve as another 
means of exiting a stasis epoch.   
In either case, 
such non-trivial dynamics will produce a non-trivial time evolution for 
the background cosmology.
This in turn could affect all sectors of the corresponding theory, including its dark sector,
and thereby leave interesting signatures in 
the matter power spectrum~\cite{Dienes:2020bmn} and corresponding halo-mass distributions~\cite{Dienes:2021itb}. 

In this context, we emphasize that we have not assumed in this work that 
our $\phi_\ell$ states are scalars.   Indeed, these fields may well be fermionic, which would then permit decays into final states which are collectively fermionic and which might include light matter fields such as neutrinos which could potentially act as radiation, as discussed above.    Of course, the spins of the
parent and daughter states can potentially affect the set of relevant exponents $\gamma$ which govern the scalings of the decay widths $\Gamma_\ell$ in Eq.~(\ref{scalings}) and which ultimately feed into the matter abundance $\barOmega_M$ during stasis.

In a similar vein, in this work we have implicitly assumed that the energy density $\rho_\ell$ associated with each massive field $\phi_\ell$ results from particle-like excitations of that field.  However, if the $\phi_\ell$ states are scalars, these fields may also have
homogeneous zero-modes whose coherent oscillations
give rise to energy densities which scale in exactly the same way. (An example of this phenomenon is the well-known coherent oscillation of the axion field.)   It therefore follows that a collection of scalar fields $\phi_\ell$ with coherently oscillating zero-modes can likewise support stasis.  This is an intriguing possibility, since large numbers of scalar fields are a generic prediction of string theory and a variety of other extensions of the Standard Model.  Moreover, vacuum-misalignment production provides a natural mechanism through which a spectrum of energy densities $\rho_\ell$ with power-law scalings can be generated for such scalars at early times~\cite{Dienes:2011ja}.

The non-trivial dynamics of such $\phi_\ell$ zero-modes can nevertheless play an interesting role in determining the associated scaling exponents.
Recall that if the misalignment production time at which $\rho_\ell$ is produced is sufficiently early that   $H(t) \gsim 2 m_\ell/3$ for a given $\ell$, the corresponding
$\phi_\ell$ zero mode will be overdamped and the corresponding energy density $\rho_\ell$  will behave like vacuum energy.   Indeed, it is only after we reach a time
at which $H(t)\sim 2 m_\ell/3$ that the $\phi_\ell$ zero-mode will ``turn on'' and become underdamped;  the corresponding energy density $\rho_\ell$ is then associated with the resulting zero-mode oscillations
and begins to scale like massive matter.
This non-trivial process by which such states turn on can have  non-trivial implications for stasis.
If the production time occurs at a late time when
$H(t) \lsim 3 m_\ell/2$ for all $\ell$ in the tower, 
all of the $\phi_\ell$ will immediately behave as matter and contribute to $\Omega_M(t)$ from the moment when they are produced.    We would therefore have realized the initial conditions for our stasis model, and a stasis epoch would merge as long as the relevant parameters satisfy the appropriate constraints, such as those in Eq.~(\ref{range}).   However, if the production time occurs earlier, it is possible that only the heaviest $\phi_\ell$ states will be ``turned on'' and immediately contribute to $\Omega_M(t)$ at the production time.   By contrast, the lighter states will only experience a subsequent ``staggered'' turn-on as time progresses~\cite{Dienes:2011ja}.   This would then have three effects:  the effective number of states in the tower contributing to $\Omega_M(t)$ would increase with time as increasing numbers of states turn on;
our system would initially contain a vacuum-energy component which eventually drops as a function of time, ultimately becoming relatively small;  and the contributions $\Omega_\ell(t)$ from the lighter states in the tower will be enhanced relative our usual expectations at the time when they turn on and begin to contribute to the total matter abundance $\Omega_M(t)$.  This enhancement is the result of the fact that these $\Omega_\ell(t)$ would have grown as vacuum energy rather than as matter during the intervening time after production but prior to turning on.

At times after all $\phi_\ell$ states have turned on, our entire tower
acts as matter and there is no remaining vacuum-energy component.   At such times, we would also expect to have stasis, but with a modified scaling exponent $\alpha$ which reflects the enhancement of the light abundances $\Omega_\ell$  that occurred during the period wherein the lighter $\phi_\ell$ were still experiencing their staggered turn-ons.
However, if the heavier $\phi_\ell$ begin decaying during this staggered turn-on phase,
the situation is much more complicated.   At late times during this phase, we expect the vacuum-energy component, though non-zero, to be relatively small.  It is possible that this will therefore not disrupt the emergence of stasis during this period, even though the effective number of matter states in the tower is still evolving.  This case requires further investigation.

Another assumption made in our work is that the values of the scaling exponents $(\alpha,\gamma,\delta)$ and mass parameters $(m_0,\Delta m)$ remain constant throughout our tower of $\phi_\ell$ states.
While this is true for the towers of states which emerge in many extensions to the Standard Model, there do exist scenarios in which these exponents themselves deform from one value to another as one passes between the high-mass and low-mass portions of the same $\phi_\ell$ tower~\cite{Dienes:2011ja,Dienes:2016zfr}.   In such cases,
the properties of the corresponding stasis epoch can also shift slightly as the stasis proceeds through successive $\phi_\ell$ decays.   For example, one stasis value of $\barOmega_M$ may persist for many $e$-folds before gradually shifting to another value which then persists through an additional extended epoch. It is also possible that some portions of the $\phi_\ell$ will have parameters $(\alpha,\gamma,\delta)$ that lie outside the limits in Eq.~(\ref{range}).   In such cases, stasis would persist only during the decays of those states for which Eq.~(\ref{range}) is valid, and thus the {\it effective}\/ size of the tower may appear truncated relative to $N$.

In this paper, we have paid significant attention to the ``edge'' effects which occur at early and late times and which are ultimately responsible for the entrance into as well as the exit from our stasis epoch.   However, it may also be possible to {\it exploit}\/ these edge effects in order to construct variations of the cosmologies that we have discussed here.   For example, in our ``stasis spliced into RD'' scenario in Eq.~(\ref{insertion2}), we indicated that an EMDE must immediately precede our stasis epoch.   Indeed, given that our stasis model in Sect.~\ref{model} begins with a fully matter-dominated universe as an initial condition, such an EMDE is required if we wish to adopt the model in Sect.~\ref{model} wholesale when splicing it into the RD epoch of the $\Lambda$CDM timeline.   However, we have seen that the edge effects associated with this model can quickly deform such an early matter-dominated initial condition into one in which most of the initial matter abundance is quickly dissipated.  Indeed, we have already seen from the orange/yellow curves in Fig.~\ref{fig5} that such a  rapid depletion can occur as an initial edge effect for sufficiently large values of $\Gamma_{N-1}/H^{(0)}$.   One could therefore contemplate splicing a previous RD-dominated epoch directly onto the time at which much of the initial matter abundance $\Omega_M$ is already depleted.   Alternatively (but with similar phenomenological consequences), we can imagine that the creation of the $\phi_\ell$ states occurs at the beginning of the inserted epoch, but that the chosen value of $\Gamma_{N-1}/H^{(0)}$ is sufficiently large that any interval of early matter domination is extremely short.  

As discussed in Sect.~\ref{model}, theories involving large extra spacetime dimensions naturally give rise to infinite towers of Kaluza-Klein states which can serve as the $\phi_\ell$.    The spectrum of such states depends on the dimensionality and geometry of the compactification manifold, and the simplest case in which a single extra dimension is compactified on a circle (or a $\mathbb{Z}_2$ orbifold thereof) results in KK spectra with $(m_0, \Delta m, \delta) = (m,1/R,1)$ or $(m, 1/(2 m^2 R), 2)$, where $R$ is the compactification radius and $m$ is the four-dimensional mass of the compactified field.  Indeed, the first possibility occurs if $mR\ll 1$, while the second arises if $mR\gg 1$.  Both of these values of $\delta$ are within the ranges that produce viable stasis values of $\barOmega_M$.  

Clearly, the space of models involving such large extra dimensions is huge, and thus the set of corresponding  values of $\alpha$ and $\gamma$ is also huge, depending on the model-specific details of how these states are produced in the early universe and how they decay to the visible sector.   However, one critical {\it model-independent}\/ issue concerns the extent to which a viable period of stasis can even arise within such a framework.   As we have seen in Eq.~(\ref{NN}), a bona-fide period of stasis requires a relatively large tower of $\phi_\ell$ states --- \ie, a relatively large value of $N$.   Given this, and given that KK theories are essentially effective field theories (EFTs), one might therefore worry that there might be an intrinsic upper limit to the size of $N$.   However, this worry ultimately turns out to be unfounded because the ultraviolet (UV) cutoff for such theories is set not by the compactification radius but by other factors which relate to the onset of new physics (such as the possible emergence of a grand-unified theory or quantum gravity). Indeed, there exist numerous explicit examples in the literature in which the compactification scale can be separated by many orders of magnitude from the scale of new physics.  For example, there is a vast literature, initiated in Refs.~\cite{Arkani-Hamed:1998jmv, Dienes:1998vh,  Antoniadis:1998ig, Dienes:1998vg, Arkani-Hamed:1998sfv}, in which the size of extra spacetime dimensions is set far below such scales in a self-consistent way.  Of course, one might worry that the emergence of KK states would drive gauge couplings towards Landau poles (through the same couplings to the visible sector that induce the $\phi_\ell$ decays), but there exist numerous examples where this does not happen, even within theories whose low-energy limits include the Standard Model~\cite{Dienes:2002bg, Abel:2017vos}.   

Within such frameworks, then, the number of KK states is therefore essentially unbounded and there is no model-independent obstruction to having a large tower of states and thus a stasis epoch of long duration.   Of course, it will be interesting to actually construct phenomenologically viable models of KK-induced stasis.  Doing so will ultimately depend on a number of further model-specific factors, such as the desired placement of this epoch within the cosmological timeline and the assumption of an appropriate theory of particle physics during that time (be it the Standard Model or a supersymmetric or grand-unified extension thereof).   

Another framework for physics beyond the Standard Model which naturally gives rise to large towers of states is string theory.  String theories generally have critical spacetime dimensions exceeding four, and thus typically involve geometric compactifications down to four dimensions.   For this reason, many of the different string states (particularly those in the so-called ``bulk'') are endowed with towers of additional KK excitations.   Moreover, these KK states will generally be unstable and decay, although their particular decay phenomenologies depend on the properties of the specific string model under discussion.  All of this is therefore consistent with the possibility of achieving KK-induced stasis, as discussed above.

However, string theories include not only towers of KK states but also towers of Regge excitations.   These excitations are independent of the spacetime compactification and are a consequence of the extended nature of the string itself.  Unlike the KK states (and their closed-string cousins, the winding states), these Regge states have mass spectra which correspond to the scaling exponent $\delta=1/2$.   In principle, this is not a problem for stasis;  we have kept $\delta$ arbitrary in our analysis, and nothing precludes $\delta=1/2$.   However, these towers of Regge excitations also have a degeneracy of states at each mass level which grows {\it exponentially}\/ with the mass of the state.  This is the well-known Hagedorn phenomenon~\cite{Hagedorn:1965st}.  As a result, such theories tend to have effective energy densities $\rho_\ell$ and abundances $\Omega_\ell$ which grow exponentially with the masses $m_\ell$ before other effects (such as those due to Boltzmann suppression~\cite{Dienes:2016vei}) are included.    It would therefore be  interesting to explore the extent to which stasis can arise in such scenarios, or more generally in theories with non-power-law scaling relations.

The above comments regarding Regge excitations are independent of the relevant scales in these theories.   For this reason we have not worried about states which might be super-Planckian and which might therefore form black holes.
However, one important special class of string theories consists of those strings whose radii of compactification are very large compared to the inverse string scale.  These are precisely the strings that are capable of yielding reduced GUT~\cite{Dienes:1998vh,Dienes:1998vg}, Planck~\cite{Arkani-Hamed:1998jmv, Antoniadis:1998ig, Arkani-Hamed:1998sfv}, and string~\cite{Lykken:1996fj, Dienes:1998vh, Kakushadze:1998wp} scales.  Depending on the details of their constructions, such strings may even be effectively stable without spacetime supersymmetry~\cite{Abel:2015oxa,Abel:2017vos}.   In general, such string theories have densely populated towers of light KK states whose masses lie well below these reduced GUT, Planck, or string scales.   Such strings are thus prime candidates for producing not only a KK-induced stasis, as described above, but one which is realized within a full string-theoretic (and therefore UV-complete) framework.  Moreover, the Regge excitations within such frameworks play no role at scales below the string scale and can therefore be disregarded as far as stasis is concerned.

A third framework for physics beyond the Standard Model which gives rise to an infinite tower 
of $\phi_\ell$ states consists of  QCD-like theories at strong coupling.   In such cases, the 
fundamental QCD-like degrees of freedom (\eg, ``quarks'') are bound into an infinite 
spectrum of composite objects (``hadrons'').   It turns out that such theories can be 
analyzed via the so-called AdS/CFT correspondence~\cite{Maldacena:1997re}.  Through this 
correspondence, such strongly-coupled theories map onto five-dimensional gravitational 
theories whose low-energy limits are KK theories on anti-de Sitter (AdS) spacetimes.  
For example, the scale $\Lambda_{\rm IR}$ below which the strongly-coupled theory becomes 
confining and the cutoff scale $\Lambda_{\rm UV}$ above which the effective theory involving the
``quarks'' breaks down are related by $\Lambda_{\rm IR} = \Lambda_{\rm UV}e^{-\pi k R}$, where 
$R$ is the radius of the extra dimension and $k$ is the AdS curvature scale.
Thus, within the $\pi k R \gg 1$ regime,  there is a significant 
separation $\Lambda_{\rm UV} \gg \Lambda_{\rm IR}$ between these scales.  Moreover,
within this same regime, the mass $m_0 \sim \pi m_{\rm KK}$ of the lightest
composite state in the theory and the difference $m_{\ell + 1} - m_{\ell} \sim \pi m_{\rm KK}$ 
between the masses of adjacent states in the tower are both set by the
scale $m_{\rm KK} \equiv k e^{-\pi k R}$~\cite{Gherghetta:2000qt}.  Thus, one finds that 
for $\pi k R \gg 1$, there can exist a large number of states with masses below the cutoff scale 
$\Lambda_{\rm UV}$, as required for stasis.

Vacuum-misalignment production provides a natural 
mechanism for generating a spectrum of abundances for the composite states in 
strongly-coupled theories of this sort.  In particular, if there also exists a fundamental
scalar in the theory which dynamically acquires a mass as the result of a phase transition at 
a critical temperature $T_c \ll \Lambda_{\rm IR}$, the composite states can acquire 
abundances by mixing with this fundamental scalar~\cite{Buyukdag:2019lhh}.  The 
scaling behavior of $\Omega_\ell^{(0)}$ with $m_\ell$ depends on the details of the model 
and the background cosmology.  For example, one finds that $\alpha = -1$ in situations in 
which all of the mass-eigenstate fields of the theory at temperatures below $T_c$ are 
sufficiently heavy that their zero-modes begin oscillating immediately at the time of
the phase transition.  By contrast, in situations in which the zero-modes for these fields 
begin oscillating only at later times, one finds that $\alpha < -1$. 

In general, in order to assess whether a cosmological model involving a large tower of states 
might potentially 
give rise to a stasis epoch, one must examine whether the effective 
scaling exponents $\alpha$, $\gamma$, and $\delta$ obtained for that cosmology satisfy the 
criterion in Eq.~(\ref{range}).  This remains true even when these states are composite.     
In the regime in which the dynamically-generated mass of the fundamental 
scalar is small compared to all other relevant scales in the theory, the decay widths
typically scale with $m_\ell$ such that $\gamma \approx 4$~\cite{Buyukdag:2019lhh}.  On the 
other hand, mixing with the fundamental scalar does not dramatically alter the mass spectrum 
of the theory within the $\pi k R \gg 1$ regime, whereupon $m_{\ell + 1} - m_{\ell}$ remains 
roughly constant~\cite{Batell:2007jv}.  This implies that $\delta \approx 1$.  
However, a more detailed analysis reveals that this mass splitting decreases slightly with 
$\ell$, implying that the effective value of $\delta$ for such a tower of states is actually 
slightly less than $1$.  As a result, for the case in which $\alpha = -1$, one finds that a 
tower of partially composite ``hadrons'' is at least approximately consistent with the 
criterion in Eq.~(\ref{range}) and thus appears promising as a potential model for cosmic stasis.  
It will be interesting to further explore possible realizations of stasis along these lines.

Finally, let us discuss the possibility that the $\phi_\ell$ are actually
primordial black holes (PBHs).~ 
The abundances of PBHs scale as matter, and thus an early PBH-dominated epoch can serve as the initial matter-dominated epoch required for stasis.
Of course, such PBHs do not produce radiation via particle-like decay --- they do so via evaporation. This is nevertheless a process that effectively converts matter into radiation, and thus it is possible that this too could yield a stasis-like solution that counterbalances the redshifting effects of cosmic expansion.
One novel feature is that lighter PBHs tend to evaporate more rapidly than heavier PBHs.  However, as long as the abundances of these lighter PBHs are greater than those of the heavier PBHs, one could still potentially achieve a stasis;   this stasis would simply proceed  {\it up}\/ the tower, from lighter to heavier PBHs, rather than down.
Another novel feature is that black-hole evaporation does not follow an exponential decay law;  rather, the mass of an evaporating black hole (and therefore the energy density of the corresponding PBH population) drops with time according to a non-trivial function  which can in various regimes be approximated as a power-law. It is nevertheless possible that stasis can be achieved even with such a change in this functional form.   We leave this question for further study~\cite{visogenesis}.

Many aspects of our work 
are tangentially related to ideas which have already
been discussed in the literature.
For example, the idea that the Friedmann equations can admit attractor solutions which cause certain 
cosmological components to evolve in predictable ways is of course not a new one.  
Within the context of quintessence, for example, so-called ``tracker solutions'' have been 
identified wherein the equation-of-state parameter for the scalar sector evolves toward 
the equation-of-state parameter for the dominant background component, be it either 
matter or radiation~\cite{Zlatev:1998tr}.  Certain types of interactions between the 
dark-energy sector and the matter sector at the level of Friedmann equations have also 
been exploited in order to engineer attractor solutions to these equations, with 
potential implications for the cosmic coincidence 
problem~\cite{Chimento:2000kq,Copeland:2006wr,Tsujikawa:2010sc,Kremer:2011cd}.  
It can also be shown that there exist 
certain types of interactions which can be added to the equations of motion for the 
cosmological components without disturbing the Lotka-Volterra structure of these
equations~\cite{Perez:2013zya,Simon-Petit:2016tud}.  Such ``jungle universe''
models often exhibit non-trivial attractor solutions.  

Despite these similarities, 
the stasis scenario presented in this paper 
differs from these other cosmological scenarios in several crucial ways.  
First, we do not assume any unorthodox equation of state for any of the particle species involved
in our scenario.  We also do not introduce any {\it ad hoc}\/ suppositions 
concerning the form of the scalar potential in our theory, nor do we posit the existence 
of any additional interaction terms in the Friedmann equations.  
Instead, we demonstrated that 
stasis emerges as the result of a subtle and complex interplay between 
the effects of cosmic expansion and the conversion of 
matter to radiation through particle decays.
This is typically not the case for the ``tracking'' solutions
that arise in quintessence models, wherein the scalar sector simply mimics the 
background rather than modifying the equation of state of the universe as a whole.  
Jungle-universe models, by contrast, are more akin to our stasis scenario in this 
respect.  Nevertheless, the structure of the dynamics which gives rise to a stasis 
epoch is fundamentally different from the dynamics which governs such models.  Likewise,
the dynamics underlying stasis does not require the introduction of any interaction 
terms within the Friedmann equation that 
do not, {\em a priori}\/, originate from fundamental particle interactions.

The above ideas concerning the physical implications of stasis are likely only the tip of the iceberg,
and new phenomenological possibilities  involving stasis are likely to continue to present themselves.
Of course, this is to be expected.  In general, the expansion of the universe has far-reaching implications 
for almost everything contained  within it.
As a result, there are many possible ``clocks'' that can be used to measure the passage 
of cosmological time.
One of these clocks, for example, is based directly on the expansion itself, 
tracking the number of $e$-folds of growth in the cosmological scale factor.
However, another clock is based on the abundances of the 
different energy components and the passage between cosmological epochs.
Viewed from this perspective, stasis represents a way of suspending the passage of time
for the second clock while allowing the first clock to keep ticking.
This in turn causes the different clocks
to fall out of alignment, implying that standard abundance-based time-markers (such as the moment of 
matter-radiation equality)
may now occur within a universe whose overall size is quite different than normally assumed.
Even more importantly, this method of misaligning the clocks introduces long-lived {\it mixed-component}\/
epochs
whose equations-of-state parameters $w$
 lie between 0 and 1/3 and remain {\it fixed}\/ throughout the entire interval.
Indeed, as we have seen, this implies that we can have long-lived epochs which are not radiation-dominated
or matter-dominated, and in fact are not dominated by any particular component at all.
This decoupling of the different clocks, together with the existence of such stable mixed-component epochs,
 thus introduces new degrees of flexibility into 
 early-universe model-building,
and it is likely that these features can be exploited to address a number of cosmological puzzles.
The implications of stasis are thus ripe for future exploration.


\begin{acknowledgments}

The research activities of KRD are supported in part by the U.S.\ Department of Energy
under Grant DE-FG02-13ER41976 / DE-SC0009913, and also by the U.S.\ National Science Foundation through its employee IR/D program.
The work of LH is supported in part by the U.K.\ Science and Technology Facilities Council (STFC) under Grant ST/P001246/1.
The work of FH is supported in part by the International Postdoctoral Exchange Fellowship Program; by the National Natural Science Foundation of China under Grants 12025507, 11690022, 11947302, 12022514, and 11875003;  by the Strategic Priority Research Program and Key Research Program of Frontier Science of the Chinese Academy of Sciences under Grants XDB21010200, XDB23010000, and ZDBS-LY-7003; and by
the CAS project for Young Scientists in Basic Research YSBR-006.
The work of DK is supported in part by the U.S.\ Department of Energy under Grant DE-SC0010813.
The work of TMPT is supported in part by the U.S.\ National Science Foundation under Grant PHY-1915005.
The research activities of BT are supported in part
by the U.S.\ National Science Foundation under Grant PHY-2014104.
The opinions and conclusions
expressed herein are those of the authors, and do not represent any funding agencies.

\end{acknowledgments}

\appendix
\FloatBarrier

\begin{table*}[t!]
\begin{center}
\begin{tabular}{||c|c||}
\hline
\hline
{\bf QUANTUM MECHANICS} &  {\bf COSMOLOGY} \\ \hline\hline
wavepacket broadening & ~~increasing $\Omega_M$ from cosmological expansion (due to $\Omega_\gamma\to \Omega_M$)~~\\ \hline
confining harmonic-oscillator potential & ~~counterbalancing effect $\Omega_M\to \Omega_\gamma$ due to particle decays ~~ \\ \hline
width/shape of wavepacket & abundances $\Omega_M$, $\Omega_\gamma$\\   \hline
coherent state (fixed width/shape) &  stasis state (fixed $\Omega_M,\Omega_\gamma$) \\   \hline
tower of oscillator states $|\ell\rangle$   &  tower of decaying states $\phi_\ell$ \\   \hline
specific linear combination in Eq.~(\ref{coherent}) & specific scaling relations in Eq.~(\ref{scalings}) \\  \hline
lowering operator $a$ &  time evolution of sequential decays down the tower \\ \hline
coherent state as eigenstate of $a$   &   stasis state invariant under time evolution of sequential decays \\ \hline
true (stable) ground state ($\eta=0$) & state with vanishing $\Omega_M$ or $\Omega_\gamma$ (trivially stable epoch)\\ \hline
excited coherent state ($\eta\not=0$) & stasis state with both $\Omega_M,\Omega_\gamma\not=0$
        \\ \hline
infinite tower of $|\ell\rangle$ states &   no edge effects  \\
with $|\ell=0\rangle$ properly annihilated by $a$ &     at top or bottom of tower  \\
  $\Longrightarrow$ ~ exact coherent state &  ~  $\Longrightarrow$ ~ stasis state eternal \\ \hline
truncate $|\ell\rangle$ tower at top  &         initial edge effects  \\
  ~~$\Longrightarrow$~ small-scale fluctuations of packet shape~~ &   
      $\Longrightarrow$~ stasis state not valid to arbitrarily early times \\ \hline
  stable orbit in phase space   &   local attractor in $(\Omega_M,\langle \Omega_M\rangle)$ space \\
\hline
  minimum uncertainty relation &  global attractor in  $(\Omega_M,\langle \Omega_M\rangle)$ space\\
\hline
\hline
\end{tabular}
\end{center}
\caption{A proposed analogy between quantum-mechanical coherent states and cosmological stasis states.}
\label{table:analogy}
\end{table*}

\section{Stasis as a cosmological coherent state? \label{sec:coherent}}

In this Appendix we present a ``bonus track'' --- an interesting analogy between stasis states
and quantum-mechanical coherent states.
Our purpose here is merely to highlight some similarities
which, although purely speculative, might eventually be developed into a more
rigorous correspondence.

One hallmark of quantum-mechanical systems is that wavepackets broaden.
This expectation is true in free space, but it is possible to counterbalance this
broadening by considering our system under the influence of an appropriate
confining potential.    With this potential taken to be that of a harmonic oscillator,
one can construct a wavepacket in which this tendency towards broadening
is exactly cancelled.   This wavepacket is that of a so-called ``coherent'' state.
Indeed, if $a^\dagger$ and $a$ are the harmonic-oscillator 
raising and lowering operators, the coherent state $|\eta\rangle$ parametrized by 
any $\eta\in \mathbb{C}$ is given as
\beq
              |\eta\rangle ~\sim ~ \sum_{\ell=0}^\infty \frac{\eta^\ell}{\sqrt{\ell!}} |\ell \rangle
\label{coherent}
\eeq
where $|\ell\rangle \equiv (a^\dagger)^\ell |0\rangle/\sqrt{\ell!}$.
One defining property of a coherent state is that it is an eigenstate of the lowering operator:
$a|\eta\rangle = \eta|\eta\rangle$.   Another is that this state is a spatially translated version of the true ground state ($\eta=0$).     Just like the true ground state of the system, coherent states have stable orbits in phase space and satisfy the minimum uncertainty relations.

Ultimately the coherent state maintains its coherence as a result of the careful balancing of the different contributing
states $|\ell\rangle$ within Eq.~(\ref{coherent}).
The lowering operator $a$ takes us from each state $|\ell\rangle$ to the state immediately below it within the tower, namely $|\ell-1\rangle$.    The particular combination of $|\ell\rangle$ states within Eq.~(\ref{coherent}) is then an eigenstate
of $a$ because our tower of $|\ell\rangle$ states is infinite, with the lowest state annihilated by $a$.

Given these observations, it is possible to propose an analogy, as shown in Table~\ref{table:analogy}, between quantum-mechanical coherent states and our cosmological stasis states.   Of course, such an analogy is at best highly speculative and awaits more rigorous development.

\FloatBarrier

\vfill\eject

\bigskip
\bigskip

\bibliography{TheLiterature2}

\begin{thebibliography}{72}%
\makeatletter
\providecommand \@ifxundefined [1]{%
 \@ifx{#1\undefined}
}%
\providecommand \@ifnum [1]{%
 \ifnum #1\expandafter \@firstoftwo
 \else \expandafter \@secondoftwo
 \fi
}%
\providecommand \@ifx [1]{%
 \ifx #1\expandafter \@firstoftwo
 \else \expandafter \@secondoftwo
 \fi
}%
\providecommand \natexlab [1]{#1}%
\providecommand \enquote  [1]{``#1''}%
\providecommand \bibnamefont  [1]{#1}%
\providecommand \bibfnamefont [1]{#1}%
\providecommand \citenamefont [1]{#1}%
\providecommand \href@noop [0]{\@secondoftwo}%
\providecommand \href [0]{\begingroup \@sanitize@url \@href}%
\providecommand \@href[1]{\@@startlink{#1}\@@href}%
\providecommand \@@href[1]{\endgroup#1\@@endlink}%
\providecommand \@sanitize@url [0]{\catcode `\\12\catcode `\$12\catcode
  `\&12\catcode `\#12\catcode `\^12\catcode `\_12\catcode `\%12\relax}%
\providecommand \@@startlink[1]{}%
\providecommand \@@endlink[0]{}%
\providecommand \url  [0]{\begingroup\@sanitize@url \@url }%
\providecommand \@url [1]{\endgroup\@href {#1}{\urlprefix }}%
\providecommand \urlprefix  [0]{URL }%
\providecommand \Eprint [0]{\href }%
\providecommand \doibase [0]{http://dx.doi.org/}%
\providecommand \selectlanguage [0]{\@gobble}%
\providecommand \bibinfo  [0]{\@secondoftwo}%
\providecommand \bibfield  [0]{\@secondoftwo}%
\providecommand \translation [1]{[#1]}%
\providecommand \BibitemOpen [0]{}%
\providecommand \bibitemStop [0]{}%
\providecommand \bibitemNoStop [0]{.\EOS\space}%
\providecommand \EOS [0]{\spacefactor3000\relax}%
\providecommand \BibitemShut  [1]{\csname bibitem#1\endcsname}%
\let\auto@bib@innerbib\@empty
\bibitem [{\citenamefont {Dienes}\ and\ \citenamefont
  {Thomas}(2012{\natexlab{a}})}]{Dienes:2011ja}%
  \BibitemOpen
  \bibfield  {author} {\bibinfo {author} {\bibfnamefont {K.~R.}\ \bibnamefont
  {Dienes}}\ and\ \bibinfo {author} {\bibfnamefont {B.}~\bibnamefont
  {Thomas}},\ }\href {\doibase 10.1103/PhysRevD.85.083523} {\bibfield
  {journal} {\bibinfo  {journal} {Phys. Rev. D}\ }\textbf {\bibinfo {volume}
  {85}},\ \bibinfo {pages} {083523} (\bibinfo {year} {2012}{\natexlab{a}})},\
  \Eprint {http://arxiv.org/abs/1106.4546} {arXiv:1106.4546 [hep-ph]}
  \BibitemShut {NoStop}%
\bibitem [{\citenamefont {Dienes}\ and\ \citenamefont
  {Thomas}(2012{\natexlab{b}})}]{Dienes:2011sa}%
  \BibitemOpen
  \bibfield  {author} {\bibinfo {author} {\bibfnamefont {K.~R.}\ \bibnamefont
  {Dienes}}\ and\ \bibinfo {author} {\bibfnamefont {B.}~\bibnamefont
  {Thomas}},\ }\href {\doibase 10.1103/PhysRevD.85.083524} {\bibfield
  {journal} {\bibinfo  {journal} {Phys. Rev. D}\ }\textbf {\bibinfo {volume}
  {85}},\ \bibinfo {pages} {083524} (\bibinfo {year} {2012}{\natexlab{b}})},\
  \Eprint {http://arxiv.org/abs/1107.0721} {arXiv:1107.0721 [hep-ph]}
  \BibitemShut {NoStop}%
\bibitem [{\citenamefont {Dienes}\ and\ \citenamefont
  {Thomas}(2012{\natexlab{c}})}]{Dienes:2012jb}%
  \BibitemOpen
  \bibfield  {author} {\bibinfo {author} {\bibfnamefont {K.~R.}\ \bibnamefont
  {Dienes}}\ and\ \bibinfo {author} {\bibfnamefont {B.}~\bibnamefont
  {Thomas}},\ }\href {\doibase 10.1103/PhysRevD.86.055013} {\bibfield
  {journal} {\bibinfo  {journal} {Phys. Rev. D}\ }\textbf {\bibinfo {volume}
  {86}},\ \bibinfo {pages} {055013} (\bibinfo {year} {2012}{\natexlab{c}})},\
  \Eprint {http://arxiv.org/abs/1203.1923} {arXiv:1203.1923 [hep-ph]}
  \BibitemShut {NoStop}%
\bibitem [{\citenamefont {Dienes}\ \emph
  {et~al.}(2017{\natexlab{a}})\citenamefont {Dienes}, \citenamefont {Huang},
  \citenamefont {Su},\ and\ \citenamefont {Thomas}}]{Dienes:2016vei}%
  \BibitemOpen
  \bibfield  {author} {\bibinfo {author} {\bibfnamefont {K.~R.}\ \bibnamefont
  {Dienes}}, \bibinfo {author} {\bibfnamefont {F.}~\bibnamefont {Huang}},
  \bibinfo {author} {\bibfnamefont {S.}~\bibnamefont {Su}}, \ and\ \bibinfo
  {author} {\bibfnamefont {B.}~\bibnamefont {Thomas}},\ }\href {\doibase
  10.1103/PhysRevD.95.043526} {\bibfield  {journal} {\bibinfo  {journal} {Phys.
  Rev. D}\ }\textbf {\bibinfo {volume} {95}},\ \bibinfo {pages} {043526}
  (\bibinfo {year} {2017}{\natexlab{a}})},\ \Eprint
  {http://arxiv.org/abs/1610.04112} {arXiv:1610.04112 [hep-ph]} \BibitemShut
  {NoStop}%
\bibitem [{\citenamefont {Dienes}\ \emph {et~al.}(2018)\citenamefont {Dienes},
  \citenamefont {Fennick}, \citenamefont {Kumar},\ and\ \citenamefont
  {Thomas}}]{Dienes:2017zjq}%
  \BibitemOpen
  \bibfield  {author} {\bibinfo {author} {\bibfnamefont {K.~R.}\ \bibnamefont
  {Dienes}}, \bibinfo {author} {\bibfnamefont {J.}~\bibnamefont {Fennick}},
  \bibinfo {author} {\bibfnamefont {J.}~\bibnamefont {Kumar}}, \ and\ \bibinfo
  {author} {\bibfnamefont {B.}~\bibnamefont {Thomas}},\ }\href {\doibase
  10.1103/PhysRevD.97.063522} {\bibfield  {journal} {\bibinfo  {journal} {Phys.
  Rev. D}\ }\textbf {\bibinfo {volume} {97}},\ \bibinfo {pages} {063522}
  (\bibinfo {year} {2018})},\ \Eprint {http://arxiv.org/abs/1712.09919}
  {arXiv:1712.09919 [hep-ph]} \BibitemShut {NoStop}%
\bibitem [{\citenamefont {Turner}(1983)}]{Turner:1983he}%
  \BibitemOpen
  \bibfield  {author} {\bibinfo {author} {\bibfnamefont {M.~S.}\ \bibnamefont
  {Turner}},\ }\href {\doibase 10.1103/PhysRevD.28.1243} {\bibfield  {journal}
  {\bibinfo  {journal} {Phys. Rev. D}\ }\textbf {\bibinfo {volume} {28}},\
  \bibinfo {pages} {1243} (\bibinfo {year} {1983})}\BibitemShut {NoStop}%
\bibitem [{\citenamefont {Dienes}\ \emph {et~al.}()\citenamefont {Dienes},
  \citenamefont {Heurtier}, \citenamefont {Huang}, \citenamefont {Kim},
  \citenamefont {Park}, \citenamefont {Shin}, \citenamefont {Tait},\ and\
  \citenamefont {Thomas}}]{visogenesis}%
  \BibitemOpen
  \bibfield  {author} {\bibinfo {author} {\bibfnamefont {K.~R.}\ \bibnamefont
  {Dienes}}, \bibinfo {author} {\bibfnamefont {L.}~\bibnamefont {Heurtier}},
  \bibinfo {author} {\bibfnamefont {F.}~\bibnamefont {Huang}}, \bibinfo
  {author} {\bibfnamefont {D.}~\bibnamefont {Kim}}, \bibinfo {author}
  {\bibfnamefont {J.-C.}\ \bibnamefont {Park}}, \bibinfo {author}
  {\bibfnamefont {S.}~\bibnamefont {Shin}}, \bibinfo {author} {\bibfnamefont
  {T.~M.~P.}\ \bibnamefont {Tait}}, \ and\ \bibinfo {author} {\bibfnamefont
  {B.}~\bibnamefont {Thomas}},\ }\href@noop {} {\ }\Eprint
  {http://arxiv.org/abs/to appear} {to appear} \BibitemShut {NoStop}%
\bibitem [{\citenamefont {Easther}\ \emph {et~al.}(2014)\citenamefont
  {Easther}, \citenamefont {Galvez}, \citenamefont {Ozsoy},\ and\ \citenamefont
  {Watson}}]{Easther:2013nga}%
  \BibitemOpen
  \bibfield  {author} {\bibinfo {author} {\bibfnamefont {R.}~\bibnamefont
  {Easther}}, \bibinfo {author} {\bibfnamefont {R.}~\bibnamefont {Galvez}},
  \bibinfo {author} {\bibfnamefont {O.}~\bibnamefont {Ozsoy}}, \ and\ \bibinfo
  {author} {\bibfnamefont {S.}~\bibnamefont {Watson}},\ }\href {\doibase
  10.1103/PhysRevD.89.023522} {\bibfield  {journal} {\bibinfo  {journal} {Phys.
  Rev. D}\ }\textbf {\bibinfo {volume} {89}},\ \bibinfo {pages} {023522}
  (\bibinfo {year} {2014})},\ \Eprint {http://arxiv.org/abs/1307.2453}
  {arXiv:1307.2453 [hep-ph]} \BibitemShut {NoStop}%
\bibitem [{\citenamefont {Das}\ \emph {et~al.}(2015)\citenamefont {Das},
  \citenamefont {Dutta},\ and\ \citenamefont {Maharana}}]{Das:2015uwa}%
  \BibitemOpen
  \bibfield  {author} {\bibinfo {author} {\bibfnamefont {K.}~\bibnamefont
  {Das}}, \bibinfo {author} {\bibfnamefont {K.}~\bibnamefont {Dutta}}, \ and\
  \bibinfo {author} {\bibfnamefont {A.}~\bibnamefont {Maharana}},\ }\href
  {\doibase 10.1016/j.physletb.2015.10.041} {\bibfield  {journal} {\bibinfo
  {journal} {Phys. Lett. B}\ }\textbf {\bibinfo {volume} {751}},\ \bibinfo
  {pages} {195} (\bibinfo {year} {2015})},\ \Eprint
  {http://arxiv.org/abs/1506.05745} {arXiv:1506.05745 [hep-ph]} \BibitemShut
  {NoStop}%
\bibitem [{\citenamefont {Allahverdi}\ \emph {et~al.}(2018)\citenamefont
  {Allahverdi}, \citenamefont {Dutta},\ and\ \citenamefont
  {Maharana}}]{Allahverdi:2018iod}%
  \BibitemOpen
  \bibfield  {author} {\bibinfo {author} {\bibfnamefont {R.}~\bibnamefont
  {Allahverdi}}, \bibinfo {author} {\bibfnamefont {K.}~\bibnamefont {Dutta}}, \
  and\ \bibinfo {author} {\bibfnamefont {A.}~\bibnamefont {Maharana}},\ }\href
  {\doibase 10.1088/1475-7516/2018/10/038} {\bibfield  {journal} {\bibinfo
  {journal} {JCAP}\ }\textbf {\bibinfo {volume} {10}},\ \bibinfo {pages} {038}
  (\bibinfo {year} {2018})},\ \Eprint {http://arxiv.org/abs/1808.02659}
  {arXiv:1808.02659 [astro-ph.CO]} \BibitemShut {NoStop}%
\bibitem [{\citenamefont {Heurtier}\ and\ \citenamefont
  {Huang}(2019)}]{Heurtier:2019eou}%
  \BibitemOpen
  \bibfield  {author} {\bibinfo {author} {\bibfnamefont {L.}~\bibnamefont
  {Heurtier}}\ and\ \bibinfo {author} {\bibfnamefont {F.}~\bibnamefont
  {Huang}},\ }\href {\doibase 10.1103/PhysRevD.100.043507} {\bibfield
  {journal} {\bibinfo  {journal} {Phys. Rev. D}\ }\textbf {\bibinfo {volume}
  {100}},\ \bibinfo {pages} {043507} (\bibinfo {year} {2019})},\ \Eprint
  {http://arxiv.org/abs/1905.05191} {arXiv:1905.05191 [hep-ph]} \BibitemShut
  {NoStop}%
\bibitem [{\citenamefont {Erickcek}\ and\ \citenamefont
  {Sigurdson}(2011)}]{Erickcek:2011us}%
  \BibitemOpen
  \bibfield  {author} {\bibinfo {author} {\bibfnamefont {A.~L.}\ \bibnamefont
  {Erickcek}}\ and\ \bibinfo {author} {\bibfnamefont {K.}~\bibnamefont
  {Sigurdson}},\ }\href {\doibase 10.1103/PhysRevD.84.083503} {\bibfield
  {journal} {\bibinfo  {journal} {Phys. Rev. D}\ }\textbf {\bibinfo {volume}
  {84}},\ \bibinfo {pages} {083503} (\bibinfo {year} {2011})},\ \Eprint
  {http://arxiv.org/abs/1106.0536} {arXiv:1106.0536 [astro-ph.CO]} \BibitemShut
  {NoStop}%
\bibitem [{\citenamefont {Fan}\ \emph {et~al.}(2014)\citenamefont {Fan},
  \citenamefont {\"Ozsoy},\ and\ \citenamefont {Watson}}]{Fan:2014zua}%
  \BibitemOpen
  \bibfield  {author} {\bibinfo {author} {\bibfnamefont {J.}~\bibnamefont
  {Fan}}, \bibinfo {author} {\bibfnamefont {O.}~\bibnamefont {\"Ozsoy}}, \ and\
  \bibinfo {author} {\bibfnamefont {S.}~\bibnamefont {Watson}},\ }\href
  {\doibase 10.1103/PhysRevD.90.043536} {\bibfield  {journal} {\bibinfo
  {journal} {Phys. Rev. D}\ }\textbf {\bibinfo {volume} {90}},\ \bibinfo
  {pages} {043536} (\bibinfo {year} {2014})},\ \Eprint
  {http://arxiv.org/abs/1405.7373} {arXiv:1405.7373 [hep-ph]} \BibitemShut
  {NoStop}%
\bibitem [{\citenamefont {Georg}\ \emph {et~al.}(2016)\citenamefont {Georg},
  \citenamefont {\c{S}eng\"or},\ and\ \citenamefont {Watson}}]{Georg:2016yxa}%
  \BibitemOpen
  \bibfield  {author} {\bibinfo {author} {\bibfnamefont {J.}~\bibnamefont
  {Georg}}, \bibinfo {author} {\bibfnamefont {G.}~\bibnamefont {\c{S}eng\"or}},
  \ and\ \bibinfo {author} {\bibfnamefont {S.}~\bibnamefont {Watson}},\ }\href
  {\doibase 10.1103/PhysRevD.93.123523} {\bibfield  {journal} {\bibinfo
  {journal} {Phys. Rev. D}\ }\textbf {\bibinfo {volume} {93}},\ \bibinfo
  {pages} {123523} (\bibinfo {year} {2016})},\ \Eprint
  {http://arxiv.org/abs/1603.00023} {arXiv:1603.00023 [hep-ph]} \BibitemShut
  {NoStop}%
\bibitem [{\citenamefont {Green}\ \emph {et~al.}(1997)\citenamefont {Green},
  \citenamefont {Liddle},\ and\ \citenamefont {Riotto}}]{Green:1997pr}%
  \BibitemOpen
  \bibfield  {author} {\bibinfo {author} {\bibfnamefont {A.~M.}\ \bibnamefont
  {Green}}, \bibinfo {author} {\bibfnamefont {A.~R.}\ \bibnamefont {Liddle}}, \
  and\ \bibinfo {author} {\bibfnamefont {A.}~\bibnamefont {Riotto}},\ }\href
  {\doibase 10.1103/PhysRevD.56.7559} {\bibfield  {journal} {\bibinfo
  {journal} {Phys. Rev. D}\ }\textbf {\bibinfo {volume} {56}},\ \bibinfo
  {pages} {7559} (\bibinfo {year} {1997})},\ \Eprint
  {http://arxiv.org/abs/astro-ph/9705166} {arXiv:astro-ph/9705166} \BibitemShut
  {NoStop}%
\bibitem [{\citenamefont {Barenboim}\ and\ \citenamefont
  {Rasero}(2014)}]{Barenboim:2013gya}%
  \BibitemOpen
  \bibfield  {author} {\bibinfo {author} {\bibfnamefont {G.}~\bibnamefont
  {Barenboim}}\ and\ \bibinfo {author} {\bibfnamefont {J.}~\bibnamefont
  {Rasero}},\ }\href {\doibase 10.1007/JHEP04(2014)138} {\bibfield  {journal}
  {\bibinfo  {journal} {JHEP}\ }\textbf {\bibinfo {volume} {04}},\ \bibinfo
  {pages} {138} (\bibinfo {year} {2014})},\ \Eprint
  {http://arxiv.org/abs/1311.4034} {arXiv:1311.4034 [hep-ph]} \BibitemShut
  {NoStop}%
\bibitem [{\citenamefont {D'Eramo}\ \emph {et~al.}(2017)\citenamefont
  {D'Eramo}, \citenamefont {Fernandez},\ and\ \citenamefont
  {Profumo}}]{DEramo:2017gpl}%
  \BibitemOpen
  \bibfield  {author} {\bibinfo {author} {\bibfnamefont {F.}~\bibnamefont
  {D'Eramo}}, \bibinfo {author} {\bibfnamefont {N.}~\bibnamefont {Fernandez}},
  \ and\ \bibinfo {author} {\bibfnamefont {S.}~\bibnamefont {Profumo}},\ }\href
  {\doibase 10.1088/1475-7516/2017/05/012} {\bibfield  {journal} {\bibinfo
  {journal} {JCAP}\ }\textbf {\bibinfo {volume} {05}},\ \bibinfo {pages} {012}
  (\bibinfo {year} {2017})},\ \Eprint {http://arxiv.org/abs/1703.04793}
  {arXiv:1703.04793 [hep-ph]} \BibitemShut {NoStop}%
\bibitem [{\citenamefont {Hamdan}\ and\ \citenamefont
  {Unwin}(2018)}]{Hamdan:2017psw}%
  \BibitemOpen
  \bibfield  {author} {\bibinfo {author} {\bibfnamefont {S.}~\bibnamefont
  {Hamdan}}\ and\ \bibinfo {author} {\bibfnamefont {J.}~\bibnamefont {Unwin}},\
  }\href {\doibase 10.1142/S021773231850181X} {\bibfield  {journal} {\bibinfo
  {journal} {Mod. Phys. Lett. A}\ }\textbf {\bibinfo {volume} {33}},\ \bibinfo
  {pages} {1850181} (\bibinfo {year} {2018})},\ \Eprint
  {http://arxiv.org/abs/1710.03758} {arXiv:1710.03758 [hep-ph]} \BibitemShut
  {NoStop}%
\bibitem [{\citenamefont {Drees}\ and\ \citenamefont
  {Hajkarim}(2018)}]{Drees:2017iod}%
  \BibitemOpen
  \bibfield  {author} {\bibinfo {author} {\bibfnamefont {M.}~\bibnamefont
  {Drees}}\ and\ \bibinfo {author} {\bibfnamefont {F.}~\bibnamefont
  {Hajkarim}},\ }\href {\doibase 10.1088/1475-7516/2018/02/057} {\bibfield
  {journal} {\bibinfo  {journal} {JCAP}\ }\textbf {\bibinfo {volume} {02}},\
  \bibinfo {pages} {057} (\bibinfo {year} {2018})},\ \Eprint
  {http://arxiv.org/abs/1711.05007} {arXiv:1711.05007 [hep-ph]} \BibitemShut
  {NoStop}%
\bibitem [{\citenamefont {D'Eramo}\ \emph {et~al.}(2018)\citenamefont
  {D'Eramo}, \citenamefont {Fernandez},\ and\ \citenamefont
  {Profumo}}]{DEramo:2017ecx}%
  \BibitemOpen
  \bibfield  {author} {\bibinfo {author} {\bibfnamefont {F.}~\bibnamefont
  {D'Eramo}}, \bibinfo {author} {\bibfnamefont {N.}~\bibnamefont {Fernandez}},
  \ and\ \bibinfo {author} {\bibfnamefont {S.}~\bibnamefont {Profumo}},\ }\href
  {\doibase 10.1088/1475-7516/2018/02/046} {\bibfield  {journal} {\bibinfo
  {journal} {JCAP}\ }\textbf {\bibinfo {volume} {02}},\ \bibinfo {pages} {046}
  (\bibinfo {year} {2018})},\ \Eprint {http://arxiv.org/abs/1712.07453}
  {arXiv:1712.07453 [hep-ph]} \BibitemShut {NoStop}%
\bibitem [{\citenamefont {Allahverdi}\ and\ \citenamefont
  {Osi\'nski}(2019)}]{Allahverdi:2018aux}%
  \BibitemOpen
  \bibfield  {author} {\bibinfo {author} {\bibfnamefont {R.}~\bibnamefont
  {Allahverdi}}\ and\ \bibinfo {author} {\bibfnamefont {J.~K.}\ \bibnamefont
  {Osi\'nski}},\ }\href {\doibase 10.1103/PhysRevD.99.083517} {\bibfield
  {journal} {\bibinfo  {journal} {Phys. Rev. D}\ }\textbf {\bibinfo {volume}
  {99}},\ \bibinfo {pages} {083517} (\bibinfo {year} {2019})},\ \Eprint
  {http://arxiv.org/abs/1812.10522} {arXiv:1812.10522 [hep-ph]} \BibitemShut
  {NoStop}%
\bibitem [{\citenamefont {Arias}\ \emph {et~al.}(2019)\citenamefont {Arias},
  \citenamefont {Bernal}, \citenamefont {Herrera},\ and\ \citenamefont
  {Maldonado}}]{Arias:2019uol}%
  \BibitemOpen
  \bibfield  {author} {\bibinfo {author} {\bibfnamefont {P.}~\bibnamefont
  {Arias}}, \bibinfo {author} {\bibfnamefont {N.}~\bibnamefont {Bernal}},
  \bibinfo {author} {\bibfnamefont {A.}~\bibnamefont {Herrera}}, \ and\
  \bibinfo {author} {\bibfnamefont {C.}~\bibnamefont {Maldonado}},\ }\href
  {\doibase 10.1088/1475-7516/2019/10/047} {\bibfield  {journal} {\bibinfo
  {journal} {JCAP}\ }\textbf {\bibinfo {volume} {10}},\ \bibinfo {pages} {047}
  (\bibinfo {year} {2019})},\ \Eprint {http://arxiv.org/abs/1906.04183}
  {arXiv:1906.04183 [hep-ph]} \BibitemShut {NoStop}%
\bibitem [{\citenamefont {Allahverdi}\ and\ \citenamefont
  {Osi\'nski}(2020)}]{Allahverdi:2019jsc}%
  \BibitemOpen
  \bibfield  {author} {\bibinfo {author} {\bibfnamefont {R.}~\bibnamefont
  {Allahverdi}}\ and\ \bibinfo {author} {\bibfnamefont {J.~K.}\ \bibnamefont
  {Osi\'nski}},\ }\href {\doibase 10.1103/PhysRevD.101.063503} {\bibfield
  {journal} {\bibinfo  {journal} {Phys. Rev. D}\ }\textbf {\bibinfo {volume}
  {101}},\ \bibinfo {pages} {063503} (\bibinfo {year} {2020})},\ \Eprint
  {http://arxiv.org/abs/1909.01457} {arXiv:1909.01457 [hep-ph]} \BibitemShut
  {NoStop}%
\bibitem [{\citenamefont {Sten~Delos}\ \emph {et~al.}(2019)\citenamefont
  {Sten~Delos}, \citenamefont {Linden},\ and\ \citenamefont
  {Erickcek}}]{StenDelos:2019xdk}%
  \BibitemOpen
  \bibfield  {author} {\bibinfo {author} {\bibfnamefont {M.}~\bibnamefont
  {Sten~Delos}}, \bibinfo {author} {\bibfnamefont {T.}~\bibnamefont {Linden}},
  \ and\ \bibinfo {author} {\bibfnamefont {A.~L.}\ \bibnamefont {Erickcek}},\
  }\href {\doibase 10.1103/PhysRevD.100.123546} {\bibfield  {journal} {\bibinfo
   {journal} {Phys. Rev. D}\ }\textbf {\bibinfo {volume} {100}},\ \bibinfo
  {pages} {123546} (\bibinfo {year} {2019})},\ \Eprint
  {http://arxiv.org/abs/1910.08553} {arXiv:1910.08553 [astro-ph.CO]}
  \BibitemShut {NoStop}%
\bibitem [{\citenamefont {Garcia}\ \emph {et~al.}(2020)\citenamefont {Garcia},
  \citenamefont {Kaneta}, \citenamefont {Mambrini},\ and\ \citenamefont
  {Olive}}]{Garcia:2020eof}%
  \BibitemOpen
  \bibfield  {author} {\bibinfo {author} {\bibfnamefont {M.~A.~G.}\
  \bibnamefont {Garcia}}, \bibinfo {author} {\bibfnamefont {K.}~\bibnamefont
  {Kaneta}}, \bibinfo {author} {\bibfnamefont {Y.}~\bibnamefont {Mambrini}}, \
  and\ \bibinfo {author} {\bibfnamefont {K.~A.}\ \bibnamefont {Olive}},\ }\href
  {\doibase 10.1103/PhysRevD.101.123507} {\bibfield  {journal} {\bibinfo
  {journal} {Phys. Rev. D}\ }\textbf {\bibinfo {volume} {101}},\ \bibinfo
  {pages} {123507} (\bibinfo {year} {2020})},\ \Eprint
  {http://arxiv.org/abs/2004.08404} {arXiv:2004.08404 [hep-ph]} \BibitemShut
  {NoStop}%
\bibitem [{\citenamefont {Cheek}\ \emph {et~al.}(2021)\citenamefont {Cheek},
  \citenamefont {Heurtier}, \citenamefont {Perez-Gonzalez},\ and\ \citenamefont
  {Turner}}]{Cheek:2021cfe}%
  \BibitemOpen
  \bibfield  {author} {\bibinfo {author} {\bibfnamefont {A.}~\bibnamefont
  {Cheek}}, \bibinfo {author} {\bibfnamefont {L.}~\bibnamefont {Heurtier}},
  \bibinfo {author} {\bibfnamefont {Y.~F.}\ \bibnamefont {Perez-Gonzalez}}, \
  and\ \bibinfo {author} {\bibfnamefont {J.}~\bibnamefont {Turner}},\
  }\href@noop {} {\  (\bibinfo {year} {2021})},\ \Eprint
  {http://arxiv.org/abs/2107.00016} {arXiv:2107.00016 [hep-ph]} \BibitemShut
  {NoStop}%
\bibitem [{\citenamefont {Joyce}(1997)}]{Joyce:1996cp}%
  \BibitemOpen
  \bibfield  {author} {\bibinfo {author} {\bibfnamefont {M.}~\bibnamefont
  {Joyce}},\ }\href {\doibase 10.1103/PhysRevD.55.1875} {\bibfield  {journal}
  {\bibinfo  {journal} {Phys. Rev. D}\ }\textbf {\bibinfo {volume} {55}},\
  \bibinfo {pages} {1875} (\bibinfo {year} {1997})},\ \Eprint
  {http://arxiv.org/abs/hep-ph/9606223} {arXiv:hep-ph/9606223} \BibitemShut
  {NoStop}%
\bibitem [{\citenamefont {Joyce}\ and\ \citenamefont
  {Prokopec}(1998)}]{Joyce:1997fc}%
  \BibitemOpen
  \bibfield  {author} {\bibinfo {author} {\bibfnamefont {M.}~\bibnamefont
  {Joyce}}\ and\ \bibinfo {author} {\bibfnamefont {T.}~\bibnamefont
  {Prokopec}},\ }\href {\doibase 10.1103/PhysRevD.57.6022} {\bibfield
  {journal} {\bibinfo  {journal} {Phys. Rev. D}\ }\textbf {\bibinfo {volume}
  {57}},\ \bibinfo {pages} {6022} (\bibinfo {year} {1998})},\ \Eprint
  {http://arxiv.org/abs/hep-ph/9709320} {arXiv:hep-ph/9709320} \BibitemShut
  {NoStop}%
\bibitem [{\citenamefont {Davidson}\ \emph {et~al.}(2000)\citenamefont
  {Davidson}, \citenamefont {Losada},\ and\ \citenamefont
  {Riotto}}]{Davidson:2000dw}%
  \BibitemOpen
  \bibfield  {author} {\bibinfo {author} {\bibfnamefont {S.}~\bibnamefont
  {Davidson}}, \bibinfo {author} {\bibfnamefont {M.}~\bibnamefont {Losada}}, \
  and\ \bibinfo {author} {\bibfnamefont {A.}~\bibnamefont {Riotto}},\ }\href
  {\doibase 10.1103/PhysRevLett.84.4284} {\bibfield  {journal} {\bibinfo
  {journal} {Phys. Rev. Lett.}\ }\textbf {\bibinfo {volume} {84}},\ \bibinfo
  {pages} {4284} (\bibinfo {year} {2000})},\ \Eprint
  {http://arxiv.org/abs/hep-ph/0001301} {arXiv:hep-ph/0001301} \BibitemShut
  {NoStop}%
\bibitem [{\citenamefont {Servant}(2002)}]{Servant:2001jh}%
  \BibitemOpen
  \bibfield  {author} {\bibinfo {author} {\bibfnamefont {G.}~\bibnamefont
  {Servant}},\ }\href {\doibase 10.1088/1126-6708/2002/01/044} {\bibfield
  {journal} {\bibinfo  {journal} {JHEP}\ }\textbf {\bibinfo {volume} {01}},\
  \bibinfo {pages} {044} (\bibinfo {year} {2002})},\ \Eprint
  {http://arxiv.org/abs/hep-ph/0112209} {arXiv:hep-ph/0112209} \BibitemShut
  {NoStop}%
\bibitem [{\citenamefont {Barenboim}\ and\ \citenamefont
  {Rasero}(2012)}]{Barenboim:2012nh}%
  \BibitemOpen
  \bibfield  {author} {\bibinfo {author} {\bibfnamefont {G.}~\bibnamefont
  {Barenboim}}\ and\ \bibinfo {author} {\bibfnamefont {J.}~\bibnamefont
  {Rasero}},\ }\href {\doibase 10.1007/JHEP07(2012)028} {\bibfield  {journal}
  {\bibinfo  {journal} {JHEP}\ }\textbf {\bibinfo {volume} {07}},\ \bibinfo
  {pages} {028} (\bibinfo {year} {2012})},\ \Eprint
  {http://arxiv.org/abs/1202.6070} {arXiv:1202.6070 [hep-ph]} \BibitemShut
  {NoStop}%
\bibitem [{\citenamefont {Nardini}\ and\ \citenamefont
  {Sahu}(2011)}]{Nardini:2011hu}%
  \BibitemOpen
  \bibfield  {author} {\bibinfo {author} {\bibfnamefont {G.}~\bibnamefont
  {Nardini}}\ and\ \bibinfo {author} {\bibfnamefont {N.}~\bibnamefont {Sahu}},\
  }\href@noop {} {\  (\bibinfo {year} {2011})},\ \Eprint
  {http://arxiv.org/abs/1109.2829} {arXiv:1109.2829 [hep-ph]} \BibitemShut
  {NoStop}%
\bibitem [{\citenamefont {Bramante}\ and\ \citenamefont
  {Unwin}(2017)}]{Bramante:2017obj}%
  \BibitemOpen
  \bibfield  {author} {\bibinfo {author} {\bibfnamefont {J.}~\bibnamefont
  {Bramante}}\ and\ \bibinfo {author} {\bibfnamefont {J.}~\bibnamefont
  {Unwin}},\ }\href {\doibase 10.1007/JHEP02(2017)119} {\bibfield  {journal}
  {\bibinfo  {journal} {JHEP}\ }\textbf {\bibinfo {volume} {02}},\ \bibinfo
  {pages} {119} (\bibinfo {year} {2017})},\ \Eprint
  {http://arxiv.org/abs/1701.05859} {arXiv:1701.05859 [hep-ph]} \BibitemShut
  {NoStop}%
\bibitem [{\citenamefont {Affleck}\ and\ \citenamefont
  {Dine}(1985)}]{Affleck:1984fy}%
  \BibitemOpen
  \bibfield  {author} {\bibinfo {author} {\bibfnamefont {I.}~\bibnamefont
  {Affleck}}\ and\ \bibinfo {author} {\bibfnamefont {M.}~\bibnamefont {Dine}},\
  }\href {\doibase 10.1016/0550-3213(85)90021-5} {\bibfield  {journal}
  {\bibinfo  {journal} {Nucl. Phys. B}\ }\textbf {\bibinfo {volume} {249}},\
  \bibinfo {pages} {361} (\bibinfo {year} {1985})}\BibitemShut {NoStop}%
\bibitem [{\citenamefont {Dine}\ \emph {et~al.}(1996)\citenamefont {Dine},
  \citenamefont {Randall},\ and\ \citenamefont {Thomas}}]{Dine:1995kz}%
  \BibitemOpen
  \bibfield  {author} {\bibinfo {author} {\bibfnamefont {M.}~\bibnamefont
  {Dine}}, \bibinfo {author} {\bibfnamefont {L.}~\bibnamefont {Randall}}, \
  and\ \bibinfo {author} {\bibfnamefont {S.~D.}\ \bibnamefont {Thomas}},\
  }\href {\doibase 10.1016/0550-3213(95)00538-2} {\bibfield  {journal}
  {\bibinfo  {journal} {Nucl. Phys. B}\ }\textbf {\bibinfo {volume} {458}},\
  \bibinfo {pages} {291} (\bibinfo {year} {1996})},\ \Eprint
  {http://arxiv.org/abs/hep-ph/9507453} {arXiv:hep-ph/9507453} \BibitemShut
  {NoStop}%
\bibitem [{\citenamefont {Dine}\ and\ \citenamefont
  {Kusenko}(2003)}]{Dine:2003ax}%
  \BibitemOpen
  \bibfield  {author} {\bibinfo {author} {\bibfnamefont {M.}~\bibnamefont
  {Dine}}\ and\ \bibinfo {author} {\bibfnamefont {A.}~\bibnamefont {Kusenko}},\
  }\href {\doibase 10.1103/RevModPhys.76.1} {\bibfield  {journal} {\bibinfo
  {journal} {Rev. Mod. Phys.}\ }\textbf {\bibinfo {volume} {76}},\ \bibinfo
  {pages} {1} (\bibinfo {year} {2003})},\ \Eprint
  {http://arxiv.org/abs/hep-ph/0303065} {arXiv:hep-ph/0303065} \BibitemShut
  {NoStop}%
\bibitem [{\citenamefont {Nelson}\ and\ \citenamefont
  {Xiao}(2018)}]{Nelson:2018via}%
  \BibitemOpen
  \bibfield  {author} {\bibinfo {author} {\bibfnamefont {A.~E.}\ \bibnamefont
  {Nelson}}\ and\ \bibinfo {author} {\bibfnamefont {H.}~\bibnamefont {Xiao}},\
  }\href {\doibase 10.1103/PhysRevD.98.063516} {\bibfield  {journal} {\bibinfo
  {journal} {Phys. Rev. D}\ }\textbf {\bibinfo {volume} {98}},\ \bibinfo
  {pages} {063516} (\bibinfo {year} {2018})},\ \Eprint
  {http://arxiv.org/abs/1807.07176} {arXiv:1807.07176 [astro-ph.CO]}
  \BibitemShut {NoStop}%
\bibitem [{\citenamefont {Bernal}\ and\ \citenamefont
  {Hajkarim}(2019)}]{Bernal:2019lpc}%
  \BibitemOpen
  \bibfield  {author} {\bibinfo {author} {\bibfnamefont {N.}~\bibnamefont
  {Bernal}}\ and\ \bibinfo {author} {\bibfnamefont {F.}~\bibnamefont
  {Hajkarim}},\ }\href {\doibase 10.1103/PhysRevD.100.063502} {\bibfield
  {journal} {\bibinfo  {journal} {Phys. Rev. D}\ }\textbf {\bibinfo {volume}
  {100}},\ \bibinfo {pages} {063502} (\bibinfo {year} {2019})},\ \Eprint
  {http://arxiv.org/abs/1905.10410} {arXiv:1905.10410 [astro-ph.CO]}
  \BibitemShut {NoStop}%
\bibitem [{\citenamefont {Allahverdi}\ \emph {et~al.}(2020)\citenamefont
  {Allahverdi} \emph {et~al.}}]{Allahverdi:2020bys}%
  \BibitemOpen
  \bibfield  {author} {\bibinfo {author} {\bibfnamefont {R.}~\bibnamefont
  {Allahverdi}} \emph {et~al.},\ }\href {\doibase 10.21105/astro.2006.16182} {\
   (\bibinfo {year} {2020}),\ 10.21105/astro.2006.16182},\ \Eprint
  {http://arxiv.org/abs/2006.16182} {arXiv:2006.16182 [astro-ph.CO]}
  \BibitemShut {NoStop}%
\bibitem [{\citenamefont {Guo}\ \emph {et~al.}(2021)\citenamefont {Guo},
  \citenamefont {Sinha}, \citenamefont {Vagie},\ and\ \citenamefont
  {White}}]{Guo:2020grp}%
  \BibitemOpen
  \bibfield  {author} {\bibinfo {author} {\bibfnamefont {H.-K.}\ \bibnamefont
  {Guo}}, \bibinfo {author} {\bibfnamefont {K.}~\bibnamefont {Sinha}}, \bibinfo
  {author} {\bibfnamefont {D.}~\bibnamefont {Vagie}}, \ and\ \bibinfo {author}
  {\bibfnamefont {G.}~\bibnamefont {White}},\ }\href {\doibase
  10.1088/1475-7516/2021/01/001} {\bibfield  {journal} {\bibinfo  {journal}
  {JCAP}\ }\textbf {\bibinfo {volume} {01}},\ \bibinfo {pages} {001} (\bibinfo
  {year} {2021})},\ \Eprint {http://arxiv.org/abs/2007.08537} {arXiv:2007.08537
  [hep-ph]} \BibitemShut {NoStop}%
\bibitem [{\citenamefont {Giudice}\ \emph {et~al.}(2001)\citenamefont
  {Giudice}, \citenamefont {Kolb},\ and\ \citenamefont
  {Riotto}}]{Giudice:2000ex}%
  \BibitemOpen
  \bibfield  {author} {\bibinfo {author} {\bibfnamefont {G.~F.}\ \bibnamefont
  {Giudice}}, \bibinfo {author} {\bibfnamefont {E.~W.}\ \bibnamefont {Kolb}}, \
  and\ \bibinfo {author} {\bibfnamefont {A.}~\bibnamefont {Riotto}},\ }\href
  {\doibase 10.1103/PhysRevD.64.023508} {\bibfield  {journal} {\bibinfo
  {journal} {Phys. Rev. D}\ }\textbf {\bibinfo {volume} {64}},\ \bibinfo
  {pages} {023508} (\bibinfo {year} {2001})},\ \Eprint
  {http://arxiv.org/abs/hep-ph/0005123} {arXiv:hep-ph/0005123} \BibitemShut
  {NoStop}%
\bibitem [{\citenamefont {Gelmini}\ and\ \citenamefont
  {Gondolo}(2006)}]{Gelmini:2006pw}%
  \BibitemOpen
  \bibfield  {author} {\bibinfo {author} {\bibfnamefont {G.~B.}\ \bibnamefont
  {Gelmini}}\ and\ \bibinfo {author} {\bibfnamefont {P.}~\bibnamefont
  {Gondolo}},\ }\href {\doibase 10.1103/PhysRevD.74.023510} {\bibfield
  {journal} {\bibinfo  {journal} {Phys. Rev. D}\ }\textbf {\bibinfo {volume}
  {74}},\ \bibinfo {pages} {023510} (\bibinfo {year} {2006})},\ \Eprint
  {http://arxiv.org/abs/hep-ph/0602230} {arXiv:hep-ph/0602230} \BibitemShut
  {NoStop}%
\bibitem [{\citenamefont {Gelmini}\ \emph {et~al.}(2006)\citenamefont
  {Gelmini}, \citenamefont {Gondolo}, \citenamefont {Soldatenko},\ and\
  \citenamefont {Yaguna}}]{Gelmini:2006pq}%
  \BibitemOpen
  \bibfield  {author} {\bibinfo {author} {\bibfnamefont {G.}~\bibnamefont
  {Gelmini}}, \bibinfo {author} {\bibfnamefont {P.}~\bibnamefont {Gondolo}},
  \bibinfo {author} {\bibfnamefont {A.}~\bibnamefont {Soldatenko}}, \ and\
  \bibinfo {author} {\bibfnamefont {C.~E.}\ \bibnamefont {Yaguna}},\ }\href
  {\doibase 10.1103/PhysRevD.74.083514} {\bibfield  {journal} {\bibinfo
  {journal} {Phys. Rev. D}\ }\textbf {\bibinfo {volume} {74}},\ \bibinfo
  {pages} {083514} (\bibinfo {year} {2006})},\ \Eprint
  {http://arxiv.org/abs/hep-ph/0605016} {arXiv:hep-ph/0605016} \BibitemShut
  {NoStop}%
\bibitem [{\citenamefont {Erickcek}(2015)}]{Erickcek:2015jza}%
  \BibitemOpen
  \bibfield  {author} {\bibinfo {author} {\bibfnamefont {A.~L.}\ \bibnamefont
  {Erickcek}},\ }\href {\doibase 10.1103/PhysRevD.92.103505} {\bibfield
  {journal} {\bibinfo  {journal} {Phys. Rev. D}\ }\textbf {\bibinfo {volume}
  {92}},\ \bibinfo {pages} {103505} (\bibinfo {year} {2015})},\ \Eprint
  {http://arxiv.org/abs/1504.03335} {arXiv:1504.03335 [astro-ph.CO]}
  \BibitemShut {NoStop}%
\bibitem [{\citenamefont {Berlin}\ \emph
  {et~al.}(2016{\natexlab{a}})\citenamefont {Berlin}, \citenamefont {Hooper},\
  and\ \citenamefont {Krnjaic}}]{Berlin:2016vnh}%
  \BibitemOpen
  \bibfield  {author} {\bibinfo {author} {\bibfnamefont {A.}~\bibnamefont
  {Berlin}}, \bibinfo {author} {\bibfnamefont {D.}~\bibnamefont {Hooper}}, \
  and\ \bibinfo {author} {\bibfnamefont {G.}~\bibnamefont {Krnjaic}},\ }\href
  {\doibase 10.1016/j.physletb.2016.06.037} {\bibfield  {journal} {\bibinfo
  {journal} {Phys. Lett. B}\ }\textbf {\bibinfo {volume} {760}},\ \bibinfo
  {pages} {106} (\bibinfo {year} {2016}{\natexlab{a}})},\ \Eprint
  {http://arxiv.org/abs/1602.08490} {arXiv:1602.08490 [hep-ph]} \BibitemShut
  {NoStop}%
\bibitem [{\citenamefont {Berlin}\ \emph
  {et~al.}(2016{\natexlab{b}})\citenamefont {Berlin}, \citenamefont {Hooper},\
  and\ \citenamefont {Krnjaic}}]{Berlin:2016gtr}%
  \BibitemOpen
  \bibfield  {author} {\bibinfo {author} {\bibfnamefont {A.}~\bibnamefont
  {Berlin}}, \bibinfo {author} {\bibfnamefont {D.}~\bibnamefont {Hooper}}, \
  and\ \bibinfo {author} {\bibfnamefont {G.}~\bibnamefont {Krnjaic}},\ }\href
  {\doibase 10.1103/PhysRevD.94.095019} {\bibfield  {journal} {\bibinfo
  {journal} {Phys. Rev. D}\ }\textbf {\bibinfo {volume} {94}},\ \bibinfo
  {pages} {095019} (\bibinfo {year} {2016}{\natexlab{b}})},\ \Eprint
  {http://arxiv.org/abs/1609.02555} {arXiv:1609.02555 [hep-ph]} \BibitemShut
  {NoStop}%
\bibitem [{\citenamefont {Miller}\ \emph {et~al.}(2019)\citenamefont {Miller},
  \citenamefont {Erickcek},\ and\ \citenamefont {Murgia}}]{Miller:2019pss}%
  \BibitemOpen
  \bibfield  {author} {\bibinfo {author} {\bibfnamefont {C.}~\bibnamefont
  {Miller}}, \bibinfo {author} {\bibfnamefont {A.~L.}\ \bibnamefont
  {Erickcek}}, \ and\ \bibinfo {author} {\bibfnamefont {R.}~\bibnamefont
  {Murgia}},\ }\href {\doibase 10.1103/PhysRevD.100.123520} {\bibfield
  {journal} {\bibinfo  {journal} {Phys. Rev. D}\ }\textbf {\bibinfo {volume}
  {100}},\ \bibinfo {pages} {123520} (\bibinfo {year} {2019})},\ \Eprint
  {http://arxiv.org/abs/1908.10369} {arXiv:1908.10369 [astro-ph.CO]}
  \BibitemShut {NoStop}%
\bibitem [{\citenamefont {Dienes}\ \emph {et~al.}(2020)\citenamefont {Dienes},
  \citenamefont {Huang}, \citenamefont {Kost}, \citenamefont {Su},\ and\
  \citenamefont {Thomas}}]{Dienes:2020bmn}%
  \BibitemOpen
  \bibfield  {author} {\bibinfo {author} {\bibfnamefont {K.~R.}\ \bibnamefont
  {Dienes}}, \bibinfo {author} {\bibfnamefont {F.}~\bibnamefont {Huang}},
  \bibinfo {author} {\bibfnamefont {J.}~\bibnamefont {Kost}}, \bibinfo {author}
  {\bibfnamefont {S.}~\bibnamefont {Su}}, \ and\ \bibinfo {author}
  {\bibfnamefont {B.}~\bibnamefont {Thomas}},\ }\href {\doibase
  10.1103/PhysRevD.101.123511} {\bibfield  {journal} {\bibinfo  {journal}
  {Phys. Rev. D}\ }\textbf {\bibinfo {volume} {101}},\ \bibinfo {pages}
  {123511} (\bibinfo {year} {2020})},\ \Eprint
  {http://arxiv.org/abs/2001.02193} {arXiv:2001.02193 [astro-ph.CO]}
  \BibitemShut {NoStop}%
\bibitem [{\citenamefont {Dienes}\ \emph {et~al.}(2021)\citenamefont {Dienes},
  \citenamefont {Huang}, \citenamefont {Kost}, \citenamefont {Manogue},\ and\
  \citenamefont {Thomas}}]{Dienes:2021itb}%
  \BibitemOpen
  \bibfield  {author} {\bibinfo {author} {\bibfnamefont {K.~R.}\ \bibnamefont
  {Dienes}}, \bibinfo {author} {\bibfnamefont {F.}~\bibnamefont {Huang}},
  \bibinfo {author} {\bibfnamefont {J.}~\bibnamefont {Kost}}, \bibinfo {author}
  {\bibfnamefont {K.}~\bibnamefont {Manogue}}, \ and\ \bibinfo {author}
  {\bibfnamefont {B.}~\bibnamefont {Thomas}},\ }\href@noop {} {\  (\bibinfo
  {year} {2021})},\ \Eprint {http://arxiv.org/abs/2101.10337} {arXiv:2101.10337
  [astro-ph.CO]} \BibitemShut {NoStop}%
\bibitem [{\citenamefont {Dienes}\ \emph
  {et~al.}(2017{\natexlab{b}})\citenamefont {Dienes}, \citenamefont {Kost},\
  and\ \citenamefont {Thomas}}]{Dienes:2016zfr}%
  \BibitemOpen
  \bibfield  {author} {\bibinfo {author} {\bibfnamefont {K.~R.}\ \bibnamefont
  {Dienes}}, \bibinfo {author} {\bibfnamefont {J.}~\bibnamefont {Kost}}, \ and\
  \bibinfo {author} {\bibfnamefont {B.}~\bibnamefont {Thomas}},\ }\href
  {\doibase 10.1103/PhysRevD.95.123539} {\bibfield  {journal} {\bibinfo
  {journal} {Phys. Rev. D}\ }\textbf {\bibinfo {volume} {95}},\ \bibinfo
  {pages} {123539} (\bibinfo {year} {2017}{\natexlab{b}})},\ \Eprint
  {http://arxiv.org/abs/1612.08950} {arXiv:1612.08950 [hep-ph]} \BibitemShut
  {NoStop}%
\bibitem [{\citenamefont {Arkani-Hamed}\ \emph {et~al.}(1998)\citenamefont
  {Arkani-Hamed}, \citenamefont {Dimopoulos},\ and\ \citenamefont
  {Dvali}}]{Arkani-Hamed:1998jmv}%
  \BibitemOpen
  \bibfield  {author} {\bibinfo {author} {\bibfnamefont {N.}~\bibnamefont
  {Arkani-Hamed}}, \bibinfo {author} {\bibfnamefont {S.}~\bibnamefont
  {Dimopoulos}}, \ and\ \bibinfo {author} {\bibfnamefont {G.~R.}\ \bibnamefont
  {Dvali}},\ }\href {\doibase 10.1016/S0370-2693(98)00466-3} {\bibfield
  {journal} {\bibinfo  {journal} {Phys. Lett. B}\ }\textbf {\bibinfo {volume}
  {429}},\ \bibinfo {pages} {263} (\bibinfo {year} {1998})},\ \Eprint
  {http://arxiv.org/abs/hep-ph/9803315} {arXiv:hep-ph/9803315} \BibitemShut
  {NoStop}%
\bibitem [{\citenamefont {Dienes}\ \emph {et~al.}(1998)\citenamefont {Dienes},
  \citenamefont {Dudas},\ and\ \citenamefont {Gherghetta}}]{Dienes:1998vh}%
  \BibitemOpen
  \bibfield  {author} {\bibinfo {author} {\bibfnamefont {K.~R.}\ \bibnamefont
  {Dienes}}, \bibinfo {author} {\bibfnamefont {E.}~\bibnamefont {Dudas}}, \
  and\ \bibinfo {author} {\bibfnamefont {T.}~\bibnamefont {Gherghetta}},\
  }\href {\doibase 10.1016/S0370-2693(98)00977-0} {\bibfield  {journal}
  {\bibinfo  {journal} {Phys. Lett. B}\ }\textbf {\bibinfo {volume} {436}},\
  \bibinfo {pages} {55} (\bibinfo {year} {1998})},\ \Eprint
  {http://arxiv.org/abs/hep-ph/9803466} {arXiv:hep-ph/9803466} \BibitemShut
  {NoStop}%
\bibitem [{\citenamefont {Antoniadis}\ \emph {et~al.}(1998)\citenamefont
  {Antoniadis}, \citenamefont {Arkani-Hamed}, \citenamefont {Dimopoulos},\ and\
  \citenamefont {Dvali}}]{Antoniadis:1998ig}%
  \BibitemOpen
  \bibfield  {author} {\bibinfo {author} {\bibfnamefont {I.}~\bibnamefont
  {Antoniadis}}, \bibinfo {author} {\bibfnamefont {N.}~\bibnamefont
  {Arkani-Hamed}}, \bibinfo {author} {\bibfnamefont {S.}~\bibnamefont
  {Dimopoulos}}, \ and\ \bibinfo {author} {\bibfnamefont {G.~R.}\ \bibnamefont
  {Dvali}},\ }\href {\doibase 10.1016/S0370-2693(98)00860-0} {\bibfield
  {journal} {\bibinfo  {journal} {Phys. Lett. B}\ }\textbf {\bibinfo {volume}
  {436}},\ \bibinfo {pages} {257} (\bibinfo {year} {1998})},\ \Eprint
  {http://arxiv.org/abs/hep-ph/9804398} {arXiv:hep-ph/9804398} \BibitemShut
  {NoStop}%
\bibitem [{\citenamefont {Dienes}\ \emph {et~al.}(1999)\citenamefont {Dienes},
  \citenamefont {Dudas},\ and\ \citenamefont {Gherghetta}}]{Dienes:1998vg}%
  \BibitemOpen
  \bibfield  {author} {\bibinfo {author} {\bibfnamefont {K.~R.}\ \bibnamefont
  {Dienes}}, \bibinfo {author} {\bibfnamefont {E.}~\bibnamefont {Dudas}}, \
  and\ \bibinfo {author} {\bibfnamefont {T.}~\bibnamefont {Gherghetta}},\
  }\href {\doibase 10.1016/S0550-3213(98)00669-5} {\bibfield  {journal}
  {\bibinfo  {journal} {Nucl. Phys. B}\ }\textbf {\bibinfo {volume} {537}},\
  \bibinfo {pages} {47} (\bibinfo {year} {1999})},\ \Eprint
  {http://arxiv.org/abs/hep-ph/9806292} {arXiv:hep-ph/9806292} \BibitemShut
  {NoStop}%
\bibitem [{\citenamefont {Arkani-Hamed}\ \emph {et~al.}(1999)\citenamefont
  {Arkani-Hamed}, \citenamefont {Dimopoulos},\ and\ \citenamefont
  {Dvali}}]{Arkani-Hamed:1998sfv}%
  \BibitemOpen
  \bibfield  {author} {\bibinfo {author} {\bibfnamefont {N.}~\bibnamefont
  {Arkani-Hamed}}, \bibinfo {author} {\bibfnamefont {S.}~\bibnamefont
  {Dimopoulos}}, \ and\ \bibinfo {author} {\bibfnamefont {G.~R.}\ \bibnamefont
  {Dvali}},\ }\href {\doibase 10.1103/PhysRevD.59.086004} {\bibfield  {journal}
  {\bibinfo  {journal} {Phys. Rev. D}\ }\textbf {\bibinfo {volume} {59}},\
  \bibinfo {pages} {086004} (\bibinfo {year} {1999})},\ \Eprint
  {http://arxiv.org/abs/hep-ph/9807344} {arXiv:hep-ph/9807344} \BibitemShut
  {NoStop}%
\bibitem [{\citenamefont {Dienes}\ \emph {et~al.}(2003)\citenamefont {Dienes},
  \citenamefont {Dudas},\ and\ \citenamefont {Gherghetta}}]{Dienes:2002bg}%
  \BibitemOpen
  \bibfield  {author} {\bibinfo {author} {\bibfnamefont {K.~R.}\ \bibnamefont
  {Dienes}}, \bibinfo {author} {\bibfnamefont {E.}~\bibnamefont {Dudas}}, \
  and\ \bibinfo {author} {\bibfnamefont {T.}~\bibnamefont {Gherghetta}},\
  }\href {\doibase 10.1103/PhysRevLett.91.061601} {\bibfield  {journal}
  {\bibinfo  {journal} {Phys. Rev. Lett.}\ }\textbf {\bibinfo {volume} {91}},\
  \bibinfo {pages} {061601} (\bibinfo {year} {2003})},\ \Eprint
  {http://arxiv.org/abs/hep-th/0210294} {arXiv:hep-th/0210294} \BibitemShut
  {NoStop}%
\bibitem [{\citenamefont {Abel}\ \emph {et~al.}(2018)\citenamefont {Abel},
  \citenamefont {Dienes},\ and\ \citenamefont {Mavroudi}}]{Abel:2017vos}%
  \BibitemOpen
  \bibfield  {author} {\bibinfo {author} {\bibfnamefont {S.}~\bibnamefont
  {Abel}}, \bibinfo {author} {\bibfnamefont {K.~R.}\ \bibnamefont {Dienes}}, \
  and\ \bibinfo {author} {\bibfnamefont {E.}~\bibnamefont {Mavroudi}},\ }\href
  {\doibase 10.1103/PhysRevD.97.126017} {\bibfield  {journal} {\bibinfo
  {journal} {Phys. Rev. D}\ }\textbf {\bibinfo {volume} {97}},\ \bibinfo
  {pages} {126017} (\bibinfo {year} {2018})},\ \Eprint
  {http://arxiv.org/abs/1712.06894} {arXiv:1712.06894 [hep-ph]} \BibitemShut
  {NoStop}%
\bibitem [{\citenamefont {Hagedorn}(1965)}]{Hagedorn:1965st}%
  \BibitemOpen
  \bibfield  {author} {\bibinfo {author} {\bibfnamefont {R.}~\bibnamefont
  {Hagedorn}},\ }\href@noop {} {\bibfield  {journal} {\bibinfo  {journal}
  {Nuovo Cim. Suppl.}\ }\textbf {\bibinfo {volume} {3}},\ \bibinfo {pages}
  {147} (\bibinfo {year} {1965})}\BibitemShut {NoStop}%
\bibitem [{\citenamefont {Lykken}(1996)}]{Lykken:1996fj}%
  \BibitemOpen
  \bibfield  {author} {\bibinfo {author} {\bibfnamefont {J.~D.}\ \bibnamefont
  {Lykken}},\ }\href {\doibase 10.1103/PhysRevD.54.R3693} {\bibfield  {journal}
  {\bibinfo  {journal} {Phys. Rev. D}\ }\textbf {\bibinfo {volume} {54}},\
  \bibinfo {pages} {R3693} (\bibinfo {year} {1996})},\ \Eprint
  {http://arxiv.org/abs/hep-th/9603133} {arXiv:hep-th/9603133} \BibitemShut
  {NoStop}%
\bibitem [{\citenamefont {Kakushadze}\ and\ \citenamefont
  {Tye}(1999)}]{Kakushadze:1998wp}%
  \BibitemOpen
  \bibfield  {author} {\bibinfo {author} {\bibfnamefont {Z.}~\bibnamefont
  {Kakushadze}}\ and\ \bibinfo {author} {\bibfnamefont {S.~H.~H.}\ \bibnamefont
  {Tye}},\ }\href {\doibase 10.1016/S0550-3213(99)00082-6} {\bibfield
  {journal} {\bibinfo  {journal} {Nucl. Phys. B}\ }\textbf {\bibinfo {volume}
  {548}},\ \bibinfo {pages} {180} (\bibinfo {year} {1999})},\ \Eprint
  {http://arxiv.org/abs/hep-th/9809147} {arXiv:hep-th/9809147} \BibitemShut
  {NoStop}%
\bibitem [{\citenamefont {Abel}\ \emph {et~al.}(2015)\citenamefont {Abel},
  \citenamefont {Dienes},\ and\ \citenamefont {Mavroudi}}]{Abel:2015oxa}%
  \BibitemOpen
  \bibfield  {author} {\bibinfo {author} {\bibfnamefont {S.}~\bibnamefont
  {Abel}}, \bibinfo {author} {\bibfnamefont {K.~R.}\ \bibnamefont {Dienes}}, \
  and\ \bibinfo {author} {\bibfnamefont {E.}~\bibnamefont {Mavroudi}},\ }\href
  {\doibase 10.1103/PhysRevD.91.126014} {\bibfield  {journal} {\bibinfo
  {journal} {Phys. Rev. D}\ }\textbf {\bibinfo {volume} {91}},\ \bibinfo
  {pages} {126014} (\bibinfo {year} {2015})},\ \Eprint
  {http://arxiv.org/abs/1502.03087} {arXiv:1502.03087 [hep-th]} \BibitemShut
  {NoStop}%
\bibitem [{\citenamefont {Maldacena}(1998)}]{Maldacena:1997re}%
  \BibitemOpen
  \bibfield  {author} {\bibinfo {author} {\bibfnamefont {J.~M.}\ \bibnamefont
  {Maldacena}},\ }\href {\doibase 10.1023/A:1026654312961} {\bibfield
  {journal} {\bibinfo  {journal} {Adv. Theor. Math. Phys.}\ }\textbf {\bibinfo
  {volume} {2}},\ \bibinfo {pages} {231} (\bibinfo {year} {1998})},\ \Eprint
  {http://arxiv.org/abs/hep-th/9711200} {arXiv:hep-th/9711200} \BibitemShut
  {NoStop}%
\bibitem [{\citenamefont {Gherghetta}\ and\ \citenamefont
  {Pomarol}(2000)}]{Gherghetta:2000qt}%
  \BibitemOpen
  \bibfield  {author} {\bibinfo {author} {\bibfnamefont {T.}~\bibnamefont
  {Gherghetta}}\ and\ \bibinfo {author} {\bibfnamefont {A.}~\bibnamefont
  {Pomarol}},\ }\href {\doibase 10.1016/S0550-3213(00)00392-8} {\bibfield
  {journal} {\bibinfo  {journal} {Nucl. Phys. B}\ }\textbf {\bibinfo {volume}
  {586}},\ \bibinfo {pages} {141} (\bibinfo {year} {2000})},\ \Eprint
  {http://arxiv.org/abs/hep-ph/0003129} {arXiv:hep-ph/0003129} \BibitemShut
  {NoStop}%
\bibitem [{\citenamefont {Buyukdag}\ \emph {et~al.}(2020)\citenamefont
  {Buyukdag}, \citenamefont {Dienes}, \citenamefont {Gherghetta},\ and\
  \citenamefont {Thomas}}]{Buyukdag:2019lhh}%
  \BibitemOpen
  \bibfield  {author} {\bibinfo {author} {\bibfnamefont {Y.}~\bibnamefont
  {Buyukdag}}, \bibinfo {author} {\bibfnamefont {K.~R.}\ \bibnamefont
  {Dienes}}, \bibinfo {author} {\bibfnamefont {T.}~\bibnamefont {Gherghetta}},
  \ and\ \bibinfo {author} {\bibfnamefont {B.}~\bibnamefont {Thomas}},\ }\href
  {\doibase 10.1103/PhysRevD.101.075054} {\bibfield  {journal} {\bibinfo
  {journal} {Phys. Rev. D}\ }\textbf {\bibinfo {volume} {101}},\ \bibinfo
  {pages} {075054} (\bibinfo {year} {2020})},\ \Eprint
  {http://arxiv.org/abs/1912.10588} {arXiv:1912.10588 [hep-ph]} \BibitemShut
  {NoStop}%
\bibitem [{\citenamefont {Batell}\ and\ \citenamefont
  {Gherghetta}(2007)}]{Batell:2007jv}%
  \BibitemOpen
  \bibfield  {author} {\bibinfo {author} {\bibfnamefont {B.}~\bibnamefont
  {Batell}}\ and\ \bibinfo {author} {\bibfnamefont {T.}~\bibnamefont
  {Gherghetta}},\ }\href {\doibase 10.1103/PhysRevD.76.045017} {\bibfield
  {journal} {\bibinfo  {journal} {Phys. Rev. D}\ }\textbf {\bibinfo {volume}
  {76}},\ \bibinfo {pages} {045017} (\bibinfo {year} {2007})},\ \Eprint
  {http://arxiv.org/abs/0706.0890} {arXiv:0706.0890 [hep-th]} \BibitemShut
  {NoStop}%
\bibitem [{\citenamefont {Zlatev}\ \emph {et~al.}(1999)\citenamefont {Zlatev},
  \citenamefont {Wang},\ and\ \citenamefont {Steinhardt}}]{Zlatev:1998tr}%
  \BibitemOpen
  \bibfield  {author} {\bibinfo {author} {\bibfnamefont {I.}~\bibnamefont
  {Zlatev}}, \bibinfo {author} {\bibfnamefont {L.-M.}\ \bibnamefont {Wang}}, \
  and\ \bibinfo {author} {\bibfnamefont {P.~J.}\ \bibnamefont {Steinhardt}},\
  }\href {\doibase 10.1103/PhysRevLett.82.896} {\bibfield  {journal} {\bibinfo
  {journal} {Phys. Rev. Lett.}\ }\textbf {\bibinfo {volume} {82}},\ \bibinfo
  {pages} {896} (\bibinfo {year} {1999})},\ \Eprint
  {http://arxiv.org/abs/astro-ph/9807002} {arXiv:astro-ph/9807002} \BibitemShut
  {NoStop}%
\bibitem [{\citenamefont {Chimento}\ \emph {et~al.}(2000)\citenamefont
  {Chimento}, \citenamefont {Jakubi},\ and\ \citenamefont
  {Pavon}}]{Chimento:2000kq}%
  \BibitemOpen
  \bibfield  {author} {\bibinfo {author} {\bibfnamefont {L.~P.}\ \bibnamefont
  {Chimento}}, \bibinfo {author} {\bibfnamefont {A.~S.}\ \bibnamefont
  {Jakubi}}, \ and\ \bibinfo {author} {\bibfnamefont {D.}~\bibnamefont
  {Pavon}},\ }\href {\doibase 10.1103/PhysRevD.62.063508} {\bibfield  {journal}
  {\bibinfo  {journal} {Phys. Rev. D}\ }\textbf {\bibinfo {volume} {62}},\
  \bibinfo {pages} {063508} (\bibinfo {year} {2000})},\ \Eprint
  {http://arxiv.org/abs/astro-ph/0005070} {arXiv:astro-ph/0005070} \BibitemShut
  {NoStop}%
\bibitem [{\citenamefont {Copeland}\ \emph {et~al.}(2006)\citenamefont
  {Copeland}, \citenamefont {Sami},\ and\ \citenamefont
  {Tsujikawa}}]{Copeland:2006wr}%
  \BibitemOpen
  \bibfield  {author} {\bibinfo {author} {\bibfnamefont {E.~J.}\ \bibnamefont
  {Copeland}}, \bibinfo {author} {\bibfnamefont {M.}~\bibnamefont {Sami}}, \
  and\ \bibinfo {author} {\bibfnamefont {S.}~\bibnamefont {Tsujikawa}},\ }\href
  {\doibase 10.1142/S021827180600942X} {\bibfield  {journal} {\bibinfo
  {journal} {Int. J. Mod. Phys. D}\ }\textbf {\bibinfo {volume} {15}},\
  \bibinfo {pages} {1753} (\bibinfo {year} {2006})},\ \Eprint
  {http://arxiv.org/abs/hep-th/0603057} {arXiv:hep-th/0603057} \BibitemShut
  {NoStop}%
\bibitem [{\citenamefont {Tsujikawa}(2010)}]{Tsujikawa:2010sc}%
  \BibitemOpen
  \bibfield  {author} {\bibinfo {author} {\bibfnamefont {S.}~\bibnamefont
  {Tsujikawa}},\ }\href {\doibase 10.1007/978-90-481-8685-3-8} {\  (\bibinfo
  {year} {2010}),\ 10.1007/978-90-481-8685-3-8},\ \Eprint
  {http://arxiv.org/abs/1004.1493} {arXiv:1004.1493 [astro-ph.CO]} \BibitemShut
  {NoStop}%
\bibitem [{\citenamefont {Kremer}\ and\ \citenamefont
  {Sobreiro}(2012)}]{Kremer:2011cd}%
  \BibitemOpen
  \bibfield  {author} {\bibinfo {author} {\bibfnamefont {G.~M.}\ \bibnamefont
  {Kremer}}\ and\ \bibinfo {author} {\bibfnamefont {O.~A.~S.}\ \bibnamefont
  {Sobreiro}},\ }\href {\doibase 10.1007/s13538-011-0051-0} {\bibfield
  {journal} {\bibinfo  {journal} {Braz. J. Phys.}\ }\textbf {\bibinfo {volume}
  {42}},\ \bibinfo {pages} {77} (\bibinfo {year} {2012})},\ \Eprint
  {http://arxiv.org/abs/1109.5068} {arXiv:1109.5068 [gr-qc]} \BibitemShut
  {NoStop}%
\bibitem [{\citenamefont {Perez}\ \emph {et~al.}(2014)\citenamefont {Perez},
  \citenamefont {F\"uzfa}, \citenamefont {Carletti}, \citenamefont {M\'elot},\
  and\ \citenamefont {Guedezounme}}]{Perez:2013zya}%
  \BibitemOpen
  \bibfield  {author} {\bibinfo {author} {\bibfnamefont {J.}~\bibnamefont
  {Perez}}, \bibinfo {author} {\bibfnamefont {A.}~\bibnamefont {F\"uzfa}},
  \bibinfo {author} {\bibfnamefont {T.}~\bibnamefont {Carletti}}, \bibinfo
  {author} {\bibfnamefont {L.}~\bibnamefont {M\'elot}}, \ and\ \bibinfo
  {author} {\bibfnamefont {L.}~\bibnamefont {Guedezounme}},\ }\href {\doibase
  10.1007/s10714-014-1753-8} {\bibfield  {journal} {\bibinfo  {journal} {Gen.
  Rel. Grav.}\ }\textbf {\bibinfo {volume} {46}},\ \bibinfo {pages} {1753}
  (\bibinfo {year} {2014})},\ \Eprint {http://arxiv.org/abs/1306.1037}
  {arXiv:1306.1037 [gr-qc]} \BibitemShut {NoStop}%
\bibitem [{\citenamefont {Simon-Petit}\ \emph {et~al.}(2016)\citenamefont
  {Simon-Petit}, \citenamefont {Yap},\ and\ \citenamefont
  {Perez}}]{Simon-Petit:2016tud}%
  \BibitemOpen
  \bibfield  {author} {\bibinfo {author} {\bibfnamefont {A.}~\bibnamefont
  {Simon-Petit}}, \bibinfo {author} {\bibfnamefont {H.-H.}\ \bibnamefont
  {Yap}}, \ and\ \bibinfo {author} {\bibfnamefont {J.}~\bibnamefont {Perez}}\
  }(\bibinfo {year} {2016})\ \Eprint {http://arxiv.org/abs/1603.02267}
  {arXiv:1603.02267 [gr-qc]} \BibitemShut {NoStop}%
\end{thebibliography}%

\end{document}